\documentclass[aps,prb,onecolumn,nofootinbib,citeautoscript,10pt]{revtex4-2}

\usepackage{amsmath,amssymb}
\usepackage{comment}
\usepackage[tight]{subfigure}
\usepackage[dvipsnames]{xcolor}
\usepackage[papersize={8.5in,11in}]{geometry}
\usepackage[colorlinks=true]{hyperref}
\hypersetup{
    bookmarks=true,
    unicode=false,
    pdftoolbar=true,
    pdfmenubar=true,
    pdffitwindow=false,
    pdfstartview={FitH},
    pdfkeywords={plasmons} {Rarita-Schwinger-Weyl semimetal} {Kramers-Weyl semimetal} {random phase approximation},
    pdfnewwindow=true,
    colorlinks=true,
    linkcolor=magenta,
    citecolor=blue,
    filecolor=magenta,
    urlcolor=blue
}
\usepackage{dcolumn}
\usepackage{color}
\usepackage{amssymb,amsmath}
\usepackage{tabularx,graphicx,makecell}
\usepackage{epstopdf}
\usepackage{latexsym}
\usepackage{colortbl}
\usepackage{psfrag}
\usepackage{bbm,bm,array,physics}
\usepackage{dsfont}
\usepackage{float,mathrsfs}
\newcommand{\bs}[1]{\boldsymbol{#1}}
\def\nn{\nonumber \\}
\def\*#1{\boldsymbol{#1}}

\begin{document}

\title{Plasmon modes in three-dimensional multifold semimetals. II. Rarita-Schwinger-Weyl and Kramers-Weyl nodes}

\author{Ipsita Mandal}
\email{ipsita.mandal@snu.edu.in}
\affiliation{Department of Physics, Shiv Nadar Institution of Eminence (SNIoE), Gautam Buddha Nagar, Uttar Pradesh 201314, India}

\begin{abstract}
We investigate collective charge excitations in three-dimensional multifold semimetals in which multiple bands cross the Fermi level at a positive chemical potential, focusing on pseudospin-3/2 Rarita–Schwinger–Weyl semimetals and on Kramers–Weyl semimetals. Both systems host multiple concentric Fermi surfaces at positive chemical potential. Within the random-phase approximation, we analyse the density response, Drude weights, dielectric function, plasmon modes and their Landau damping, and Friedel oscillations, considering both small and arbitrary momentum transfers. In our companion work, \href{https://arxiv.org/abs/2610.18629}{arXiv:2610.18629}, we investigated pseudospin-1 triple-point semimetals, with and without a quadratic correction to their dispersion, as well as Hopf and Weyl semimetals. Comparing these systems allows us to elucidate the distinct roles of band dispersions and eigenstates in shaping collective charge dynamics. Our results demonstrate that collective charge excitations can distinguish multifold nodes with multiple Fermi surfaces and reveal information encoded in their eigenstates beyond that contained in the single-particle dispersion.
\end{abstract}

\maketitle



\section{Introduction}
\label{secint}

Three-dimensional (3d) semimetals have been studied intensively on both the theoretical and the experimental fronts. They host symmetry-protected band-crossing points in the Brillouin zone (BZ) near the Fermi level~\cite{burkov11_Weyl, armitage_review, yan17_topological, bernevig, ips-biref, grushin-multifold, ips-hermann-review, ips-kush-review, ips-internode}. The density-of-states (DOS) vanishes exactly at a nodal point, which distinguishes semimetals from insulators, whose bands are always gapped, and from conventional metals, in which the bands overlap over finite regions of the BZ and the DOS remains finite. The simplest nodal points are the twofold band crossings of the Weyl semimetals (WSMs). Each Weyl node is a singular point of the Berry-curvature (BC) field and carries a BC-monopole~\cite{polash-review, xiao_review}. The monopole charge is the Chern number, and its sign is the chirality $\chi$ of the node, which distinguishes right-handed ($\chi = 1$) from left-handed ($\chi = -1$) quasiparticles. The chiral charges of all nodes in the BZ add up to zero, as required by the Nielsen-Ninomiya theorem~\cite{nielsen81_no}. The nodes therefore occur in groups of vanishing net chirality, the simplest being a pair of opposite chirality.

\begin{figure}[t!]
\centering
\subfigure[]{\includegraphics[width=0.25\linewidth]{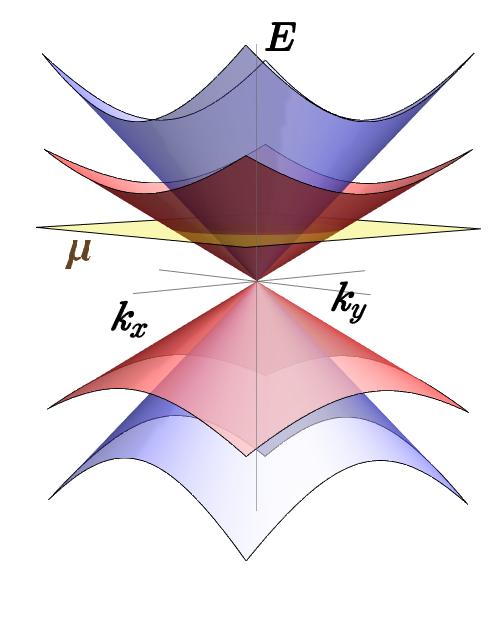}} \hspace{ 2 cm}
\subfigure[]{\includegraphics[width=0.25\linewidth]{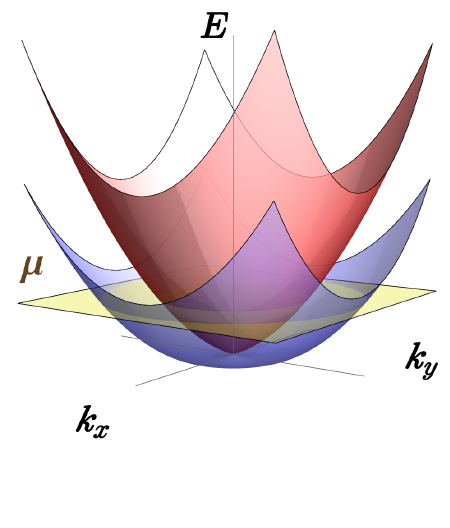}}
\caption{\label{fig:rsw-kw-bands}
Band-dispersions ($E$) over the $k_x\,k_y$-plane for (a) an RSW node and (b) a KW node. The yellow plane marks the chemical potential ($\mu$).}
\end{figure}

Nodes of higher degeneracy are described by a low-energy Hamiltonian $\bs{d} (\bs{k}) \cdot \bs{\mathcal{S}}$, where $\bs{d} (\bs{k}) = \{d_x (\bs{k}), d_y (\bs{k}), d_z (\bs{k})\}$ is a function of the momentum $\bs{k} = \{k_x, k_y, k_z\}$ and $\bs{\mathcal{S}}$ collects the generators of SU(2) in the spin-$\varsigma$ representation. The resulting $(2\,\varsigma + 1)$-fold node is said to carry pseudospin-$\varsigma$~\cite{bernevig, ips-cd1, grushin-multifold, tang2017_multiple}, a terminology that distinguishes this quantum number, which arises from crystal symmetries, from the relativistic spin, which arises from Lorentz invariance. Multifold crossing points have been identified in the 65 chiral space groups, which contain only orientation-preserving symmetries~\cite{grushin-multifold}. For dominant terms linear in momentum, $\bs{d} (\bs{k}) \propto \bs{k}$, and the WSM corresponds to $\varsigma = 1/2$. The case $\varsigma = 3/2$ defines the pseudospin-3/2 Rarita-Schwinger-Weyl (RSW) node~\cite{bernevig, grushin-multifold, ips-exact-rsw, ips-shreya, ips-rsw-ph, ips_jj_rsw}, a fourfold crossing at a high-symmetry point of the BZ of a chiral crystal with a strong spin-orbit coupling (SOC). Examples include materials like AlPt~\cite{schroter2019_chiral}, RhGe \cite{bahadur-RhGe}, RhSi \cite{sanchez} Li$_2$Pd$_3$B \cite{law-LiPdB}, Li$_2$Pt$_3$B \cite{law-LiPdB}, and PtGa \cite{mengyu}. Its four bands, shown in Fig.~\ref{fig:rsw-kw-bands}(a), have linear dispersions and a positive chemical potential cuts the two upper bands.

A distinct type of twofold node is the Kramers–Weyl (KW) node~\cite{law-ksm, chang2018, titus-ksm, ips-exact-kwn}, pinned to a time-reversal-invariant momentum of a chiral crystal in the presence of spin–orbit coupling (SOC)~\cite{bernevig}. When the SOC energy scale is much smaller than the interband separation in the absence of SOC, the KW quasiparticles are described by a two-component spinor in the electron-spin space spanned by the Pauli matrices~\cite{chang2018}, allowing spin-polarisation measurements to probe the Chern numbers of the associated bands directly. Time-reversal symmetry forbids a linear-in-momentum term proportional to the identity matrix, ruling out tilting, which is otherwise allowed for conventional Weyl nodes (and other semimetals) at generic points in the BZ~\cite{yadav23_magneto, staalhammar20_magneto, ips-ruiz, ips-tilted, ips-nr-plasmon, ips-vnr-plasmon}. However, quadratic-in-momentum terms are permitted and can modify the band structure such that, for an appropriate sign, a positive chemical potential intersects both bands [cf. Fig.~\ref{fig:rsw-kw-bands}(b)]. Characteristic signatures of KW nodes have been observed in $\beta$-Ag$_2$Se~\cite{long-ksm}. Here, we consider isotropic KW nodes associated with the cubic point groups $\mathrm{T}$ and $\mathrm{O}$, with candidate-materials including K$_2$Sn$_2$O$_3$~\cite{chang2018}, RhSi and CoSi~\cite{sanchez}, AlPt~\cite{schroter2019_chiral}, and PtGa~\cite{mengyu}.

Plasmons, the collective oscillations of the charge density, probe these band structures. In the doped regime, the long-wavelength plasmon frequency is controlled by the Drude weight, which is fixed by the Fermi surface (FS) and the band-velocities, whereas the damping is governed by the interband continuum, as in graphene~\cite{hwang-dassarma-graphene}. Collective modes have been studied for multi-Weyl~\cite{ahn_mWeyl} and Luttinger semimetals~\cite{ips-plasmon-luttinger-epjb, ips-plasmon-luttinger-aop}, and for nodal-ring semimetals, whose band crossings form closed curves in the BZ~\cite{balents-nodal, fu_nlsm, yang_review_nlsm}, in two~\cite{rahmipoor-2d-nlsm} and three dimensions~\cite{rhim-kim-friedel, ips-nr-plasmon, ips-vnr-plasmon}. For the WSM, Ref.~\cite{ips-nr-plasmon} derived the plasmons of the untilted and type-I tilted nodes at low doping and long wavelengths. They are isotropic without tilt, and the interband screening depends logarithmically on the momentum cut-off of the linearised description. For the Hopf semimetal (HSM), the plasmons are anisotropic although the band structure is isotropic, and the anisotropy was attributed to the dipolar BC~\cite{hopf-plasmon}. The full dielectric function of the HSM was subsequently derived in Ref.~\cite{ips-vnr-plasmon}. For the pseudospin-1 triple-point semimetal (TSM), Ref.~\cite{ips-tsm-plasmon} obtained the plasmons, the screening, and the Friedel oscillations for the linear model with a flat-band and for its quadratic-corrected counterpart (QTSM), together with finite-wavevector extensions of the WSM and HSM results. A corresponding analysis for the RSW and KW nodes is still missing. The RSW node has four bands and two concentric FSs, and the quadratic term of the KW node breaks the particle-hole symmetry and gives rise to two FSs of opposite BC. These features motivate the present study.

In this paper, we study the collective charge excitations of the RSW and KW nodes within the random-phase approximation (RPA). We compute the density response, the Drude weights, the dielectric function, and the plasmon dispersion. For both nodes, we first treat small momentum-transfers and then go beyond the long-wavelength limit, which gives access to the particle-hole continua, the plasmon damping and its critical wavevector, the static screening, and the Friedel oscillations. For the KW node, our expressions at arbitrary momentum-transfers hold for $c\,q < 1$, where $c$ parametrises the quadratic term. The Friedel oscillations are therefore obtained only for the band with the smaller Fermi momentum, since the anomaly of the other band lies outside this range. The quadratic term of the KW node is proportional to the identity matrix. It leaves the eigenstates unchanged and modifies only the FSs and the band-velocities. The KW node thus has the eigenstates of the WSM but a different spectrum, whereas the RSW node and the TSM have the same structure of the Hamiltonian and differ in the spin representation. We compare the RSW node with the KW node, the TSM, the QTSM, and the WSM, and the KW node with the QTSM and the WSM, using the results of Ref.~\cite{ips-tsm-plasmon} for the TSM, the QTSM, and the WSM. The comparison separates the roles of the spectrum and of the eigenstates in the plasmon response.

The notions of low doping and long wavelength require a precise statement for point nodes. A point node has no analogue of the lattice scale of a nodal ring, and its linearised Hamiltonian is limited only by the cut-off $\Lambda$ beyond which the linear dispersion fails. Low doping means that the Fermi momenta of all the bands are much smaller than $\Lambda$, namely $k_F \ll \Lambda$, and the cut-off then enters the response only through logarithms of the ratio of $\Lambda$ to the chemical potential. The momentum exchanged between the probe and the system is the momentum-transfer $\bs{q}$. Small momentum-transfers means that $q$ is much smaller than every Fermi momentum $k_F$. The plasmon frequency is then set by the Drude weight of the bands that cross the Fermi level, and the interband transitions reduce to a background dielectric constant that lowers the bare frequency. The overlap factors can be expanded in powers of $q/k_F$, which brings the quantum metric of the relevant bands to the foreground, and the radial integrals become elementary. Arbitrary momentum-transfers means that $q$ is not restricted to be much smaller than $k_F$, and the full range up to the cut-off is covered. The overlap factors must then be retained exactly, the quantum metric is no longer explicit, and the response is assembled channel by channel. Both regimes rely on the same RPA formalism. For the RSW node, the doped part of the polarisability is independent of $\Lambda$ and depends on $q$ and $\omega$ only through $q/k_F$ and $\omega/\mu$, up to an overall factor. For the KW node, the quadratic term introduces the additional dimensionless parameter $c\,\mu/v_0$. No further condition on the doping is required in either case.

The paper is organised as follows. Sec.~\ref{secmodel} defines the overlap factors, the quantum geometry, and the density-response formalism for generic Bloch bands, and it presents the Hamiltonians of the RSW and KW nodes together with their BC, orbital magnetic moment (OMM), eigenstates, and overlap factors. The RSW node is treated at small and at arbitrary momentum-transfers in Secs.~\ref{secrswlin} and~\ref{secrswfull}, respectively, and the KW node in Secs.~\ref{seckwlin} and~\ref{seckwfull}. Sec.~\ref{seccompare} compares the two nodes with each other and with the TSM, the QTSM, and the WSM, and it collects the results in tables. Sec.~\ref{secsum} contains the summary and outlook, and the appendices elucidate the intermediate steps.

We use natural units and set $\hbar = k_B = e = 1$, but we retain the symbol $e$ in the analytical expressions for book-keeping. The low-energy models are valid for momenta below the lattice cut-off $\Lambda \sim 1/a$ (where $a$ is the lattice constant) and for energies below $v_0\,\Lambda$ (where $v_0$ is proportional to the Fermi velocity). For typical nodal-point semimetals, $\Lambda \sim 150$ to $200\,\text{eV}$~\cite{lv_weyl,spin13d1}. We set $v_0 = 0.0004$ and $\Lambda = 160\,\text{eV}$ for generating all the figures, and these values are not repeated in the figure captions.

\section{Model}
\label{secmodel}

\subsection{Overlap factors, quantum geometry, and density response of Bloch bands}
\label{sec:geometry}

Consider a Bloch Hamiltonian $\mathcal{H} (\bs{k})$ with the eigenvalues $\varepsilon_s (\bs{k})$ and the normalised eigenstates $|s, \bs{k} \rangle$, where $s$ labels the bands. The Hamiltonians of this paper are low-energy continuum models, and $|s, \bs{k} \rangle$ denotes the pseudospin part of the Bloch state, with the plane-wave factor understood. The density response depends on the eigenstates only through the overlap factor
\begin{align}
\label{eq:overlap-def}
\mathcal{F}_{ss'} (\bs{k}, \bs{k}') = |\langle s', \bs{k}' | s, \bs{k} \rangle|^2,
\end{align}
which is the squared modulus of the inner product of two states and measures how much they resemble each other. The phase of each state drops out of Eq.~\eqref{eq:overlap-def}, and the overlap factor is therefore independent of the gauge. By completeness, $\sum_{s'} \mathcal{F}_{ss'} (\bs{k}, \bs{k}') = 1$ for each $s$. For $\bs{k}' = \bs{k}$, the overlap factor reduces to $\delta_{ss'}$, because the states of different bands are orthogonal at the same momentum.

For $\bs{k}' = \bs{k} + \bs{q}$ with $q \ll k$, the overlap factors can be expanded in $\bs{q}$. For $s' \neq s$, the orthogonality of the states at the same momentum gives $\langle s', \bs{k} + \bs{q} | s, \bs{k} \rangle \simeq -\,q_i\,\langle s', \bs{k} | \partial_{k_i} s, \bs{k} \rangle$, and hence
\begin{align}
\label{eq:ff-expansion}
\mathcal{F}_{ss'} (\bs{k}, \bs{k} + \bs{q}) \simeq q_i\,q_j\,Q^{(ss')}_{ij} (\bs{k})\,, \quad
Q^{(ss')}_{ij} (\bs{k}) = \Re \left[ \langle \partial_{k_i} s, \bs{k} | s', \bs{k} \rangle \langle s', \bs{k} | \partial_{k_j} s, \bs{k} \rangle \right], \quad s' \neq s\,,
\end{align}
with a sum over repeated indices. The intraband factor follows from the sum rule as $\mathcal{F}_{ss} \simeq 1 - q_i\,q_j\,Q^{(s)}_{ij}$, where
\begin{align}
\label{eq:qmetric}
Q^{(s)}_{ij} (\bs{k}) = \sum_{s' \neq s} Q^{(ss')}_{ij} (\bs{k}) = \Re \left[ \langle \partial_{k_i} s, \bs{k} | \left( 1 - |s, \bs{k}\rangle \langle s, \bs{k}| \right) | \partial_{k_j} s, \bs{k} \rangle \right].
\end{align}
The second equality follows from completeness. The tensor $Q^{(s)}_{ij}$ is the quantum metric~\cite{provost-vallee, berry} of the band $s$. It is real, symmetric, and positive semidefinite, and it measures the distance between the states at neighbouring momenta. The pair-resolved metric $Q^{(ss')}_{ij}$ is the part of $Q^{(s)}_{ij}$ that connects the band $s$ to the band $s'$, and it fixes the weight of the interband transitions $s \to s'$ at order $q^2$. Equation~\eqref{eq:ff-expansion} is the only input from the eigenstates that the interband part of the response needs at small momentum-transfers, and this part is then proportional to the pair-resolved quantum metric.

The metric is the real part of the quantum geometric tensor,
\begin{align}
\mathcal{T}^{(s)}_{ij} (\bs{k}) = \langle \partial_{k_i} s, \bs{k} | \left( 1 - |s, \bs{k}\rangle \langle s, \bs{k}| \right) | \partial_{k_j} s, \bs{k} \rangle = Q^{(s)}_{ij} (\bs{k}) - \frac{i}{2}\,\Omega^{(s)}_{ij} (\bs{k})\,,
\end{align}
whose imaginary part is the BC $\Omega^{(s)}_{ij} = \partial_{k_i} A^{(s)}_j - \partial_{k_j} A^{(s)}_i$, with the Berry connection $\bs{A}^{(s)} = i \langle s, \bs{k} | \bs{\nabla}_{\bs{k}} | s, \bs{k} \rangle$. In 3d, the BC is conveniently written as the vector $\bs{\Omega}_s = \bs{\nabla}_{\bs{k}} \times \bs{A}^{(s)}$, for which $\Omega^{(s)}_{ij} = \varepsilon_{ijl}\,\Omega_{s,l}$. The metric measures the distance between neighbouring states and enters the overlap factors, and hence the density response. The BC measures the phase that a state acquires around a closed loop in momentum space, and it does not enter the overlap factors, which are insensitive to the phases of the states. Neither the BC nor the OMM therefore appears in the calculations of this paper. We quote them for each node below, because they characterise the topology of the nodes and enter the comparisons of Sec.~\ref{seccompare}.

We now collect the density-response formalism, which is common to all the nodes. At temperature $T = 0$, the noninteracting density-density response function at chemical potential $\mu$ is given by the retarded bare bubble diagram,
\begin{align}
\label{eq:pol}
\Pi(\bs{q},\omega) = g \int \frac{d^3\bs{k}}{(2\,\pi)^3} \sum_{s,s'} \mathcal{F}_{ss'}(\bs{k},\bs{k}')
\frac{f_0(\varepsilon_s(\bs{k})) - f_0(\varepsilon_{s'}(\bs{k}'))}{\omega + \varepsilon_s(\bs{k}) - \varepsilon_{s'}(\bs{k}') + i\,0^+}\,,
\quad \bs{k}' = \bs{k} + \bs{q}\,,
\end{align}
where $g$ is the degeneracy factor, $f_0$ is the Fermi-Dirac distribution, and $\mathcal{F}_{ss'}$ is the overlap factor. The polarisability splits as $\Pi = \Pi_{\rm intra} + \Pi_{\rm inter}$, where the first term collects the contributions with $s = s'$. The imaginary part of the retarded polarisability,
\begin{align}
\Im \Pi (\bs{q}, \omega) \propto \sum_{s,s'} \int d^3 \bs{k}\;\mathcal{F}_{ss'}\left[f_0 (\varepsilon_s) - f_0 (\varepsilon_{s'})\right]\delta\!\left(\omega + \varepsilon_s (\bs{k}) - \varepsilon_{s'} (\bs{k}')\right),
\end{align}
is nonzero precisely when a particle-hole pair with momentum transfer $\bs{q}$ and energy $\omega$ can be created. The set of $(\bs{q}, \omega)$ for which such a pair exists is the particle-hole continuum. The real part of the polarisability determines the frequency of the collective mode. Within the RPA, the dielectric function is
\begin{align}
\label{eq:epsrpa}
\epsilon^{\rm RPA}(\bs{q},\omega) = 1 - \mathcal{V} (q)\,\Pi(\bs{q},\omega)\,, \quad \mathcal{V} (q) = \frac{4\,\pi\,e^2}{q^2}\,,
\end{align}
where $\mathcal{V} (q)$ is the Fourier transform of the 3d Coulomb interaction. For the nodes considered below, we define the dimensionless parameter $\alpha \equiv g\,e^2/v_0$, where $v_0$ is the Fermi velocity. An undamped collective mode is a zero of $\Re \epsilon^{\rm RPA}$ outside the particle-hole continuum, where $\Im \Pi$ vanishes. Inside the continuum, the mode acquires a finite lifetime.

The Drude-weight tensor is
\begin{align}
\label{eq:drude}
D_{ij} = \pi\,g\,e^2 \sum_s \int \frac{d^3\bs{k}}{(2\,\pi)^3}\, \delta(\varepsilon_s - \mu)\, v_i^{(s)}\, v_j^{(s)}\,,
\end{align}
which fixes the Drude peak of the optical conductivity, $\Re\tilde\sigma_{ij}(\omega) = D_{ij}\,\delta(\omega)$. Taking the $q \to 0$ limit and then $\omega \to 0$, the intraband part gives
\begin{align}
\label{eq:intra}
\Pi_{\rm intra}(\bs{q},\omega) \simeq \frac{q_i\,q_j\,D_{ij}}{\pi\,e^2\,\omega^2}\,.
\end{align}
At long wavelengths, $\Re \Pi$ is dominated by the intraband part. Inserting Eq.~\eqref{eq:intra} into $\Re \epsilon^{\rm RPA} = 0$ gives
\begin{align}
\label{eq:plasmon}
\omega_{p,i}^2 = 4\,D_{ii}\,,
\end{align}
with no sum over $i$. The nodes of this paper are isotropic, with $D_{ij} = D\,\delta_{ij}$, and the subscript $i$ can be dropped. The interband part is real and negative below the interband threshold. It acts as a background dielectric constant $\epsilon_b$ and lowers the bare frequency to $\tilde\omega_p^2 = \omega_p^2/\epsilon_b$.

\subsection{Rarita-Schwinger-Weyl semimetal}
\label{sec:rsw}

The effective low-energy Hamiltonian in the vicinity of an RSW node is~\cite{ips-exact-rsw, ips-rsw-ph}
\begin{align}
\label{eq:rsw-ham}
\mathcal{H}_{\rm RSW, \chi} (\bs{k}) = v_0\,\bs{\tilde{k}} \cdot \bs{\mathcal{J}}\,,
\quad
\bs{\tilde{k}} = \left\lbrace k_x, k_y, \chi\,k_z \right\rbrace,
\quad
k = \sqrt{k_x^2 + k_y^2 + k_z^2}\,,
\end{align}
where $\chi = \pm 1$ is the chirality of the node, $v_0$ is the isotropic Fermi velocity, and $\bs{\mathcal{J}} = \{\mathcal{J}_x, \mathcal{J}_y, \mathcal{J}_z\}$ are the spin-$3/2$ matrices. In the basis of the eigenstates of $\mathcal{J}_z$ with the eigenvalues $3/2$, $1/2$, $-1/2$, and $-3/2$, they are given by $\mathcal{J}_x = (\mathcal{J}_+ + \mathcal{J}_-)/2$ and $\mathcal{J}_y = (\mathcal{J}_+ - \mathcal{J}_-)/(2\,i)$, with
\begin{align}
\mathcal{J}_+ = \mathcal{J}_-^\dagger = \begin{pmatrix} 0 & \sqrt{3} & 0 & 0 \\ 0 & 0 & 2 & 0 \\ 0 & 0 & 0 & \sqrt{3} \\ 0 & 0 & 0 & 0 \end{pmatrix}, \quad
\mathcal{J}_z = \operatorname{diag} \left( \frac{3}{2}, \frac{1}{2}, -\frac{1}{2}, -\frac{3}{2} \right).
\end{align}
The eigenvalue and group-velocity of the band $s$ are
\begin{align}
\label{eq:rsw-ev}
\varepsilon_s (\bs{k}) = s\,v_0\,k\,, \quad
\bs{v}_s (\bs{k}) = \bs{\nabla}_{\bs{k}} \varepsilon_s (\bs{k}) = s\,v_0\,\hat{\bs{k}}\,, \quad
s \in \left\lbrace -\frac{3}{2}, -\frac{1}{2}, \frac{1}{2}, \frac{3}{2} \right\rbrace.
\end{align}
The spectrum is symmetric about the nodal energy. For a chemical potential $\mu > 0$, the bands $s = 1/2$ and $s = 3/2$ cross the Fermi level, and the FSs are two concentric spheres with the Fermi momenta
\begin{align}
\label{eq:rsw-kf}
k_F^{(s)} = \frac{\mu}{s\,v_0}\,, \quad k_F^{(1/2)} = \frac{2\,\mu}{v_0}\,, \quad k_F^{(3/2)} = \frac{2\,\mu}{3\,v_0}\,.
\end{align}
The faster band thus has the smaller FS. The bands with $s < 0$ lie below the Fermi level, are fully occupied, and contribute only to interband transitions.

The BC and the OMM of the band $s$ are~\cite{ips-rsw-ph, ips-exact-rsw}
\begin{align}
\label{eq:rsw-bcomm}
\bs{\Omega}_s (\bs{k}) = -\,\frac{\chi\,s}{k^3}\,\bs{k}\,, \quad
\bs{m}_s (\bs{k}) = -\,\frac{\chi\,e\,v_0\,\tilde{\mathcal{G}}_s}{k^2}\,\bs{k}\,, \quad
\tilde{\mathcal{G}}_{\pm 1/2} = \frac{7}{4}\,, \quad \tilde{\mathcal{G}}_{\pm 3/2} = \frac{3}{4}\,.
\end{align}
The flux of $\bs{\Omega}_s$ through a sphere around the node gives the Chern number $\mathcal{C}_s = -\,2\,\chi\,s$ of the band $s$. The two bands below the node add up to $4\,\chi$, which is the Chern number of the node.

The eigenstates are the spin-$3/2$ coherent states, for which $v_0\,\bs{\tilde{k}} \cdot \bs{\mathcal{J}}\,|s, \bs{k} \rangle = s\,v_0\,k\,|s, \bs{k} \rangle$. Let $\vartheta$ and $\varphi$ be the polar and azimuthal angles of $\bs{\tilde{k}}$. In the basis of $\mathcal{J}_z$, the components of the eigenstates are $\langle m | s, \bs{k} \rangle = d^{\,3/2}_{m s} (\vartheta)\,e^{-\,i\,(m + 3/2)\,\varphi}$, where $d^{\,3/2}_{m s}$ are the Wigner small-$d$ matrices. For the bands $s = 3/2$ and $s = 1/2$, the states read
\begin{align}
\label{eq:rsw-states}
|3/2, \bs{k} \rangle = \begin{pmatrix} \cos^3 \!\left( \vartheta/2 \right) e^{-\,3\,i\,\varphi} \\ \sqrt{3}\,\cos^2 \!\left( \vartheta/2 \right) \sin \!\left( \vartheta/2 \right) e^{-\,2\,i\,\varphi} \\ \sqrt{3}\,\cos \!\left( \vartheta/2 \right) \sin^2 \!\left( \vartheta/2 \right) e^{-\,i\,\varphi} \\ \sin^3 \!\left( \vartheta/2 \right) \end{pmatrix}, \quad
|1/2, \bs{k} \rangle = \begin{pmatrix} -\,\sqrt{3}\,\cos^2 \!\left( \vartheta/2 \right) \sin \!\left( \vartheta/2 \right) e^{-\,3\,i\,\varphi} \\ \cos \!\left( \vartheta/2 \right) \left( 3\,\cos \vartheta - 1 \right) e^{-\,2\,i\,\varphi}/2 \\ \sin \!\left( \vartheta/2 \right) \left( 3\,\cos \vartheta + 1 \right) e^{-\,i\,\varphi}/2 \\ \sqrt{3}\,\cos \!\left( \vartheta/2 \right) \sin^2 \!\left( \vartheta/2 \right) \end{pmatrix},
\end{align}
and the states with $s < 0$ follow from the same expression for $d^{\,3/2}_{m s}$. The spectrum and the overlap factors of Eq.~\eqref{eq:overlap-def} are independent of $\chi$, because the latter depend on $\bs{\tilde{k}}$ and $\bs{\tilde{k}}'$ only through $\cos\theta = \hat{\bs{k}} \cdot \hat{\bs{k}}'$. The density response is therefore the same for both chiralities, and $\chi$ enters only through the signs of the BC and the OMM.

The overlap factors are the squared Wigner matrix elements $\mathcal{F}_{ss'} = |d^{\,3/2}_{s s'} (\theta)|^2$, with $\mathcal{F}_{-s,-s'} = \mathcal{F}_{ss'}$ and $\mathcal{F}_{s's} = \mathcal{F}_{ss'}$. Defining $a = (1 + \cos\theta)/2$ and $b = (1 - \cos\theta)/2$, the independent factors are
\begin{align}
\label{eq:rsw-ff}
& \mathcal{F}_{\frac{3}{2}, \frac{3}{2}} = a^3\,, \quad
\mathcal{F}_{\frac{3}{2}, \frac{1}{2}} = 3\,a^2\,b\,, \quad
\mathcal{F}_{\frac{3}{2}, -\frac{1}{2}} = 3\,a\,b^2\,, \quad
\mathcal{F}_{\frac{3}{2}, -\frac{3}{2}} = b^3\,, \nn
& \mathcal{F}_{\frac{1}{2}, \frac{1}{2}} = \frac{a\,\left( 3\,\cos\theta - 1 \right)^2}{4}\,, \quad
\mathcal{F}_{\frac{1}{2}, -\frac{1}{2}} = \frac{b\,\left( 3\,\cos\theta + 1 \right)^2}{4}\,, \quad
\mathcal{F}_{\frac{1}{2}, -\frac{3}{2}} = 3\,a\,b^2\,.
\end{align}
For $\bs{k}' = \bs{k} + \bs{q}$ with $q \ll k$, we have $\sin^2\theta \simeq |\bs{q} \times \hat{\bs{k}}|^2/k^2$, and the overlap factors become
\begin{align}
\label{eq:rsw-qm-ff}
& \mathcal{F}_{\frac{3}{2}, \frac{1}{2}} \simeq q_i\,q_j\,Q^{(3/2, 1/2)}_{ij} (\bs{k})\,, \quad
\mathcal{F}_{\frac{1}{2}, -\frac{1}{2}} \simeq q_i\,q_j\,Q^{(1/2, -1/2)}_{ij} (\bs{k})\,, \nn
& Q^{(3/2, 1/2)}_{ij} = \frac{3}{4}\,\frac{\delta_{ij} - \hat{k}_i\,\hat{k}_j}{k^2}\,, \quad
Q^{(1/2, -1/2)}_{ij} = \frac{\delta_{ij} - \hat{k}_i\,\hat{k}_j}{k^2}\,,
\end{align}
while $\mathcal{F}_{3/2, -1/2} = \mathcal{O} \left( q^4 \right)$, $\mathcal{F}_{3/2, -3/2} = \mathcal{O} \left( q^6 \right)$, and $\mathcal{F}_{ss} = 1 - \mathcal{O} \left( q^2 \right)$. The pair-resolved metrics of the pairs $(-1/2, -3/2)$ and $(-3/2, -1/2)$ follow from $\mathcal{F}_{-s,-s'} = \mathcal{F}_{ss'}$. Three interband pairs, namely $3/2 \leftrightarrow 1/2$, $1/2 \leftrightarrow -1/2$, and $-1/2 \leftrightarrow -3/2$, therefore carry weight at order $q^2$. Equation~\eqref{eq:qmetric} gives the quantum metric of the band $s$ as
\begin{align}
\label{eq:rsw-qmetric}
Q^{(s)}_{ij} (\bs{k}) = \tilde{\mathcal{G}}_s\,\frac{\delta_{ij} - \hat{k}_i\,\hat{k}_j}{k^2}\,,
\end{align}
whose coefficient coincides with that of the OMM in Eq.~\eqref{eq:rsw-bcomm}. The derivation of Eqs.~\eqref{eq:rsw-ff}--\eqref{eq:rsw-qmetric} is provided in Appendix~\ref{app:rsw-states}.

\subsection{Kramers-Weyl semimetal}
\label{sec:kw}

The effective low-energy Hamiltonian in the vicinity of an isotropic KW node, obtained through the $\bs{k} \cdot \bs{p}$ method for the point groups isomorphic to the chiral cubic groups, is~\cite{ips-exact-kwn}
\begin{align}
\label{eq:kw-ham}
\mathcal{H}_{\rm KW, \chi} (\bs{k}) = c\,v_0\,k^2\,\mathbb{I}_{2 \times 2} + v_0\,\bs{\tilde{k}} \cdot \bs{\sigma}\,,
\quad
\bs{\tilde{k}} = \left\lbrace k_x, k_y, \chi\,k_z \right\rbrace,
\end{align}
where $\bs{\sigma} = \{\sigma_x, \sigma_y, \sigma_z\}$ are the Pauli matrices, $v_0$ is the isotropic Fermi velocity of the linear term, and $c = 1/(2\,m\,v_0) > 0$ parametrises the quadratic term, with $m$ being the effective mass. The eigenvalue and group-velocity of the band $s$ are
\begin{align}
\label{eq:kw-ev}
\varepsilon_s (\bs{k}) = s\,v_0\,k + c\,v_0\,k^2\,, \quad
\bs{v}_s (\bs{k}) = v_0 \left( s + 2\,c\,k \right) \hat{\bs{k}}\,, \quad
s \in \lbrace -1, 1 \rbrace.
\end{align}
For $c = 0$, the node reduces to a conventional Weyl node. For $c > 0$, the spectrum is not symmetric about the nodal energy, and a chemical potential $\mu > 0$ cuts both bands. The FSs are two concentric spheres with the Fermi momenta determined by $\varepsilon_s (k_F^{(s)}) = \mu$, that is,
\begin{align}
\label{eq:kw-kf}
k_F^{(s)} = \frac{1}{2\,c} \left( -\,s + \sqrt{1 + \frac{4\,c\,\mu}{v_0}} \right),
\end{align}
where we have kept the physically-admissible root. The result for $s = 1$ reduces to $\mu/v_0$ in the limit $c \to 0$, and the FS of the band $s = -1$ moves to infinity in this limit.

The BC and the OMM of the band $s$ are~\cite{ips-exact-kwn}
\begin{align}
\label{eq:kw-bcomm}
\bs{\Omega}_s (\bs{k}) = -\,\frac{\chi\,s}{2\,k^3}\,\bs{k}\,, \quad
\bs{m}_s (\bs{k}) = -\,\frac{\chi\,e\,v_0}{2\,k^2}\,\bs{k}\,.
\end{align}
The BC is a monopole field with the Chern number $\mathcal{C}_s = -\,\chi\,s$, which is opposite for the two bands, and the OMM is the same for both bands. The quadratic term is proportional to the identity matrix and commutes with the linear term. The eigenstates are therefore those of a Weyl node, namely
\begin{align}
\label{eq:kw-states}
|+, \bs{k} \rangle = \begin{pmatrix} \cos \!\left( \vartheta/2 \right) \\ \sin \!\left( \vartheta/2 \right) e^{i\,\varphi} \end{pmatrix}, \quad
|-, \bs{k} \rangle = \begin{pmatrix} -\,\sin \!\left( \vartheta/2 \right) e^{-\,i\,\varphi} \\ \cos \!\left( \vartheta/2 \right) \end{pmatrix},
\end{align}
where $\vartheta$ and $\varphi$ are the polar and azimuthal angles of $\bs{\tilde{k}}$. As for the RSW node, the spectrum and the overlap factors are independent of $\chi$, which enters only through the signs of the BC and the OMM. The overlap factors of Eq.~\eqref{eq:overlap-def} are
\begin{align}
\label{eq:kw-ff}
\mathcal{F}_{ss'} (\bs{k}, \bs{k}') = \frac{1 + s\,s'\,\cos\theta}{2}\,, \quad \cos\theta = \hat{\bs{k}} \cdot \hat{\bs{k}}'\,.
\end{align}
For $\bs{k}' = \bs{k} + \bs{q}$ with $q \ll k$, we have $\mathcal{F}_{+-} = (1 - \cos\theta)/2 \simeq \sin^2\theta/4$, and Eq.~\eqref{eq:qmetric} gives the quantum metric
\begin{align}
\label{eq:kw-qmetric}
Q^{(\pm)}_{ij} (\bs{k}) = Q^{(+-)}_{ij} (\bs{k}) = \frac{\delta_{ij} - \hat{k}_i\,\hat{k}_j}{4\,k^2}
\end{align}
for both bands. The overlap factors, the quantum metric, and the angular averages of Appendix~\ref{app:kw-states} are independent of $c$, which enters the response only through the energies and the occupations.


\section{RSW at small momentum-transfers}
\label{secrswlin}

In this section, we use the Hamiltonian of Eq.~\eqref{eq:rsw-ham} and restrict to $q \ll k_F$ for every Fermi momentum $k_F$. The response is independent of $\chi$ and isotropic, and we set $\chi = 1$. The spectrum is symmetric under $\varepsilon \to -\,\varepsilon$ together with $s \to -\,s$. It is therefore sufficient to treat $\mu \geq 0$, and the results for $\mu < 0$ follow by replacing $\mu$ with $|\mu|$. The derivations are given in Appendix~\ref{app:rswlin}.

\subsection{Intrinsic limit}
\label{secrswlinintr}

In the limit $\mu \to 0^+$, the FSs shrink to the node, the bands $s = -3/2$ and $s = -1/2$ are filled, and the bands $s = 1/2$ and $s = 3/2$ are empty. The transitions $-1/2 \leftrightarrow -3/2$ and $1/2 \leftrightarrow 3/2$ are therefore blocked. The transitions $-3/2 \leftrightarrow 1/2$, $-1/2 \leftrightarrow 3/2$, and $-3/2 \leftrightarrow 3/2$ are unblocked but carry the weights of order $q^4$ and $q^6$ according to Eq.~\eqref{eq:rsw-qm-ff}. Only the transitions $-1/2 \to 1/2$ with the energy $v_0\,k$ contribute at order $q^2$. The two time orderings of these transitions combine into $1/(\omega - v_0\,k) - 1/(\omega + v_0\,k) = 2\,v_0\,k/(\omega^2 - v_0^2\,k^2)$. For $v_0\,q \ll \omega \ll v_0\,\Lambda$, Eq.~\eqref{eq:pol} then gives
\begin{align}
\label{eq:rswlin-pi0}
\Pi_{\rm inter} (\bs{q}, \omega)
= \frac{g\,q_i\,q_j}{(2\,\pi)^3} \int d\tilde\Omega\; k^2\,Q^{(1/2, -1/2)}_{ij} \int_0^{\Lambda} dk\, \frac{2\,v_0\,k}{(\omega + i\,0^+)^2 - v_0^2\,k^2}
= \frac{g\,q^2}{3\,\pi^2} \int_0^{\Lambda} dk\, \frac{2\,v_0\,k}{(\omega + i\,0^+)^2 - v_0^2\,k^2}\,.
\end{align}
Evaluating the radial integral, we obtain
\begin{align}
\label{eq:rswlin-pi0b}
\Re \Pi (\bs{q}, \omega) = -\,\frac{g\,q^2}{3\,\pi^2\,v_0}
\ln \!\left( \frac{v_0^2\,\Lambda^2}{\omega^2} \right), \quad
\Im \Pi (\bs{q}, \omega) = -\,\frac{g\,q^2}{3\,\pi\,v_0}\,,
\end{align}
which do not depend on the direction of $\bs{q}$. The dielectric function then follows from Eq.~\eqref{eq:epsrpa} as
\begin{align}
\label{eq:rswlin-eps0}
\Re \epsilon^{\rm RPA} (\bs{q}, \omega) = 1 + \frac{4\,g\,e^2}{3\,\pi\,v_0}\,\ln \!\left( \frac{v_0^2\,\Lambda^2}{\omega^2} \right), \quad
\Im \epsilon^{\rm RPA} (\bs{q}, \omega) = \frac{4\,g\,e^2}{3\,v_0}\,.
\end{align}
The real part exceeds unity for $\omega < v_0\,\Lambda$, and the imaginary part is nonzero for all $\omega > 0$. The dielectric function therefore has no zero, and the intrinsic system supports no undamped RPA plasmon. The interband onset vanishes, because the transitions $-1/2 \to 1/2$ are unblocked at every $k > 0$.

\subsection{Doped regime: DOS and Drude weights}
\label{secrswlindrude}

For $\mu > 0$, the bands $s = 1/2$ and $s = 3/2$ cross the Fermi level, and the FSs are the concentric spheres $k = k_F^{(s)} = \mu/(s\,v_0)$ of Eq.~\eqref{eq:rsw-kf}. The DOS, the density, and the Drude weights are
\begin{align}
\label{eq:rswlin-dos}
\rho (\mu) = \frac{112\,g\,\mu^2}{27\,\pi^2\,v_0^3}\,, \quad
\tilde n = \frac{112\,g\,\mu^3}{81\,\pi^2\,v_0^3}\,, \quad
D_{ij} = \pi\,e^2 \sum_{s = 1/2, 3/2} \rho_s\,\langle V^{(s)}_i\,V^{(s)}_j \rangle = \frac{4\,g\,e^2\,\mu^2}{9\,\pi\,v_0}\,\delta_{ij}\,,
\end{align}
where $\rho_s = g\,\mu^2/(2\,\pi^2\,s^3\,v_0^3)$ is the DOS of the band $s$, $\bs{V}^{(s)} = s\,v_0\,\hat{\bs{k}}$, and $\langle \hat k_i\,\hat k_j \rangle = \delta_{ij}/3$. The band $s = 1/2$ with the larger FS contributes three times as much to the Drude weight as the band $s = 3/2$. According to Eq.~\eqref{eq:plasmon}, the bare plasmon frequencies are isotropic and given by
\begin{align}
\label{eq:rswlin-wp}
\omega_p^2 = \frac{16\,g\,e^2\,\mu^2}{9\,\pi\,v_0}\,.
\end{align}
They scale as $\omega_p \propto \mu \propto \tilde n^{1/3}$, as for the WSM and the TSM.

\subsection{Interband screening and plasmons}
\label{secrswlinscreen}

For $\mu > 0$, the Pauli-blocking acts differently on the two transitions that carry weight at order $q^2$. The transitions $-1/2 \to 1/2$ are allowed only for $k > k_F^{(1/2)}$, and the transitions $1/2 \to 3/2$ are allowed only for $k_F^{(3/2)} < k < k_F^{(1/2)}$. The transitions $-3/2 \leftrightarrow -1/2$ remain blocked at every $k$. The lower limit of the radial integral in Eq.~\eqref{eq:rswlin-pi0} therefore becomes $k_F^{(1/2)}$, and the transitions $1/2 \to 3/2$ add a term with the weight $3/4$ instead of $1$ and with the finite range of $k$. After adding the Drude term, we obtain
\begin{align}
\label{eq:rswlin-eps}
\Re \epsilon^{\rm RPA} (\bs{q}, \omega) = {} & 1 - \frac{\omega_p^2}{\omega^2} + \frac{4\,g\,e^2}{3\,\pi\,v_0}\,\ln \!\left( \frac{v_0^2\,\Lambda^2}{|4\,\mu^2 - \omega^2|} \right)
+ \frac{g\,e^2}{\pi\,v_0}\,\ln \!\left( \frac{|4\,\mu^2 - \omega^2|}{|4\,\mu^2/9 - \omega^2|} \right), \nn
\Im \epsilon^{\rm RPA} (\bs{q}, \omega) = {} & \frac{g\,e^2}{v_0}\,\Theta (\omega - \omega_{\rm th})\,\Theta (2\,\mu - \omega) + \frac{4\,g\,e^2}{3\,v_0}\,\Theta (\omega - 2\,\mu)\,.
\end{align}
The lowest interband onset is $\omega_{\rm th} = v_0\,k_F^{(3/2)} = 2\,\mu/3$, which is set by the band with the smaller FS. The transitions $-1/2 \to 1/2$ set the second onset at $2\,\mu$, where the continuum of the transitions $1/2 \to 3/2$ ends. The background dielectric constant, defined by the limit $\omega \to 0$ taken after $q \to 0$, is
\begin{align}
\label{eq:rswlin-eb}
\epsilon_b = 1 + \frac{8\,g\,e^2}{3\,\pi\,v_0}\,\ln \!\left( \frac{v_0\,\Lambda}{2\,\mu} \right) + \frac{2\,g\,e^2}{\pi\,v_0}\,\ln 3\,.
\end{align}
This constant is isotropic, and only its dependence on $\mu$ is independent of the regularisation. The ratio $\omega_p^2/\omega_{\rm th}^2 = 4\,g\,e^2/(\pi\,v_0)$ is independent of $\mu$.

For $\omega < \omega_{\rm th}$, the Drude term $-\,\omega_p^2/\omega^2$ and the interband term of Eq.~\eqref{eq:rswlin-eps} both increase with $\omega$, and $\Re\epsilon^{\rm RPA}$ is therefore monotonic. It tends to $-\,\infty$ for $\omega \to 0$ and to $+\,\infty$ for $\omega \to \omega_{\rm th}^-$, and it hence has a unique zero $\omega_*$ below the lowest onset. Since the interband term is at least $\epsilon_b$, this zero obeys $\omega_*^2 \leq \omega_p^2/\epsilon_b$. The imaginary part vanishes at $\omega_*$, and the plasmon is therefore undamped. It is isotropic, because both the Drude weight and the interband background are.

\section{RSW at arbitrary momentum-transfers}
\label{secrswfull}

In this section, we lift the restriction $q \ll k_F$ of Sec.~\ref{secrswlin}. We write $k_\mu = \mu/v_0$ with $\mu > 0$, which gives the Fermi momenta $k_F^{(s)} = k_\mu/s$ of Eq.~\eqref{eq:rsw-kf}. We denote the frequency in momentum units by $\varpi = \omega/v_0$. The polarisability is then $g/v_0$ times a function of $q/k_\mu$ and $\varpi/k_\mu$, and $\alpha = g\,e^2/v_0$ enters only through Eq.~\eqref{eq:epsrpa}. The cut-off of the linearised description affects only the filled bands, and we implement it as an upper limit $v_0\,\Lambda$ on the transition energy, which reproduces Eq.~\eqref{eq:rswlin-pi0b} at long wavelengths. A different regularisation adds a term proportional to $q^2$ to $\Pi$ and shifts $\epsilon_b$ by a constant, while every nonanalytic feature remains unchanged.

\begin{figure}[t!]
\centering
\subfigure[]{\includegraphics[width=0.75\linewidth]{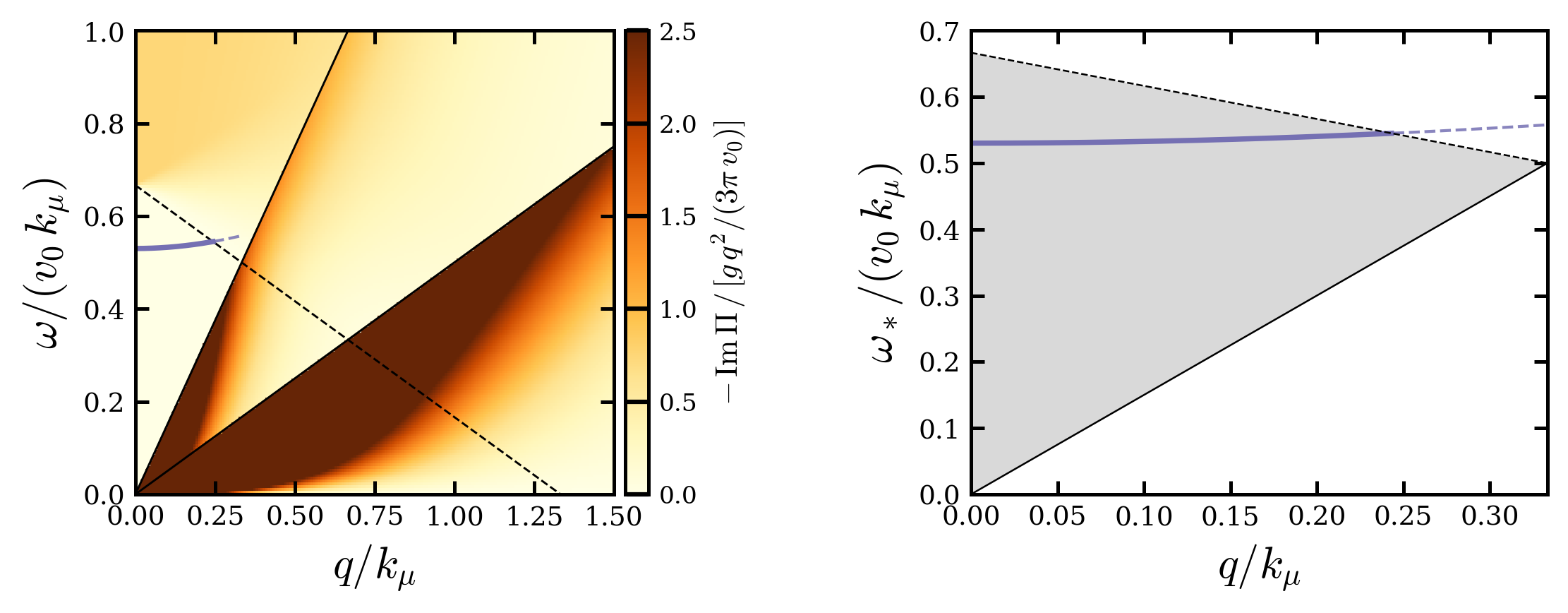}}
\subfigure[]{\includegraphics[width=0.75\linewidth]{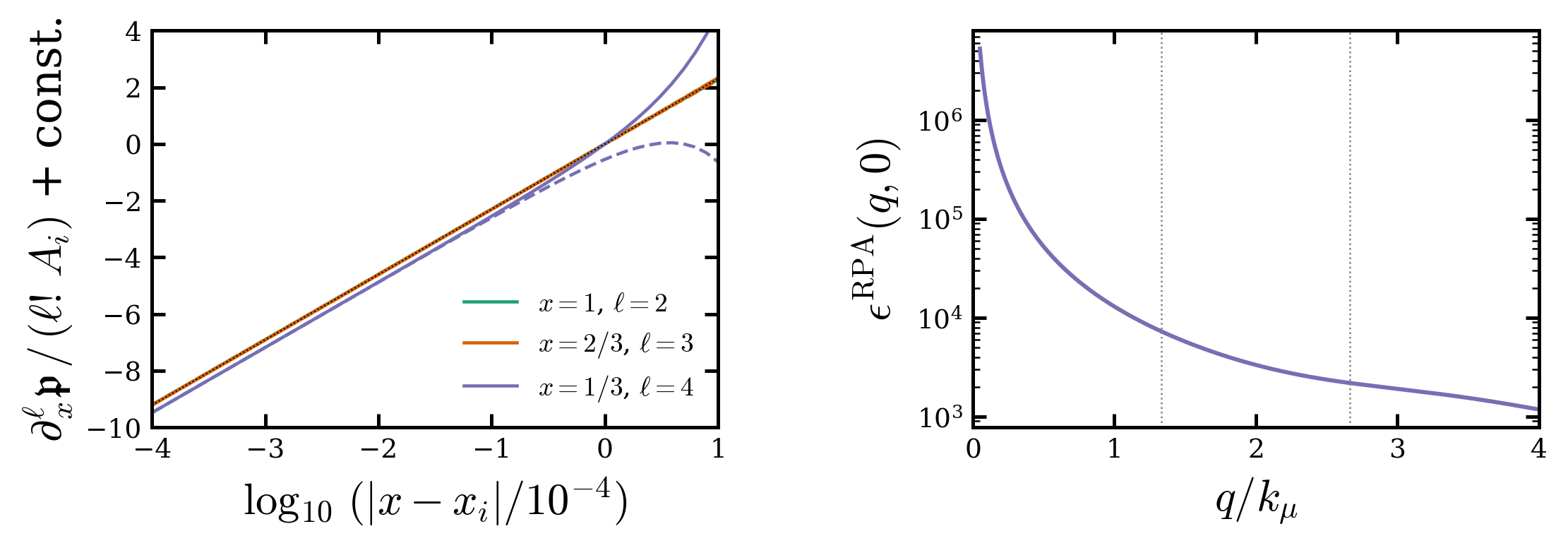}}
\caption{\label{fig:rsw-plasmon-static}
RSW characteristics. (a) Particle-hole continuum and plasmon branch. (b) Derivatives of $\mathfrak{p}$ and static dielectric function.}
\end{figure}

\subsection{Imaginary part of polarisability: Particle-hole continua and their edges}
\label{secrswim}

With $k' = |\bs{k} + \bs{q}|$, the overlap factors of Eq.~\eqref{eq:rsw-ff} depend on $\cos\theta = (k^2 + k'^2 - q^2)/(2\,k\,k')$. In terms of $A = (k + k')^2 - q^2$ and $B = q^2 - (k - k')^2$, we have $a = A/(4\,k\,k')$ and $b = B/(4\,k\,k')$, and every overlap factor takes the form $\mathcal{F}_{ss'} = \mathcal{P}_{ss'}/[64\,(k\,k')^3]$ with the cubic polynomials
\begin{align}
\label{eq:rswim-P}
& \mathcal{P}_{\frac{1}{2}, \frac{1}{2}} = A \left( A - 2\,B \right)^2, \quad
\mathcal{P}_{\frac{3}{2}, \frac{3}{2}} = A^3, \quad
\mathcal{P}_{\frac{1}{2}, \frac{3}{2}} = \mathcal{P}_{\frac{3}{2}, \frac{1}{2}} = 3\,A^2\,B, \nn
& \mathcal{P}_{-\frac{1}{2}, \frac{1}{2}} = B \left( 2\,A - B \right)^2, \quad
\mathcal{P}_{-\frac{3}{2}, \frac{3}{2}} = B^3, \quad
\mathcal{P}_{-\frac{1}{2}, \frac{3}{2}} = \mathcal{P}_{-\frac{3}{2}, \frac{1}{2}} = 3\,A\,B^2\,.
\end{align}
For $\varpi > 0$, only the transitions from an occupied state at $\bs{k}$ to an empty state at $\bs{k}'$ of higher energy contribute. The transitions within the filled bands $s = -3/2$ and $s = -1/2$ are blocked, and eight channels remain. We label them by $\lambda\,s \to s'$ with $s, s' \in \lbrace 1/2, 3/2 \rbrace$, where $\lambda = +$ denotes the four conduction channels and $\lambda = -$ the four valence channels. The conduction channels are the intraband transitions across the FS of each band and the interband transitions $1/2 \leftrightarrow 3/2$. The valence channels are the transitions from the filled bands $-1/2$ and $-3/2$ into the empty states of the bands $1/2$ and $3/2$. The delta function of Eq.~\eqref{eq:pol} fixes $k' = (\varpi + \lambda\,s\,k)/s'$, and the remaining integral over $k$ is elementary. We obtain $\Im \Pi = \sum_{\lambda\,s\,s'} \Im \Pi^{(\lambda\,s \to s')}$ with
\begin{align}
\label{eq:rswim-int}
\Im \Pi^{(\lambda\,s \to s')} = -\,\frac{g}{4\,\pi\,v_0\,q\,s'} \int_{k_-}^{k_+} dk\; k\,k'\,\mathcal{F}_{\lambda\,s,\,s'} = -\,\frac{g}{256\,\pi\,v_0\,q\,s'} \int_{k_-}^{k_+} dk\; \frac{\mathcal{P}_{\lambda\,s,\,s'}}{\left( k\,k' \right)^2}\,.
\end{align}
The limits are fixed by three conditions. The first is the triangle condition $|k - q| \leq k' \leq k + q$. The second is that the initial state is occupied, which reads $s\,k < k_\mu$ for $\lambda = +$ with $k_\mu = \mu/v_0$, and holds automatically for $\lambda = -$. The third is that the final state is empty, which reads $\varpi + \lambda\,s\,k > k_\mu$ because $s'\,k' = \varpi + \lambda\,s\,k$. The channel is nonzero only if these conditions leave a nonempty interval. The resulting limits and the ranges of $\varpi$ are collected in Table~\ref{tab:rswim-ranges}, and the closed form of Eq.~\eqref{eq:rswim-int} is given in Appendix~\ref{app:rswfull-im}. The imaginary part is odd in $\omega$.

\begin{table}[t!]
\centering
\caption{\label{tab:rswim-ranges} Ranges of $\varpi$ and limits $k_\pm$ for each channel.}
\small
\renewcommand{\arraystretch}{1.5}
\begin{tabular}{|c|c|c|c|}
\hline
Channel & Range of $\varpi$ & $k_-$ & $k_+$ \\
\hline
$\frac{1}{2} \to \frac{1}{2}$ & $\max \lbrace 0, \frac{q}{2} - 2\,k_\mu \rbrace < \varpi < \frac{q}{2}$ & $\max \lbrace \frac{q}{2} - \varpi, 2 \left( k_\mu - \varpi \right) \rbrace$ & $2\,k_\mu$ \\
\hline
$\frac{3}{2} \to \frac{3}{2}$ & $\max \lbrace 0, \frac{3\,q}{2} - 2\,k_\mu \rbrace < \varpi < \frac{3\,q}{2}$ & $\max \lbrace \frac{q}{2} - \frac{\varpi}{3}, \frac{2}{3} \left( k_\mu - \varpi \right) \rbrace$ & $\frac{2\,k_\mu}{3}$ \\
\hline
$\frac{1}{2} \to \frac{3}{2}$ & $\max \lbrace 0, \frac{2\,k_\mu}{3} - \frac{q}{2}, \frac{3\,q}{2} - 4\,k_\mu \rbrace < \varpi < 2\,k_\mu + \frac{3\,q}{2}$ & $\max \lbrace \varpi - \frac{3\,q}{2}, \frac{1}{2} \left( \frac{3\,q}{2} - \varpi \right), 2 \left( k_\mu - \varpi \right) \rbrace$ & $\min \lbrace \varpi + \frac{3\,q}{2}, 2\,k_\mu \rbrace$ \\
\hline
$\frac{3}{2} \to \frac{1}{2}$ & $\max \lbrace 0, \frac{q}{2} - \frac{4\,k_\mu}{3} \rbrace < \varpi < \min \lbrace \frac{q}{2}, \frac{3\,q}{2} - 2\,k_\mu \rbrace$ & $\max \lbrace \frac{1}{2} \left( \frac{q}{2} - \varpi \right), \frac{2}{3} \left( k_\mu - \varpi \right) \rbrace$ & $\min \lbrace \frac{q}{2} - \varpi, \frac{2\,k_\mu}{3} \rbrace$ \\
\hline
$-\frac{1}{2} \to \frac{1}{2}$ & $\varpi > \max \lbrace \frac{q}{2}, 2\,k_\mu - \frac{q}{2} \rbrace$ & $\varpi - \frac{q}{2}$ & $\min \lbrace \varpi + \frac{q}{2}, 2 \left( \varpi - k_\mu \right) \rbrace$ \\
\hline
$-\frac{1}{2} \to \frac{3}{2}$ & $\varpi > \max \lbrace \frac{4\,k_\mu}{3} - \frac{q}{2}, \frac{2\,k_\mu}{3} + \frac{q}{2} \rbrace$ & $\max \lbrace \frac{1}{2} \left( \varpi - \frac{3\,q}{2} \right), \frac{3\,q}{2} - \varpi \rbrace$ & $\min \lbrace \frac{1}{2} \left( \varpi + \frac{3\,q}{2} \right), 2 \left( \varpi - k_\mu \right) \rbrace$ \\
\hline
$-\frac{3}{2} \to \frac{1}{2}$ & $\varpi > \max \lbrace \frac{q}{2}, 4\,k_\mu - \frac{3\,q}{2} \rbrace$ & $\frac{1}{2} \left( \varpi - \frac{q}{2} \right)$ & $\min \lbrace \frac{1}{2} \left( \varpi + \frac{q}{2} \right), \varpi - \frac{q}{2}, \frac{2}{3} \left( \varpi - k_\mu \right) \rbrace$ \\
\hline
$-\frac{3}{2} \to \frac{3}{2}$ & $\varpi > \max \lbrace \frac{3\,q}{2}, 2\,k_\mu - \frac{3\,q}{2} \rbrace$ & $\frac{1}{3} \left( \varpi - \frac{3\,q}{2} \right)$ & $\min \lbrace \frac{1}{3} \left( \varpi + \frac{3\,q}{2} \right), \frac{2}{3} \left( \varpi - k_\mu \right) \rbrace$ \\
\hline
\end{tabular}
\end{table}

In the intrinsic limit $k_\mu \to 0$, only the four valence channels survive. For $\varpi > 3\,q/2$, their contributions add up to
\begin{align}
\label{eq:rswim-intr}
\Im \Pi (q, \omega) = -\,\frac{g\,q^2}{3\,\pi\,v_0}\,.
\end{align}
This is Eq.~\eqref{eq:rswlin-pi0b} without any restriction on $q/\varpi$. For $q/2 < \varpi < 3\,q/2$, the imaginary part is $\Im \Pi (q, \omega) = -\,g\,q^2\,\mathfrak{f} (\varpi/q)/(3\,\pi\,v_0)$, with
\begin{align}
\label{eq:rswim-intr2}
\mathfrak{f} (v) = {} & \frac{3136\,v^5 - 4496\,v^4 - 1296\,v^3 + 1776\,v^2 - 81}{288\,v^2}
- \frac{9 \left( 4\,v^2 - 1 \right)^2 \left( 28\,v^2 + 3 \right)}{256\,v^3}\,\ln \!\left( \frac{2 \left( 2\,v - 1 \right)}{2\,v + 1} \right) \nn
& - \frac{\left( 9 - 4\,v^2 \right)^2 \left( 20\,v^2 + 9 \right)}{6912\,v^3}\,\ln \!\left( \frac{2\,v + 3}{2 \left( 3 - 2\,v \right)} \right),
\end{align}
and it vanishes for $\varpi < q/2$. The function $\mathfrak{f}$ jumps from $0$ to $1/4$ at $\varpi = q/2$ and from $11/12$ to $1$ at $\varpi = 3\,q/2$. The two steps have the heights $-\,g\,q^2/(12\,\pi\,v_0)$ and $-\,g\,q^2/(36\,\pi\,v_0)$. They are finite because the overlap factors of the channels $-1/2 \to 1/2$ and $-3/2 \to 3/2$ equal unity when the two states have opposite momenta.

In the doped system, the ranges of Table~\ref{tab:rswim-ranges} define the edges of the continuum in the $(q, \varpi)$ plane. The intraband channels end at the largest intraband energies $s\,v_0\,q$ of the two bands, namely $\varpi = q/2$ and $\varpi = 3\,q/2$. The interband channels $1/2 \to 3/2$ and $-1/2 \to 1/2$ begin at $2\,k_\mu/3 - q/2$ and $2\,k_\mu - q/2$, respectively, which reduce to the thresholds of Sec.~\ref{secrswlin} for $q \to 0$. No occupation constraint is active for $\varpi > 3\,q/2 + 4\,k_\mu$, and the imaginary part has reached its intrinsic value of Eq.~\eqref{eq:rswim-intr} there. Only two regions are free of single-particle excitations, namely,
\begin{align}
\label{eq:rswim-free}
\frac{3\,q}{2} < \varpi < \frac{2\,k_\mu}{3} - \frac{q}{2} \quad \text{for} \quad q < \frac{k_\mu}{3} \text{ and } 
\varpi < \frac{q}{2} - 2\,k_\mu \quad \text{for} \quad q > 4\,k_\mu\,.
\end{align}
The first region lies above the intraband continuum of the faster band and below the lowest interband onset. It hosts the plasmon of Sec.~\ref{secrswlinscreen} at small $q$ and closes at $q = k_\mu/3$. The second region lies below the intraband continuum of the slower band and exists for $q$ above twice the larger Fermi momentum $k_F^{(1/2)}$. The left panel of Fig.~\ref{fig:rsw-plasmon-static}(a) shows $-\,\Im \Pi$ of Eq.~\eqref{eq:rswim-int} against $q/k_\mu$ and $\omega/(v_0\,k_\mu)$, normalised by its intrinsic value. The solid and dashed black lines mark the intraband edges and the lowest interband onset, respectively, and the plasmon branch of Sec.~\ref{secrswplasmon} is drawn in both panels. The colour scale is capped at $2.5$, which saturates the intraband continuum of the slower band. The right panel zooms in on the first free region. The figure uses $\mu = 0.01$\,eV, which gives $k_\mu = 25$\,eV, and the cut-off $\Lambda = 160$\,eV gives $L = \ln \!\left( \Lambda/(2\,k_\mu) \right) \simeq 1.16$.

\subsection{Real part of polarisability and RPA dielectric function}
\label{secrswre}

The real part follows from the imaginary part of Sec.~\ref{secrswim} by causality. The imaginary part reaches its intrinsic value for $\varpi > 3\,q/2 + 4\,k_\mu$, and the difference from the intrinsic reference vanishes beyond this point. With the cut-off $v_0\,\Lambda$ on the transition energy, the dispersion relation therefore reduces to an integral over a finite window,
\begin{align}
\label{eq:rswre-kk}
\Re \Pi (q, \omega) =  -\,\frac{g\,q^2}{3\,\pi^2\,v_0} 
\left[ \ln \!\left( \frac{v_0^2\,\Lambda^2}{\left| \omega^2 - 9\,v_0^2\,q^2/4 \right|} \right) + \mathcal{K} (W) \right] 
 + \frac{2}{\pi}\,\mathrm{P} \!\! \int_0^{\omega_{\rm w}} d\omega'\, 
 \frac{\omega' \left[ \Im \Pi (q, \omega') - \Im \Pi^{(0)} (q, \omega') \right]}{\omega'^2 - \omega^2}\,,
\end{align}
where $\omega_{\rm w} = v_0 \left( 3\,q/2 + 4\,k_\mu \right)$ and $W = \omega/(v_0\,q)$. The intrinsic imaginary part is $\Im \Pi^{(0)} = -\,g\,q^2\,\mathfrak{f}_0 (\omega'/(v_0\,q))/(3\,\pi\,v_0)$, with $\mathfrak{f}_0 = 0$ for $v < 1/2$, $\mathfrak{f}_0 = \mathfrak{f}$ of Eq.~\eqref{eq:rswim-intr2} for $1/2 < v < 3/2$, and $\mathfrak{f}_0 = 1$ for $v > 3/2$. The doping dependence is contained in the window integral, which is evaluated by a single quadrature with the closed forms of Sec.~\ref{secrswim}. The function $\mathcal{K}$ collects the intrinsic part of the dispersion relation that is not a logarithm. It is the finite integral
\begin{align}
\label{eq:rswre-K}
\mathcal{K} (W) = 2\,\mathrm{P} \!\! \int_{1/2}^{3/2} dv\, \frac{v\,\mathfrak{f} (v)}{v^2 - W^2}\,.
\end{align}
The function $\mathcal{K}$ decays as $\mathcal{K} (W) \to -\,2 \left( 1 - 3 \ln 2/5 \right)/W^2$ for $W \to \infty$, and Eq.~\eqref{eq:rswre-kk} then reduces to Eq.~\eqref{eq:rswlin-pi0b} for $\omega \gg v_0\,q$. For $W \to 0$, it tends to $\mathcal{K} (0) = 31 \ln 3/2 - 40 \ln 2/3 - 62/9 \approx 0.898$. The principal value is needed only for $1/2 < W < 3/2$. The steps of $\Im \Pi^{(0)}$ at $\omega = v_0\,q/2$ and $\omega = 3\,v_0\,q/2$, whose heights are given after Eq.~\eqref{eq:rswim-intr2}, produce the logarithmic singularities of $\Re \Pi^{(0)}$ at these two frequencies, $\Re \Pi^{(0)} \sim -\,h\,g\,q^2 \ln \!\left( 1/|\omega^2 - \omega_0^2| \right)/(3\,\pi^2\,v_0)$ with $h = 1/4$ and $\omega_0 = v_0\,q/2$, and with $h = 1/12$ and $\omega_0 = 3\,v_0\,q/2$.

The RPA dielectric function of Eq.~\eqref{eq:epsrpa} follows as
\begin{align}
\label{eq:rswre-eps}
& \Re \epsilon^{\rm RPA} (q, \omega) = 1 + \frac{4\,g\,e^2}{3\,\pi\,v_0} 
\left[ \ln \!\left( \frac{v_0^2\,\Lambda^2}{\left| \omega^2 - 9\,v_0^2\,q^2/4 \right|} \right) + \mathcal{K} (W) \right]  - \frac{8\,e^2}{q^2}\,\mathrm{P} \!\! \int_0^{\omega_{\rm w}} d\omega'\, 
\frac{\omega' \left[ \Im \Pi (q, \omega') - \Im \Pi^{(0)} (q, \omega') \right]}{\omega'^2 - \omega^2}\,, 
\nn &
\Im \epsilon^{\rm RPA} (q, \omega) =  \frac{-\,4\,\pi\,e^2}{q^2}\,\Im \Pi (q, \omega)\,.
\end{align}
For $q \to 0$ at fixed $\omega$, the window integral produces the Drude term and the two interband terms of Eq.~\eqref{eq:rswlin-eps}, and the first line tends to the intrinsic form of Eq.~\eqref{eq:rswlin-eps0}. The derivation is given in Appendix~\ref{app:rswfull-re}.

\subsection{Static polarisability and screening}
\label{secrswstat}

At $\omega = 0$, the polarisability splits as $\Pi = \Pi^{(0)} + \Pi_\mu$, as in Sec.~\ref{secrswre}. The intrinsic part is the first line of Eq.~\eqref{eq:rswre-kk} at $W = 0$, namely
\begin{align}
\label{eq:rswstat-pi0}
\Pi^{(0)} (q, 0) = -\,\frac{g\,q^2}{3\,\pi^2\,v_0} \left[ \ln \!\left( \frac{4\,\Lambda^2}{9\,q^2} \right) + \mathcal{K} (0) \right],
\end{align}
with $\mathcal{K} (0)$ given after Eq.~\eqref{eq:rswre-K}. The doping-dependent part $\Pi_\mu$ contains only the occupied states of the bands $s = 1/2$ and $s = 3/2$, and it does not depend on $\Lambda$. Instead of the window integral of Eq.~\eqref{eq:rswre-kk}, we evaluate it directly from Eq.~\eqref{eq:pol} at $\omega = 0$. The integral over $k' = |\bs{k} + \bs{q}|$ is elementary. After the sum over the final bands, the logarithms of $k'$ alone cancel, and the remaining logarithms have prefactors that are even in $k$. The integral over the Fermi spheres is then also elementary, and no dilogarithm appears, in contrast to the KW node of Eq.~\eqref{eq:kwstat-pi}. The steps are given in Appendix~\ref{app:rswfull-static}. With $x = q/(4\,k_\mu) = q/(2\,k_F^{(1/2)})$, the result for arbitrary $q \ll \Lambda$ is
\begin{align}
\label{eq:rswstat-pi}
\Pi (q, 0) = -\,\frac{g\,q^2}{3\,\pi^2\,v_0} \left[ 2 \ln \!\left( \frac{\Lambda}{2\,k_\mu} \right) + \frac{3}{2} \ln 3 - \frac{38}{9} + \mathfrak{p} (x) \right],
\end{align}
where
\begin{align}
\label{eq:rswstat-p}
\mathfrak{p} (x) = {} & \frac{2}{3\,x^2} - 7 \ln \left| 1 - x^2 \right| - \frac{2}{3} \ln \left| 1 - 9\,x^2 \right| + \frac{20}{3} \ln \left| 1 - \frac{9\,x^2}{4} \right| \nn
& + \mathfrak{a} (x) \ln \left| \frac{1 + x}{1 - x} \right| + \mathfrak{b} (x) \ln \left| \frac{1 + 3\,x}{1 - 3\,x} \right| + \mathfrak{c} (x) \ln \left| \frac{2 + 3\,x}{2 - 3\,x} \right|,
\end{align}
and
\begin{align}
\label{eq:rswstat-abc}
& \mathfrak{a} (x) = \frac{9\,x^6 - 45\,x^4 - 21\,x^2 + 1}{8\,x^3}\,, \quad
\mathfrak{b} (x) = \frac{\left( 9\,x^2 + 1 \right) \left( 81\,x^4 - 90\,x^2 + 1 \right)}{648\,x^3}\,, \nn
& \mathfrak{c} (x) = -\,\frac{729\,x^6 - 2187\,x^4 - 648\,x^2 + 16}{324\,x^3}\,.
\end{align}
The static polarisability is negative for every $q$, because $f_0 (\varepsilon_s (\bs{k})) - f_0 (\varepsilon_{s'} (\bs{k}'))$ and $\varepsilon_s (\bs{k}) - \varepsilon_{s'} (\bs{k}')$ have opposite signs in every term of Eq.~\eqref{eq:pol}. The cut-off enters only through the first term in the square bracket of Eq.~\eqref{eq:rswstat-pi}, which multiplies $q^2$ and shifts the dielectric function by a constant. The static dielectric function follows from Eq.~\eqref{eq:epsrpa} as
\begin{align}
\label{eq:rswstat-eps}
\epsilon^{\rm RPA} (q, 0) = 1 + \frac{4\,\alpha}{3\,\pi} \left[ 2 \ln \!\left( \frac{\Lambda}{2\,k_\mu} \right) + \frac{3}{2} \ln 3 - \frac{38}{9} + \mathfrak{p} (x) \right],
\end{align}
and it exceeds unity.

The function $\mathfrak{p}$ is analytic except at $x = 1/3$, $2/3$, and $1$, which correspond to $q = 2\,k_F^{(3/2)}$, $q = k_F^{(1/2)} + k_F^{(3/2)}$, and $q = 2\,k_F^{(1/2)}$, respectively. At these momentum transfers, the argument of one of the logarithms vanishes. The prefactor of that logarithm vanishes as $(q - q_0)^{\ell}$ with $\ell = 4$, $3$, and $2$, respectively, and the $\ell$th derivative of $\Pi (q, 0)$ therefore diverges logarithmically at $q_0$. The singularity at $q = 2\,k_F^{(1/2)}$ is the strongest, and the three singularities are examined in Sec.~\ref{secrswfried}.

For $q \ll k_\mu$, we expand Eq.~\eqref{eq:rswstat-p} in $x$ and obtain $\mathfrak{p} (x) = 7/(9\,x^2) + 4\,x^2/5 + \mathcal{O} (x^4)$, which gives
\begin{align}
\label{eq:rswstat-small}
\Pi (q, 0) = -\,\rho - \frac{g\,q^2}{3\,\pi^2\,v_0} \left[ 2 \ln \!\left( \frac{\Lambda}{2\,k_\mu} \right) + \frac{3}{2} \ln 3 - \frac{38}{9} \right] - \frac{g\,q^4}{60\,\pi^2\,v_0\,k_\mu^2} + \mathcal{O} \left( q^6 \right),
\end{align}
with the DOS $\rho$ of Eq.~\eqref{eq:rswlin-dos}. The expansion converges for $q < 2\,k_F^{(3/2)} = 4\,k_\mu/3$. The first term is the intraband screening of both FSs. The logarithmic part of the second term is the interband contribution, which reproduces $\epsilon_b$ of Eq.~\eqref{eq:rswlin-eb}, and the constant $-\,38/9$ is the curvature correction of the intraband response. The corresponding static dielectric function is
\begin{align}
\label{eq:rswstat-eps-small}
\epsilon^{\rm RPA} (q, 0) = \frac{q_{\rm TF}^2}{q^2} + \epsilon_b - \frac{152\,\alpha}{27\,\pi} + \frac{\alpha}{15\,\pi}\,\frac{q^2}{k_\mu^2} + \mathcal{O} \left( q^4 \right), \quad
q_{\rm TF}^2 = 4\,\pi\,e^2\,\rho = \frac{448\,\alpha\,k_\mu^2}{27\,\pi}\,.
\end{align}
The Thomas-Fermi wavevector receives contributions from both FSs. The band $s = 1/2$ with the larger FS provides $27/28$ of the DOS, and $q_{\rm TF}$ is independent of $\Lambda$.

For $q \gg k_\mu$, the doping correction to the polarisability decays as $q^{-2}$,
\begin{align}
\label{eq:rswstat-large}
\Pi (q, 0) = \Pi^{(0)} (q, 0) - \frac{6272\,g\,k_\mu^4}{1215\,\pi^2\,v_0\,q^2} + \mathcal{O} \left( \frac{k_\mu^6}{q^4} \right),
\end{align}
and the static dielectric function approaches its intrinsic form,
\begin{align}
\label{eq:rswstat-eps-large}
\epsilon^{\rm RPA} (q, 0) = 1 + \frac{4\,\alpha}{3\,\pi} \left[ \ln \!\left( \frac{4\,\Lambda^2}{9\,q^2} \right) + \mathcal{K} (0) \right] + \frac{25088\,\alpha}{1215\,\pi}\,\frac{k_\mu^4}{q^4} + \mathcal{O} \left( \frac{k_\mu^6}{q^6} \right),
\end{align}
for $k_\mu \ll q \ll \Lambda$. The screening at wavevectors well above the Fermi momenta is therefore the same as in the undoped system. In the opposite limit $\mu \to 0^+$ at fixed $q$, Eq.~\eqref{eq:rswstat-pi} reduces to Eq.~\eqref{eq:rswstat-pi0}.

The static dielectric function of Eq.~\eqref{eq:rswstat-eps} is shown in the right panel of Fig.~\ref{fig:rsw-plasmon-static}(b). It exceeds unity and diverges as $q_{\rm TF}^2/q^2$ for $q \to 0$. The Thomas-Fermi form of Eq.~\eqref{eq:rswstat-eps-small} is drawn as a dashed line for $q \leq k_\mu$. It agrees closely with the full result and is therefore not separately visible. The dotted vertical lines mark the Kohn momenta $q = 4\,k_\mu/3$ and $8\,k_\mu/3$. The third Kohn momentum, $q = 4\,k_\mu$, coincides with the right edge of the plot. The curve and its first derivative remain continuous there, since the singularities appear only in the second and higher derivatives of $\Pi (q, 0)$.


\subsection{Friedel oscillations}
\label{secrswfried}

A static impurity of charge $Z\,e$ that attracts the electrons induces a density $\delta n (\bs{r})$ in the Fermi sea. Within the RPA, its Fourier transform is
\begin{align}
\label{eq:rswfried-dnq}
\delta n (q) = Z \left[ 1 - \frac{1}{\epsilon^{\rm RPA} (q, 0)} \right],
\end{align}
with $\epsilon^{\rm RPA} (q, 0)$ of Eq.~\eqref{eq:rswstat-eps}. It tends to $Z$ for $q \to 0$, which expresses perfect screening. The response is isotropic, and the density in real space is
\begin{align}
\label{eq:rswfried-ft}
\delta n (r) = \frac{1}{2\,\pi^2\,r} \int_0^\infty dq\, q\,\sin (q\,r)\,\delta n (q)\,.
\end{align}
Equation~\eqref{eq:rswstat-small} contains only even powers of $q$, and $\delta n (q)$ is therefore analytic around $q = 0$. It is also analytic at every $q > 0$ except for the three momenta of Sec.~\ref{secrswstat}, namely $q_1 = 2\,k_F^{(1/2)} = 4\,k_\mu$, $q_2 = k_F^{(1/2)} + k_F^{(3/2)} = 8\,k_\mu/3$, and $q_3 = 2\,k_F^{(3/2)} = 4\,k_\mu/3$. For $r \gg 1/k_\mu$, the neighbourhoods of these momenta therefore determine the integral, and the smooth remainder gives no algebraic tail. The derivations are given in Appendix~\ref{app:rswfull-friedel}.

Near $x_i = q_i/(4\,k_\mu)$, the logarithm of Eq.~\eqref{eq:rswstat-p} with the vanishing argument has a prefactor with a zero of order $\ell_i$. The singular part of the function $\mathfrak{p}$ is then
\begin{align}
\label{eq:rswfried-sing}
\mathfrak{p} (x) = \mathfrak{p}_{\rm reg} (x) + A_i\,(x - x_i)^{\ell_i} \ln |x - x_i| + \dots, \quad
\left( x_i, \ell_i, A_i \right) = \left( 1, 2, -\,\frac{3}{2} \right), \left( \frac{2}{3}, 3, \frac{27}{4} \right), \left( \frac{1}{3}, 4, -\,\frac{81}{4} \right),
\end{align}
for $i = 1, 2, 3$, respectively. Here $\mathfrak{p}_{\rm reg}$ is analytic at $x_i$, and the dots denote terms of higher order in $x - x_i$. The exponents are fixed by the overlap factors at backscattering. For $\bs{k}' = -\,\bs{k}$, we have $a = 0$ and $b = 1$, and Eq.~\eqref{eq:rsw-ff} gives $\mathcal{F}_{\frac{1}{2}, \frac{1}{2}} \simeq 4\,a$, $\mathcal{F}_{\frac{1}{2}, \frac{3}{2}} \simeq 3\,a^2$, and $\mathcal{F}_{\frac{3}{2}, \frac{3}{2}} = a^3$. These channels are responsible for the singularities at $q_1$, $q_2$, and $q_3$, respectively. In each case, $\ell_i$ exceeds by one the power of $a$ with which the overlap factor vanishes, which gives the familiar value $\ell = 1$ for a parabolic band, whose overlap factor equals unity.

The left panel of Fig.~\ref{fig:rsw-plasmon-static}(b) tests Eq.~\eqref{eq:rswfried-sing}. Near $x_i$, the $\ell_i$th derivative of $\mathfrak{p}$ equals $\ell_i!\,A_i \ln |x - x_i|$ plus a finite function. After division by $\ell_i!\,A_i$ and a constant shift, the derivatives for $\ell_i = 2$, $3$, and $4$ should therefore collapse onto a line of unit slope in $\ln |x - x_i|$ on both sides of $x_i$. The figure shows them against $\log_{10} \left( |x - x_i|/10^{-4} \right)$, with solid and dashed lines for the two sides of $x_i$ and the unit-slope line dotted. The derivatives for $\ell_i = 2$ and $3$ follow this line over the whole range shown, whereas that for $\ell_i = 4$ departs from it for $|x - x_i| \gtrsim 10^{-4}$, where higher-order terms become visible.

At the three momenta, the dielectric function is finite, and we denote $\epsilon_i = \epsilon^{\rm RPA} (q_i, 0)$. It follows from Eq.~\eqref{eq:rswstat-eps} with
\begin{align}
\label{eq:rswfried-pvals}
& \mathfrak{p} \left( \frac{1}{3} \right) = 6 + \frac{55}{2} \ln 3 - 42 \ln 2 \simeq 7.0997\,, \quad
\mathfrak{p} \left( \frac{2}{3} \right) = \frac{3}{2} + \frac{815}{64} \ln 3 + \frac{40}{3} \ln 2 - \frac{2675}{192} \ln 5 \simeq 2.3089\,, \nn
& \mathfrak{p} (1) = \frac{2}{3} + \frac{2125}{162} \ln 5 - \frac{2386}{81} \ln 2 \simeq 1.3602\,.
\end{align}
The logarithm with the vanishing argument does not contribute to these values, and the cut-off enters $\epsilon_i$ only through the term $2 \ln \left( \Lambda/(2\,k_\mu) \right)$ of Eq.~\eqref{eq:rswstat-eps}.

The power-law tails follow from the singular parts of Eq.~\eqref{eq:rswfried-dnq}. Near $q_i$, the induced density behaves as $\delta n (q) \simeq C_i\,(q - q_i)^{\ell_i} \ln |q - q_i|$ with
\begin{align}
\label{eq:rswfried-C}
C_i = \frac{4\,Z\,\alpha}{3\,\pi\,\epsilon_i^2}\,\frac{A_i}{\left( 4\,k_\mu \right)^{\ell_i}}\,.
\end{align}
The Fourier transform of $t^{\ell} \ln |t|$ is proportional to $r^{-\ell-1}$ for large $r$. Inserting this into Eq.~\eqref{eq:rswfried-ft}, we find
\begin{align}
\label{eq:rswfried-dn}
\delta n (r) \simeq {} & -\,\frac{Z\,\alpha}{2\,\pi^2\,\epsilon_1^2}\,\frac{\sin (4\,k_\mu\,r)}{k_\mu\,r^4}
+ \frac{9\,Z\,\alpha}{8\,\pi^2\,\epsilon_2^2}\,\frac{\cos (8\,k_\mu\,r/3)}{k_\mu^2\,r^5}
+ \frac{27\,Z\,\alpha}{16\,\pi^2\,\epsilon_3^2}\,\frac{\sin (4\,k_\mu\,r/3)}{k_\mu^3\,r^6}\,,
\end{align}
for $r \gg 1/k_\mu$, with corrections that are smaller by the factors of $1/(q_i\,r)$.

The slowest decay comes from the backscattering across the larger FS, and it dominates the density at large distances. It is one power of $r$ faster than the $r^{-3}$ of a parabolic band in three dimensions, because the overlap factor suppresses backscattering within the band $s = 1/2$. The terms with odd $\ell_i$ are proportional to the cosine and those with even $\ell_i$ to the sine, owing to the factor $i^{\ell_i}$ in the transform. The intraband singularity of the faster band has the weakest effect, since its overlap factor $a^3$ is the most strongly suppressed. Figure~\ref{fig:friedel-rsw-kw} plots the density of Eq.~\eqref{eq:rswfried-ft} together with the asymptotic form of Eq.~\eqref{eq:rswfried-dn}, and Sec.~\ref{seccompfried} compares the other nodes.

\subsection{Plasmon dispersion and Landau damping}
\label{secrswplasmon}

We consider $q \to 0$ and the first free region of Eq.~\eqref{eq:rswim-free}, which lies between the intraband edge $\varpi = 3\,q/2$ of the faster band and the lowest interband onset $\varpi = 2\,k_\mu/3 - q/2$. The onset tends to $\omega_{\rm th}/v_0$ of Sec.~\ref{secrswlinscreen} for $q \to 0$. The polarisability contains only even powers of $q$ there, $\Pi = q^2\,\Pi_2 + q^4\,\Pi_4 + \dots$, as the steps of Appendix~\ref{app:rswfull-wdisp} show. The term $\Pi_2$ is fixed by the dielectric function of Eq.~\eqref{eq:rswlin-eps}, whose real part $\epsilon_0 (\omega) = 1 - 4\,\pi\,e^2\,\Pi_2$ consists of the Drude term and the interband screening. The new ingredient is $\Pi_4$. With $\varpi = \omega/v_0$, it is given by
\begin{align}
\label{eq:rswplas-pi4}
\Pi_4 = \frac{g}{240\,\pi^2\,v_0}\,\tilde{\mathcal{P}} (\varpi)\,,
\end{align}
where
\begin{align}
\label{eq:rswplas-P}
\tilde{\mathcal{P}} (\varpi) = {} & \frac{48\,k_\mu^2}{\varpi^4} - \frac{40}{\varpi^2} + \frac{23}{\varpi^2 - 4\,k_\mu^2} + \frac{184\,k_\mu^2}{\left( \varpi^2 - 4\,k_\mu^2 \right)^2} - \frac{27}{9\,\varpi^2 - 4\,k_\mu^2} - \frac{216\,k_\mu^2}{\left( 9\,\varpi^2 - 4\,k_\mu^2 \right)^2} \nn
& + \frac{24}{\varpi^2}\,\ln \!\left( \frac{\left( 16\,k_\mu^2 - \varpi^2 \right) \left( 16\,k_\mu^2 - 9\,\varpi^2 \right)}{16 \left( 4\,k_\mu^2 - \varpi^2 \right) \left( 4\,k_\mu^2 - 9\,\varpi^2 \right)} \right).
\end{align}
The arguments of all the logarithms are positive for $\varpi < 2\,k_\mu/3$. No cut-off enters, in contrast to $\Pi_2$. The first two terms of $\tilde{\mathcal{P}}$ originate from the intraband transitions of the two bands, and the remaining terms originate from the interband transitions. The expansion requires that $\omega$ is far from the edges of the free region on the scale of $v_0\,q$. The intermediate steps are given in Appendix~\ref{app:rswfull-wdisp}.

Within the RPA, $\Re\epsilon^{\rm RPA} = \epsilon_0 - 4\,\pi\,e^2\,q^2\,\Pi_4 + \mathcal{O} \left( q^4 \right)$, with $4\,\pi\,e^2\,g/(240\,\pi^2\,v_0) = \alpha/(60\,\pi)$. We define
\begin{align}
\label{eq:rswplas-D}
\mathcal{D} (\omega) = \partial_{\omega^2} \epsilon_0 = \frac{\omega_p^2}{\omega^4} + \frac{\alpha}{3\,\pi \left( 4\,\mu^2 - \omega^2 \right)} + \frac{\alpha}{\pi \left( 4\,\mu^2/9 - \omega^2 \right)}\,.
\end{align}
All three terms are positive for $\omega < \omega_{\rm th}$, and hence $\mathcal{D} > 0$ in the free region. The zero of $\Re\epsilon^{\rm RPA}$ obeys $\epsilon_0 = \alpha\,q^2\,\tilde{\mathcal{P}}/(60\,\pi)$, and it is an analytic function of $q^2$ near $q = 0$. To order $q^2$, it is
\begin{align}
\label{eq:rswplas-wdisp}
\omega_*^2 (q) = \omega_*^2 (0) + \frac{\alpha\,q^2}{60\,\pi}\,\frac{\tilde{\mathcal{P}} (\omega_*/v_0)}{\mathcal{D} (\omega_*)} \Bigg|_{\omega_* = \omega_* (0)},
\end{align}
where $\omega_* (0)$ is the unique zero below the lowest onset found in Sec.~\ref{secrswlinscreen}. The plasmon lies inside the free region for small $q$ and is undamped.

For $\alpha \to 0$, we have $\omega_*^2 (0) \simeq \omega_p^2 \ll \omega_{\rm th}^2$ and $\mathcal{D} \to \omega_p^2/\omega_*^4$. The first term of $\tilde{\mathcal{P}}$ dominates, since the other terms are smaller by the factor $\omega_p^2/\mu^2 \propto \alpha$, and Eq.~\eqref{eq:rswplas-wdisp} reduces to
\begin{align}
\label{eq:rswplas-bg}
\omega_*^2 (q) = \omega_p^2 + \frac{9}{20}\,v_0^2\,q^2\,.
\end{align}
This is the Bohm-Gross dispersion~\cite{bohm-gross}. The coefficient equals $3/5$ times the ratio $\sum_s \rho_s\,V_s^4/\sum_s \rho_s\,V_s^2$, where $\rho_s \propto s^{-3}$ and $V_s = s\,v_0$ are the DOS and the Fermi speed of the band $s$. This ratio is $3\,v_0^2/4$, which is smaller than the value $v_0^2$ for the WSM, because the slower band $s = 1/2$ carries three times the Drude weight of the band $s = 3/2$. The corrections to Eq.~\eqref{eq:rswplas-bg} are of relative order $\alpha$.

For larger $q$, the zero leaves the free region through the intraband edge $\varpi = 3\,q/2$ or through the onset $\varpi = 2\,k_\mu/3 - q/2$, whichever it reaches first, at the wavevector $q_{\rm max}$. It is found by solving $\Re\epsilon^{\rm RPA} = 0$ on the relevant boundary, with $\Re \Pi$ of Eq.~\eqref{eq:rswre-kk}. We write $\epsilon_0 = 1 + \alpha\,\tilde{\mathcal{F}} (\omega)$ with $\tilde{\mathcal{F}}$ independent of $\alpha$. By Appendix~\ref{app:rswlin-screen}, $\tilde{\mathcal{F}}$ increases strictly from $-\,\infty$ to $+\,\infty$ on $(0, \omega_{\rm th})$. The condition $\tilde{\mathcal{F}} (\omega_* (0)) = -\,1/\alpha$ then shows that $\omega_* (0)$ increases with $\alpha$ and stays below the unique zero $\omega_\infty < \omega_{\rm th}$ of $\tilde{\mathcal{F}}$. The plasmon therefore never starts at the onset, and for $\alpha \to 0$ it starts far below it, because $\omega_p^2/\omega_{\rm th}^2 = 4\,\alpha/\pi$. Since the free region closes at $q = k_\mu/3$, we have $q_{\rm max} \leq k_\mu/3$. The expansion in $q$ is not controlled near the edges of the free region, and Eq.~\eqref{eq:rswplas-wdisp} is therefore not used to locate $q_{\rm max}$.

For $q > q_{\rm max}$, the zero of $\Re\epsilon^{\rm RPA}$ lies inside the continuum, where $\Im\epsilon^{\rm RPA} = -\left( 4\,\pi\,e^2/q^2 \right) \Im \Pi \neq 0$, and the mode is Landau damped with the rate $\gamma = \Im\epsilon^{\rm RPA}/\partial_\omega \Re\epsilon^{\rm RPA}$ at $\omega = \Re\,\omega_*$, provided that $\gamma$ is small compared with the scales on which $\Im\epsilon^{\rm RPA}$ and $\Re\epsilon^{\rm RPA}$ vary. We do not give $\gamma$ in closed form, because $\partial_\omega \Re\epsilon^{\rm RPA}$ at the zero requires $\Re \Pi$ at finite $q$ close to the edges of the continuum, where it is available only through Eq.~\eqref{eq:rswre-kk}. For $q < k_\mu/3$, Table~\ref{tab:rswim-ranges} shows the origin of the damping. Only the intraband transitions of the faster band $s = 3/2$ contribute for $q/2 < \varpi < 3\,q/2$, those of the band $s = 1/2$ are added for $\varpi < q/2$, and the interband transitions $1/2 \to 3/2$ set in at $\varpi = 2\,k_\mu/3 - q/2$. For $q \ll k_\mu$ and $q/2 < \varpi < 3\,q/2$, we obtain
\begin{align}
\label{eq:rswplas-imeps}
\Im\epsilon^{\rm RPA} (q, \omega) = \frac{16\,\alpha}{81}\,\frac{k_\mu^2\,\varpi}{q^3}\,,
\end{align}
to leading order in $q/k_\mu$ at fixed $\varpi/q$, which is the Landau damping of a 3d Fermi gas with the Fermi speed $3\,v_0/2$ and the DOS of the band $s = 3/2$. For $k_\mu/3 < q < 4\,k_\mu$, the continua overlap, and the whole range $\varpi > 0$ is damped, as shown in Sec.~\ref{secrswim}.

The plasmon branch of Eq.~\eqref{eq:rswplas-wdisp} is shown in Fig.~\ref{fig:rsw-plasmon-static}(a). It rises with $q$ from $\omega_* (0)$, which lies below the lowest onset, and leaves the window through the onset $\omega = v_0 \left( 2\,k_\mu/3 - q/2 \right)$ for the parameters of the figure. It is drawn as a solid line inside the free window in both panels. The continuation of the $q^2$ formula beyond the window is drawn as a dashed line and serves only as a guide to the eye, because the zero of $\Re\epsilon^{\rm RPA}$ lies inside the continuum there and the mode is damped.


\section{KW at small momentum-transfers}
\label{seckwlin}

In this section, we use the Hamiltonian of Eq.~\eqref{eq:kw-ham} with $c > 0$ and restrict to $q \ll k_F$ for every Fermi momentum $k_F$. The response is independent of $\chi$ and isotropic, and we set $\chi = 1$. The spectrum is not symmetric under $\varepsilon \to -\,\varepsilon$, and we consider $\mu \geq 0$, for which both bands cross the Fermi level. It is convenient to introduce
\begin{align}
\label{eq:kwlin-R}
\mathcal{R} = \sqrt{1 + \frac{4\,c\,\mu}{v_0}}\,, \quad
k_F^{(1)} = \frac{\mathcal{R} - 1}{2\,c}\,, \quad
k_F^{(-1)} = \frac{\mathcal{R} + 1}{2\,c}\,,
\end{align}
which satisfy $k_F^{(-1)} - k_F^{(1)} = 1/c$ for every $\mu$. Both FSs have the same Fermi speed $v_0\,\mathcal{R}$, since $v_s (k_F^{(s)}) = v_0 \left( s + 2\,c\,k_F^{(s)} \right) = v_0\,\mathcal{R}$. The cut-off $\Lambda$ must satisfy $k_F^{(-1)} \ll \Lambda$, and it does not enter the results below. The derivations are given in Appendix~\ref{app:kwlin}.

\subsection{Intrinsic limit}
\label{seckwlinintr}

In the limit $\mu \to 0^+$, we have $k_F^{(1)} \to 0$ and $k_F^{(-1)} \to 1/c$. The band $s = 1$ is empty, and the band $s = -1$ is filled for $k < 1/c$, where its energy $v_0\,k\,(c\,k - 1)$ is negative. In contrast to the RSW node, the limit is therefore not an undoped semimetal, because the band $s = -1$ retains an FS at $k = 1/c$. The Drude weight of this FS is considered in Sec.~\ref{seckwlindrude}. Here we collect the interband response. The transitions $-1 \to 1$ are unblocked for $0 < k < 1/c$, and their energy $\varepsilon_1 - \varepsilon_{-1} = 2\,v_0\,k$ is independent of $c$. By Eq.~\eqref{eq:kw-qmetric}, they are the only transitions in the model, and they carry the weight $q_i\,q_j\,Q^{(+-)}_{ij}$ at order $q^2$. The two time orderings combine into $1/(\omega - 2\,v_0\,k) - 1/(\omega + 2\,v_0\,k) = 4\,v_0\,k/(\omega^2 - 4\,v_0^2\,k^2)$. For $v_0\,q \ll \omega$, Eq.~\eqref{eq:pol} then gives
\begin{align}
\label{eq:kwlin-pi0}
\Pi_{\rm inter} (\bs{q}, \omega)
= \frac{g\,q_i\,q_j}{(2\,\pi)^3} \int d\tilde\Omega\; k^2\,Q^{(+-)}_{ij} 
\int_0^{1/c} dk\, \frac{4\,v_0\,k}{(\omega + i\,0^+)^2 - 4\,v_0^2\,k^2}
= \frac{g\,q^2}{12\,\pi^2} \int_0^{1/c} dk\, \frac{4\,v_0\,k}{(\omega + i\,0^+)^2 - 4\,v_0^2\,k^2}\,.
\end{align}
Evaluating the radial integral, we obtain
\begin{align}
\label{eq:kwlin-pi0b}
\Re \Pi_{\rm inter} (\bs{q}, \omega) = -\,\frac{g\,q^2}{24\,\pi^2\,v_0}
\ln \!\left| \frac{4\,v_0^2} {c^2 \, \omega^2} -1 \right|, \quad
\Im \Pi_{\rm inter} (\bs{q}, \omega) = -\,\frac{g\,q^2}{24\,\pi\,v_0}\,
\Theta \!\left( 2\,v_0/c - \omega \right),
\end{align}
and the interband part of the dielectric function follows from Eq.~\eqref{eq:epsrpa} as
\begin{align}
\label{eq:kwlin-eps0}
\Re \epsilon^{\rm RPA}_{\rm inter} (\bs{q}, \omega) = 1 + \frac{g\,e^2}{6\,\pi\,v_0}
\ln \!\left| \frac{4\,v_0^2} {c^2 \, \omega^2} -1 \right|, \quad
\Im \epsilon^{\rm RPA}_{\rm inter} (\bs{q}, \omega) =
 \frac{g\,e^2}{6\,v_0}\,\Theta \!\left(  2\,v_0/c - \omega \right).
\end{align}
The scale $2\,v_0/c$ takes the role that the cut-off plays for the Weyl node, and the logarithmic divergence at small $\omega$ is the same. The imaginary part is nonzero for every $0 < \omega < 2\,v_0/c$, and no undamped mode exists in this range.

\subsection{Doped regime: DOS and Drude weights}
\label{seckwlindrude}

For $\mu > 0$, the FSs are the concentric spheres $k = k_F^{(1)}$ and $k = k_F^{(-1)}$ of Eq.~\eqref{eq:kwlin-R}, which are both electron-like. Because the Fermi speeds are equal, the DOS of the band $s$ is $\rho_s = g \left( k_F^{(s)} \right)^2/(2\,\pi^2\,v_0\,\mathcal{R})$. The DOS, the density measured from $\mu = 0$, and the Drude weights are
\begin{align}
\label{eq:kwlin-dos}
\rho (\mu) = \frac{g \left( \mathcal{R}^2 + 1 \right)}{4\,\pi^2\,c^2\,v_0\,\mathcal{R}}\,, \quad
\tilde n = \frac{g \left[ \mathcal{R} \left( \mathcal{R}^2 + 3 \right) - 4 \right]}{24\,\pi^2\,c^3}\,, \quad
D_{ij} = \frac{g\,e^2\,v_0\,\mathcal{R} \left( \mathcal{R}^2 + 1 \right)}{12\,\pi\,c^2}\,\delta_{ij}\,.
\end{align}
The contributions of the two bands to $D_{ij}$ are in the ratio $(\mathcal{R} - 1)^2 : (\mathcal{R} + 1)^2$, and the band $s = -1$ with the larger FS dominates. According to Eq.~\eqref{eq:plasmon}, the bare plasmon frequencies are isotropic and given by
\begin{align}
\label{eq:kwlin-wp}
\omega_p^2 = \frac{g\,e^2\,v_0\,\mathcal{R} \left( \mathcal{R}^2 + 1 \right)}{3\,\pi\,c^2}\,.
\end{align}
For $\mu \to 0^+$, we have $\mathcal{R} \to 1$, and $\omega_p^2$ tends to the finite value $2\,g\,e^2\,v_0/(3\,\pi\,c^2)$, which is contributed by the FS at $k = 1/c$. For $c\,\mu/v_0 \gg 1$, the dispersion is dominated by the quadratic term, and Eq.~\eqref{eq:kwlin-wp} reduces to $\omega_p^2 = 8\,\pi\,e^2\,c\,v_0\,\tilde n$ with $\omega_p \propto \mu^{3/4} \propto \tilde n^{1/2}$, which is the plasma frequency of a parabolic band with the effective mass $1/(2\,c\,v_0)$. The limit $c \to 0$ at fixed $\mu$ is not smooth, because the FS of the band $s = -1$ moves to infinity, and the WSM result is recovered only after discarding this band.

\subsection{Interband screening and plasmons}
\label{seckwlinscreen}

For $\mu > 0$, the transitions $-1 \to 1$ are allowed only in the window $k_F^{(1)} < k < k_F^{(-1)}$, where the band $s = -1$ is filled and the band $s = 1$ is empty. For $k < k_F^{(1)}$, both bands are filled, and for $k > k_F^{(-1)}$, both are empty. The limits of the radial integral in Eq.~\eqref{eq:kwlin-pi0} are therefore replaced by $k_F^{(1)}$ and $k_F^{(-1)}$. The window of the transition energies has the lower edge $\omega_{\rm th}$ and the upper edge $\omega_{\rm max}$,
\begin{align}
\label{eq:kwlin-onsets}
\omega_{\rm th} = 2\,v_0\,k_F^{(1)} = \frac{v_0 \left( \mathcal{R} - 1 \right)}{c}\,, \quad
\omega_{\rm max} = 2\,v_0\,k_F^{(-1)} = \frac{v_0 \left( \mathcal{R} + 1 \right)}{c}\,,
\end{align}
with the fixed width $\omega_{\rm max} - \omega_{\rm th} = 2\,v_0/c$. After adding the Drude term, we obtain
\begin{align}
\label{eq:kwlin-eps}
\Re \epsilon^{\rm RPA} (\bs{q}, \omega) = 1 - \frac{\omega_p^2}{\omega^2} + \frac{g\,e^2}{6\,\pi\,v_0}\,\ln \!\left( \frac{\left| \omega^2 - \omega_{\rm max}^2 \right|}{\left| \omega^2 - \omega_{\rm th}^2 \right|} \right), \quad
\Im \epsilon^{\rm RPA} (\bs{q}, \omega) = \frac{g\,e^2}{6\,v_0}\,\Theta (\omega - \omega_{\rm th})\,\Theta (\omega_{\rm max} - \omega)\,.
\end{align}
The background dielectric constant, defined by the limit $\omega \to 0$ taken after $q \to 0$, is
\begin{align}
\label{eq:kwlin-eb}
\epsilon_b = 1 + \frac{g\,e^2}{3\,\pi\,v_0}\,\ln \!\left( \frac{\mathcal{R} + 1}{\mathcal{R} - 1} \right).
\end{align}
It is isotropic and independent of $\Lambda$, and it diverges logarithmically for $\mu \to 0^+$. In contrast to the RSW node, the ratio $\omega_p^2/\omega_{\rm th}^2$ depends on $c\,\mu/v_0$.

Since the continuum occupies only the finite window $(\omega_{\rm th}, \omega_{\rm max})$, the real part of the dielectric function diverges logarithmically at each edge, and two undamped modes exist at long wavelengths. For $\omega < \omega_{\rm th}$, the Drude term and the interband term both increase with $\omega$, and the real part tends to $-\,\infty$ for $\omega \to 0$ and to $+\,\infty$ for $\omega \to \omega_{\rm th}^-$. It hence has a unique zero $\omega_*$ below the onset, which obeys $\omega_*^2 \leq \omega_p^2/\epsilon_b$, since the interband term is at least $\epsilon_b$. For $\omega > \omega_{\rm max}$, the real part increases monotonically from $-\,\infty$ at $\omega_{\rm max}^+$ to $1$ for $\omega \to \infty$, which gives a second unique zero $\omega_*'$. The logarithm is negative in this range, and $\omega_*' > \omega_p$ therefore holds in addition to $\omega_*' > \omega_{\rm max}$. The imaginary part vanishes at both zeros, and both plasmons are undamped. The upper mode lies within the validity of the linearised description only for $\omega_*' \ll v_0\,\Lambda$. Both modes are isotropic, because the Drude weight and the interband background are.

\section{KW at arbitrary momentum-transfers}
\label{seckwfull}

In this section, we lift the restriction $q \ll k_F$ of Sec.~\ref{seckwlin} for the KW node, with $c > 0$, $\mu > 0$, and $\varpi = \omega/v_0$. The eigenstates do not depend on $c$. The overlap factors of Eq.~\eqref{eq:kw-ff} therefore remain functions of $\cos\theta = (k^2 + k'^2 - q^2)/(2\,k\,k')$ alone, and the quadratic term enters only through the energies. We use the Fermi momenta $k_1 = k_F^{(1)}$ and $k_2 = k_F^{(-1)}$ and the quantity $\mathcal{R}$ of Eq.~\eqref{eq:kwlin-R}, which obey $k_2 - k_1 = 1/c$ and $k_1 + k_2 = \mathcal{R}/c$. We also define $\Sigma = k + k'$ and $\check{D} = k' - k$ with $k' = |\bs{k} + \bs{q}|$. These satisfy $|\check{D}| \leq q \leq \Sigma$. The occupations are $n_1 (\bs{k}) = \Theta (k_1 - k)$ and $n_{-1} (\bs{k}) = \Theta (k_2 - k)$. In contrast to the TSM, the filled states of both bands are bounded in momentum space. No cut-off enters the response, and the polarisability has the prefactor $g/v_0$ and depends on $q$, $\varpi$, $c$, and $k_1$ only.

The transition energies are
\begin{align}
\label{eq:kwfull-X}
\varepsilon_s (k') - \varepsilon_s (k) = v_0\,\check{D} \left( s + c\,\Sigma \right), \quad
\varepsilon_1 (k') - \varepsilon_{-1} (k) = v_0\,\Sigma \left( 1 + c\,\check{D} \right),
\end{align}
for the intraband transitions of the band $s = \pm 1$ and for the transitions $-1 \to 1$, respectively. The transitions from the filled states of the band $s = 1$ into the empty states of the band $s = -1$ require $k' - k > k_2 - k_1 = 1/c$ and are therefore absent for $q < 1/c$. Three channels remain, namely the intraband transitions of each of the two bands and the transitions $-1 \to 1$. We assume $c\,q < 1$ in this section, which suffices for the real part and the static response. The imaginary part of Sec.~\ref{seckwim} is derived for $2\,c\,q < 1$, which guarantees that the energies of Eq.~\eqref{eq:kwfull-X} are monotonic on the relevant domains and fixes the continuum onset for $q \geq k_1$, as shown in Appendix~\ref{app:kwfull-im}.

\subsection{Imaginary part of polarisability: Particle-hole continua and their edges}
\label{seckwim}

\subsubsection{Continuum edges}
\label{seckwedges}

The intraband transitions of the two bands have a common upper edge. For the band $s$, the largest energy at fixed $q$ belongs to the collinear transition $k \to k + q$ from the Fermi momentum $k_s$. Since $s + 2\,c\,k_s = \mathcal{R}$ for both bands, the energy is $v_0\,q \left( \mathcal{R} + c\,q \right)$ in both cases. The lower edge of the band $s = -1$ is zero for every $q$, because $q < 2\,k_2$ allows $k = k' = k_2$. For the band $s = 1$, it is zero for $q \leq 2\,k_1$ and nonzero beyond. The edges are
\begin{align}
\label{eq:kwim-edges}
\varpi_+ (q) = q \left( \mathcal{R} + c\,q \right), \quad
\varpi_-^{(1)} (q) = \left( q - 2\,k_1 \right) \left( 1 + c\,q \right) \Theta \left( q - 2\,k_1 \right), \quad
\varpi_-^{(-1)} (q) = 0\,.
\end{align}
For $q \to 0$, the upper edge is tangent to the line $\varpi = \mathcal{R}\,q$, which is the line $\omega = V_F\,q$ with the Fermi velocity $V_F = v_0\,\mathcal{R}$ of both bands.

The transitions $-1 \to 1$ connect a filled state at $k < k_2$ to an empty state at $k' > k_1$. At fixed $\check{D}$, the energy of Eq.~\eqref{eq:kwfull-X} increases with $\Sigma$, and the allowed range of $\Sigma$ is $\max \lbrace q, 2\,k_1 - \check{D} \rbrace < \Sigma < 2\,k_2 + \check{D}$. The continuum is therefore bounded from above, and its onset and upper edge are
\begin{align}
\label{eq:kwim-th}
\varpi_{\rm th} (q) = \begin{cases}
\min \left\lbrace \left( 2\,k_1 - q \right) \left( 1 + c\,q \right), \left( 2\,k_1 + q \right) \left( 1 - c\,q \right) \right\rbrace & \text{ for } q \leq k_1 \\
q \left( \mathcal{R} - c\,q \right) & \text{ for } q \geq k_1
\end{cases}, \quad
\varpi_{\rm max} (q) = \left( 2\,k_2 + q \right) \left( 1 + c\,q \right).
\end{align}
The first entry of the minimum applies for $\mathcal{R} \leq 2$ and the second for $\mathcal{R} \geq 2$, because the difference of the two entries is $2\,q \left( 2 - \mathcal{R} \right)$. For $\mathcal{R} > 2$, we have $k_1 > 1/(2\,c) > q$, and only the first line of Eq.~\eqref{eq:kwim-th} is relevant. For $q \to 0$, we obtain $\varpi_{\rm th} \to 2\,k_1$ and $\varpi_{\rm max} \to 2\,k_2$, which are the edges $\omega_{\rm th}/v_0$ and $\omega_{\rm max}/v_0$ of Eq.~\eqref{eq:kwlin-onsets}.

The intraband continuum of the band $s = -1$ reaches $\varpi \to 0$ for every $q$, which leaves no region free of excitations at low frequencies, in contrast to the TSM at $q > 2\,k_F$. Two regions are free of single-particle excitations. The first lies above the intraband continua and below the interband onset, $\varpi_+ (q) < \varpi < \varpi_{\rm th} (q)$. It hosts the lower plasmon of Sec.~\ref{seckwlinscreen}. The window closes at the wavevector where $\varpi_+ = \varpi_{\rm th}$, which is
\begin{align}
\label{eq:kwim-qc}
q_c = \begin{cases}
\dfrac{\sqrt{2\,\mathcal{R} - 1} - 1}{2\,c} & \text{ for } \mathcal{R} \leq 2 \\[8pt]
\dfrac{\sqrt{\mathcal{R}^2 - 1} - \left( \mathcal{R} - 1 \right)}{2\,c} & \text{ for } \mathcal{R} \geq 2
\end{cases}.
\end{align}
For $q > q_c$, the interband continuum overlaps the intraband continua, and the region $0 < \varpi < \varpi_{\rm max}$ is covered without a gap. The second free region lies above the interband continuum, $\varpi > \varpi_{\rm max} (q)$, and it contains the upper zero of Sec.~\ref{seckwlinscreen} at long wavelengths. It exists for every $q$ because the interband continuum has an upper edge. The intermediate steps are given in Appendix~\ref{app:kwfull-im}.

\subsubsection{Spectral weight of individual channels}
\label{seckwchan}

For $\varpi > 0$, the delta function of the imaginary part removes one integration variable in each channel, and the remaining integrand is a rational function. We define $\mathfrak{r}_\pm = \sqrt{\mathcal{R}^2 \pm 4\,c\,\varpi}$ and the primitives
\begin{align}
\label{eq:kwim-G}
& \mathcal{G}_+ (\Sigma) = \frac{\Sigma^2}{2\,c} - \frac{\Sigma}{c^2} + \frac{1 - c^2\,q^2}{c^3}\,\ln \!\left( 1 + c\,\Sigma \right), \quad
\mathcal{G}_- (\Sigma) = \frac{\Sigma^2}{2\,c} + \frac{\Sigma}{c^2} + \frac{1 - c^2\,q^2}{c^3}\,\ln \!\left( c\,\Sigma - 1 \right), \nn
& \mathcal{G}_0 (\check{D}) = -\,\frac{\check{D}^2}{2\,c} + \frac{\check{D}}{c^2} - \frac{1 - c^2\,q^2}{c^3}\,\ln \!\left( 1 + c\,\check{D} \right),
\end{align}
which are the antiderivatives of $\left( \Sigma^2 - q^2 \right)/\left( 1 + c\,\Sigma \right)$, $\left( \Sigma^2 - q^2 \right)/\left( c\,\Sigma - 1 \right)$, and $\left( q^2 - \check{D}^2 \right)/\left( 1 + c\,\check{D} \right)$, respectively. The intraband channels are
\begin{align}
\label{eq:kwim-intra}
& \Im \Pi_{1,1} (q, \omega) = -\,\frac{g}{32\,\pi\,v_0\,q} \left[ \mathcal{G}_+ \left( \Sigma_+^{(1)} \right) - \mathcal{G}_+ \left( \Sigma_-^{(1)} \right) \right], \nn
& \Sigma_+^{(1)} = \frac{\mathcal{R} - 2 + \mathfrak{r}_+}{2\,c}\,, \quad
\Sigma_-^{(1)} = \max \left\lbrace q, 2\,k_1 - q, \frac{\varpi - q}{c\,q}, \frac{\mathcal{R} - 2 + \mathfrak{r}_-}{2\,c} \right\rbrace, \nn
& \Im \Pi_{-1,-1} (q, \omega) = -\,\frac{g}{32\,\pi\,v_0\,q} \left[ \mathcal{G}_- \left( \Sigma_+^{(-1)} \right) - \mathcal{G}_- \left( \Sigma_-^{(-1)} \right) \right], \nn
& \Sigma_+^{(-1)} = \frac{\mathcal{R} + 2 + \mathfrak{r}_+}{2\,c}\,, \quad
\Sigma_-^{(-1)} = \max \left\lbrace 2\,k_2 - q, \frac{q + \varpi}{c\,q}, \frac{\mathcal{R} + 2 + \mathfrak{r}_-}{2\,c} \right\rbrace.
\end{align}
Each contribution vanishes for $\varpi \geq \varpi_+$ and whenever the upper limit does not exceed the lower limit, which reproduces $\varpi_-^{(1)}$ of Eq.~\eqref{eq:kwim-edges}. The last entry of each maximum is absent for $\varpi > \mathcal{R}^2/(4\,c)$, where $\mathfrak{r}_-$ is imaginary. The transitions $-1 \to 1$ contribute
\begin{align}
\label{eq:kwim-inter}
& \Im \Pi_{-1,1} (q, \omega) = -\,\frac{g}{32\,\pi\,v_0\,q} \Big\lbrace \left[ \mathcal{G}_0 \left( \min \lbrace \check{D}_b, \check{D}_- \rbrace \right) - \mathcal{G}_0 \left( \check{D}_a \right) \right] \Theta \left( \min \lbrace \check{D}_b, \check{D}_- \rbrace - \check{D}_a \right) \nn
& \hspace{3.0cm} + \left[ \mathcal{G}_0 \left( \check{D}_b \right) - \mathcal{G}_0 \left( \max \lbrace \check{D}_a, \check{D}_+ \rbrace \right) \right] \Theta \left( \check{D}_b - \max \lbrace \check{D}_a, \check{D}_+ \rbrace \right) \Big\rbrace, \nn
& \check{D}_a = \max \left\lbrace -\,q, \frac{-\,\mathcal{R} - 2 + \mathfrak{r}_+}{2\,c} \right\rbrace, \quad
\check{D}_b = \min \left\lbrace q, \frac{\varpi - q}{c\,q} \right\rbrace, \quad
\check{D}_\mp = \frac{\mathcal{R} - 2 \mp \mathfrak{r}_-}{2\,c}\,,
\end{align}
for $\varpi < \mathcal{R}^2/(4\,c)$. For $\varpi > \mathcal{R}^2/(4\,c)$, the excluded interval between $\check{D}_-$ and $\check{D}_+$ is absent, and only the combination $\mathcal{G}_0 (\check{D}_b) - \mathcal{G}_0 (\check{D}_a)$ with the Heaviside function $\Theta (\check{D}_b - \check{D}_a)$ remains. The interval $(\check{D}_-, \check{D}_+)$ is excluded because the final states of the transitions with these values of $\check{D}$ lie in the filled region of the band $s = 1$ and are Pauli-blocked, as shown in Appendix~\ref{app:kwfull-im}. The contribution vanishes for $\varpi < \varpi_{\rm th}$ and for $\varpi > \varpi_{\rm max}$.

\subsubsection{Total spectral function and absorption}
\label{seckwimtot}

For $\omega > 0$, the imaginary part is the sum of the three channels,
\begin{align}
\label{eq:kwim-tot}
\Im \Pi (q, \omega) = \Im \Pi_{1,1} + \Im \Pi_{-1,-1} + \Im \Pi_{-1,1}\,,
\end{align}
and it is odd in $\omega$. The imaginary part of the dielectric function is $\Im \epsilon^{\rm RPA} = -\,\mathcal{V} (q)\,\Im \Pi$, which equals $\alpha/(8\,q^3)$ times the sum of the differences of the primitives in Eqs.~\eqref{eq:kwim-intra} and~\eqref{eq:kwim-inter}. For $q \to 0$ at fixed $\varpi/q < \mathcal{R}$, the intraband channels reduce to
\begin{align}
\label{eq:kwim-landau}
\Im \Pi_{s,s} (q, \omega) \to -\,\frac{g\,k_s^2\,\varpi}{4\,\pi\,v_0\,\mathcal{R}^2\,q}\,\Theta \left( \mathcal{R}\,q - \varpi \right) = -\,\frac{\pi\,\rho_s\,\omega}{2\,V_F\,q}\,\Theta \left( V_F\,q - \omega \right),
\end{align}
with $\rho_s = g\,k_s^2/(2\,\pi^2\,v_0\,\mathcal{R})$ and $V_F = v_0\,\mathcal{R}$, which is the Landau damping of a 3d Fermi gas with the Fermi velocity $V_F$. For the transitions $-1 \to 1$, we obtain $\Im \Pi_{-1,1} \to -\,g\,q^2/(24\,\pi\,v_0)$ for $2\,k_1 < \varpi < 2\,k_2$ and zero otherwise, in agreement with Eq.~\eqref{eq:kwlin-pi0b}. At finite $q$, the absorption of the transitions $-1 \to 1$ is not constant between the edges, and it is bounded by its value for the full range of $\check{D}$,
\begin{align}
\label{eq:kwim-plateau}
\Im \Pi_{-1,1}^{\rm full} = -\,\frac{g}{32\,\pi\,v_0\,q} \left[ \frac{2\,q}{c^2} - \frac{1 - c^2\,q^2}{c^3}\,\ln \!\left( \frac{1 + c\,q}{1 - c\,q} \right) \right],
\end{align}
which is the value of Eq.~\eqref{eq:kwim-inter} when none of the constraints on $\check{D}$ is active. For $c\,q \to 0$, it reduces to $-\,g\,q^2/(24\,\pi\,v_0)$.

\begin{figure}[t!]
\centering
\subfigure[]{\includegraphics[width=0.75\linewidth]{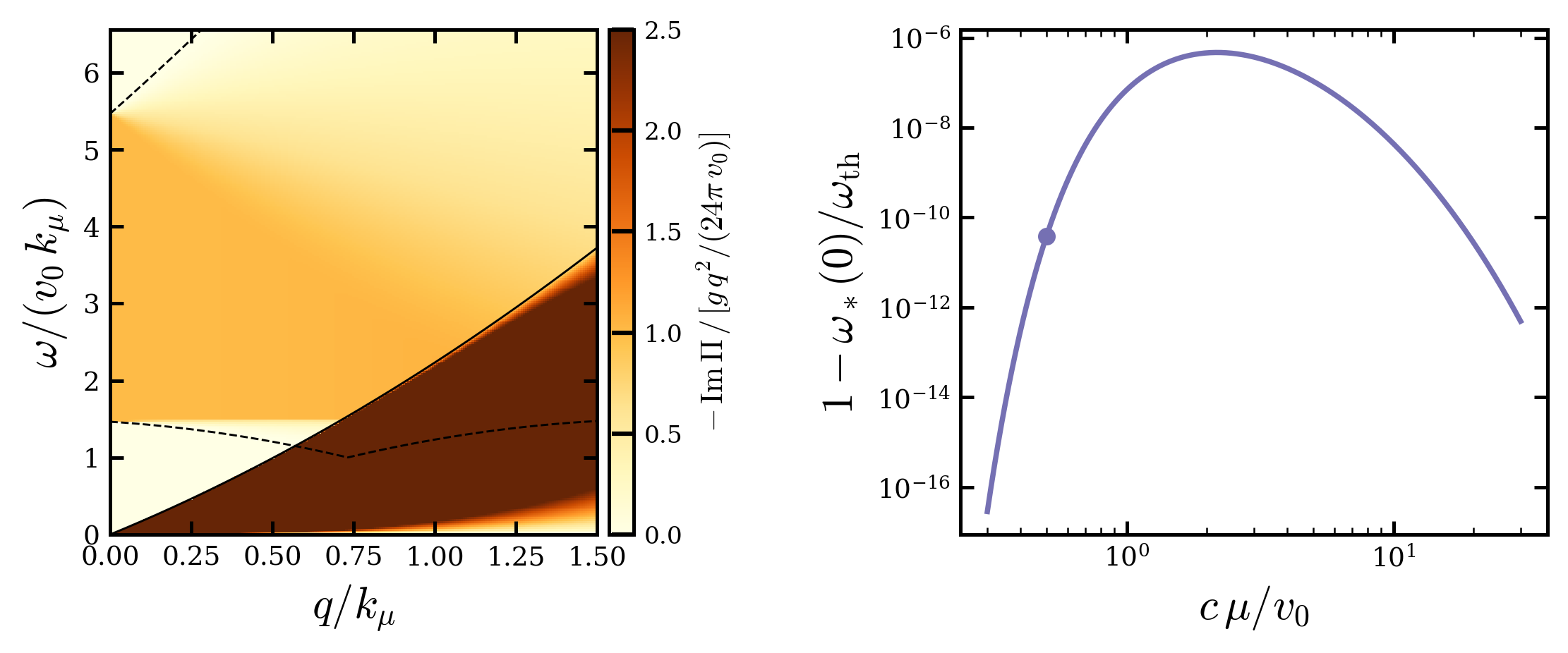}}
\subfigure[]{\includegraphics[width=0.75\linewidth]{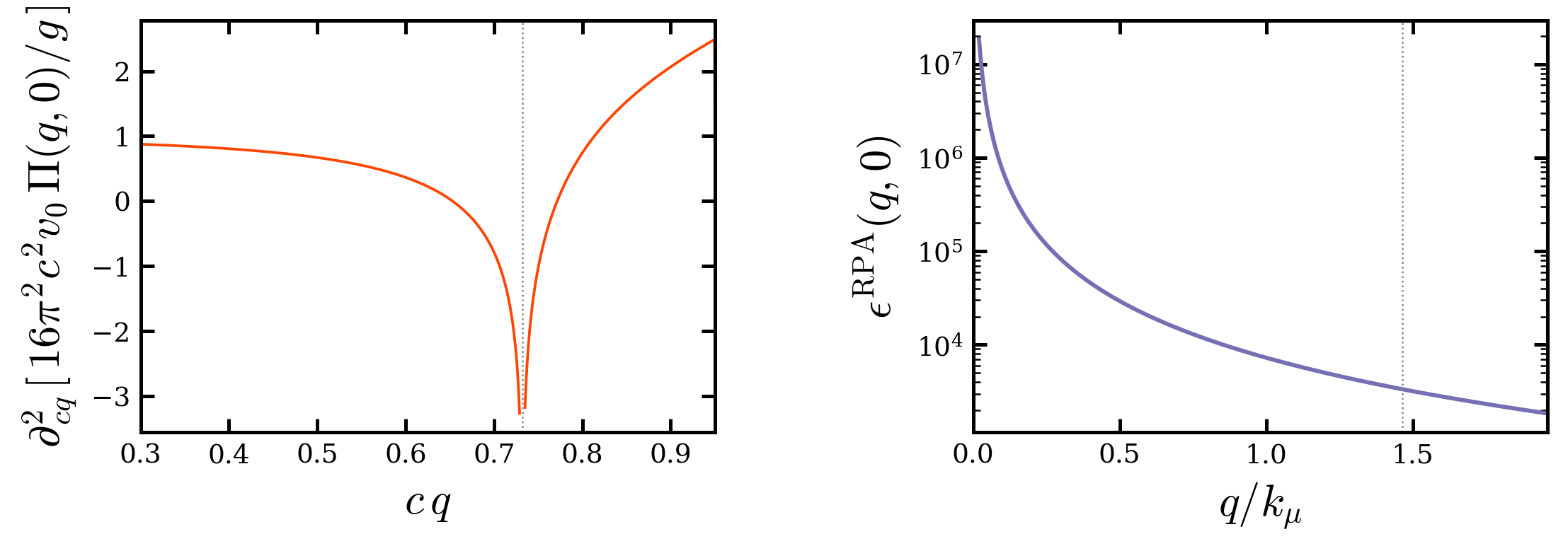}}
\caption{\label{fig:kw-plasmon-static}
KW characteristics. (a) Particle-hole continuum and distance of the lower plasmon from the onset. (b) Second derivative of $\Pi (q, 0)$ and static dielectric function.}
\end{figure}

The particle-hole continua are shown in Fig.~\ref{fig:kw-plasmon-static}(a). The left panel displays $-\,\Im \Pi$ of Eqs.~\eqref{eq:kwim-intra} and~\eqref{eq:kwim-inter} against $q/k_\mu$ and $\omega/(v_0\,k_\mu)$, normalised by the plateau value $g\,q^2/(24\,\pi\,v_0)$ of the transitions $-1 \to 1$. The colour scale is capped at $2.5$. The solid black line is the upper intraband edge $\omega = v_0\,q \left( \mathcal{R} + c\,q \right)$, which is common to both bands. The dashed black lines are the interband onset $\omega_{\rm th} (q)$ and the upper edge $\omega_{\rm max} (q)$ of Eq.~\eqref{eq:kwim-th}. In contrast to the RSW node, the interband window is finite, and the region above $\omega_{\rm max} (q)$ is free of excitations. The normalised absorption inside the window is close to unity at small $q$, as expected for the plateau. The lower plasmon discussed in Sec.~\ref{seckwplasmon} lies too close to the onset to be resolved as a branch in the left panel. The right panel shows its distance $1 - \omega_* (0)/\omega_{\rm th}$ from the onset against $c\,\mu/v_0$, with the dot marking the parameters of the left panel. This distance is very small for all the values shown. The upper plasmon lies above the cut-off $v_0\,\Lambda$ for the parameters of the figure and is not shown. The free window below the onset closes at $q_c$ of Eq.~\eqref{eq:kwim-qc}. The figure uses $\mu = 0.01$\,eV and $c = 0.02$, which give $k_\mu = 25$\,eV, $\mathcal{R} = \sqrt{3}$, $k_F^{(1)} \simeq 18.3$\,eV, $k_F^{(-1)} \simeq 68.3$\,eV, and $q_c \simeq 14.2$\,eV.

\subsection{Real part of polarisability and RPA dielectric function}
\label{seckwre}

The filled states of both bands are bounded in momentum space, and the dispersion relation with an intrinsic reference, which was needed for the TSM, is hence not required. We evaluate Eq.~\eqref{eq:pol} directly, with the notation of the opening of this section. The results of this subsection require only $c\,q < 1$.

In the variables $\Sigma$ and $\check{D}$, the measure is $d^3 \bs{k} = \pi \left( \Sigma^2 - \check{D}^2 \right) d\Sigma\,d\check{D}/(4\,q)$, and the overlap factors of Eq.~\eqref{eq:kw-ff} are $\mathcal{F}_{ss} = \left( \Sigma^2 - q^2 \right)/\left( \Sigma^2 - \check{D}^2 \right)$ and $\mathcal{F}_{-1,1} = \left( q^2 - \check{D}^2 \right)/\left( \Sigma^2 - \check{D}^2 \right)$. The factor $\Sigma^2 - \check{D}^2$ thus cancels in every channel. In the intraband channels, the integral over $\check{D}$ at fixed $\Sigma$ is elementary. In the interband channel, the integral over $\Sigma$ at fixed $\check{D}$ is elementary. Each gives a logarithm, and a single integral remains in every channel. We obtain
\begin{align}
\label{eq:kwre-pi}
\Re \Pi (q, \omega) = -\,\frac{g}{32\,\pi^2\,v_0\,q} \left[ \mathcal{N}_{1,1} + \mathcal{N}_{-1,-1} + \mathcal{N}_{-1,1} \right],
\end{align}
with
\begin{align}
\label{eq:kwre-N}
& \mathcal{N}_{1,1} = \int_{\Sigma_a}^{2\,k_1 + q} d\Sigma\; \frac{\Sigma^2 - q^2}{1 + c\,\Sigma}\,\ln \!\left( \frac{\left| \varpi^2 - q^2 \left( 1 + c\,\Sigma \right)^2 \right|}{\left| \varpi^2 - \left( \Sigma - 2\,k_1 \right)^2 \left( 1 + c\,\Sigma \right)^2 \right|} \right), \quad
\Sigma_a = \max \lbrace q, 2\,k_1 - q \rbrace, \nn
& \mathcal{N}_{-1,-1} = \int_{2\,k_2 - q}^{2\,k_2 + q} d\Sigma\; \frac{\Sigma^2 - q^2}{c\,\Sigma - 1}\,\ln \!\left( \frac{\left| \varpi^2 - q^2 \left( c\,\Sigma - 1 \right)^2 \right|}{\left| \varpi^2 - \left( \Sigma - 2\,k_2 \right)^2 \left( c\,\Sigma - 1 \right)^2 \right|} \right), \nn
& \mathcal{N}_{-1,1} = \int_{-q}^{q} d\check{D}\; \frac{q^2 - \check{D}^2}{1 + c\,\check{D}}\,\ln \!\left( \frac{\left| \varpi^2 - \left( 2\,k_2 + \check{D} \right)^2 \left( 1 + c\,\check{D} \right)^2 \right|}{\left| \varpi^2 - \Sigma_l^2 \left( 1 + c\,\check{D} \right)^2 \right|} \right), \quad
\Sigma_l = \max \lbrace q, 2\,k_1 - \check{D} \rbrace.
\end{align}
The three terms are the contributions of the intraband transitions of the band $s = 1$, those of the band $s = -1$, and the transitions $-1 \to 1$, respectively. For $q \leq k_1$, we have $\Sigma_a = 2\,k_1 - q$ and $\Sigma_l = 2\,k_1 - \check{D}$ for every $\check{D}$ in the range of integration. The integrands are regular, apart from the integrable logarithmic singularities at the frequencies of the continuum edges, and Eq.~\eqref{eq:kwre-pi} is the principal value. The result is even in $\omega$ and independent of $\Lambda$. The dielectric function follows from Eq.~\eqref{eq:epsrpa} as
\begin{align}
\label{eq:kwre-eps}
\Re \epsilon^{\rm RPA} (q, \omega) = 1 + \frac{\alpha}{8\,\pi\,q^3} \left[ \mathcal{N}_{1,1} + \mathcal{N}_{-1,-1} + \mathcal{N}_{-1,1} \right],
\end{align}
with $\alpha = g\,e^2/v_0$. For $q \to 0$ at fixed $\omega$, the intraband terms give $g\,k_1^2\,\mathcal{R}\,q^2/(6\,\pi^2\,v_0\,\varpi^2)$ and $g\,k_2^2\,\mathcal{R}\,q^2/(6\,\pi^2\,v_0\,\varpi^2)$. Their sum is the Drude term of Eq.~\eqref{eq:kwlin-wp}. The interband term gives $-\,\left( g\,q^2/(24\,\pi^2\,v_0) \right) \ln \!\left( \left| \varpi^2 - 4\,k_2^2 \right|/\left| \varpi^2 - 4\,k_1^2 \right| \right)$, which is the interband screening of Eq.~\eqref{eq:kwlin-eps}. The intermediate steps are given in Appendix~\ref{app:kwfull-re}.


\subsection{Static polarisability and screening}
\label{seckwstat}

At $\omega = 0$, the logarithms of Eq.~\eqref{eq:kwre-N} reduce to $2 \ln |x|$ with the transition energy $x$, and the factors $1 + c\,\Sigma$, $c\,\Sigma - 1$, and $1 + c\,\check{D}$ cancel inside the logarithms. After the shift $\Sigma = 2\,k_s + x$, the two intraband integrals have the common denominator $\mathcal{R} + c\,x$ and combine. The dependence on the limits $\Sigma_a$ and $\Sigma_l$ cancels between the intraband channel of the band $s = 1$ and the transitions $-1 \to 1$. The result therefore holds for every $c\,q < 1$, irrespective of the relation of $q$ to $k_1$ and to $2\,k_1$, and it is
\begin{align}
\label{eq:kwstat-pi}
& \Pi (q, 0) = \frac{-\,g}{16\,\pi^2\,v_0\,c^3\,q}\,\mathcal{B} (c\,q)\,,\quad
 \mathcal{B} (u) = 4\,\mathcal{R}\,u + \mathcal{E} (1 + u) - \mathcal{E} (1 - u) 
 + \left( 1 - u^2 \right) \left[ 2\,\Xi (u) - \Xi (1 + u) + \Xi (1 - u) \right],
\end{align}
with
\begin{align}
\label{eq:kwstat-EX}
& \mathcal{E} (t) = \frac{\left( \mathcal{R} + t \right) 
\left( 4 + \mathcal{R} - t \right) \ln \!\left| \mathcal{R} + t \right| }{2}
+ \frac{ \left( \mathcal{R} - t \right) \left( 4 - \mathcal{R} - t \right) 
\ln \!\left| \mathcal{R} - t \right| } {2}- \mathcal{R}\,t\,, \quad
 \Xi (t) = \Re\,\mathrm{Li}_2 \!\left( \frac{t}{\mathcal{R}} \right) 
- \mathrm{Li}_2 \!\left( -\,\frac{t}{\mathcal{R}} \right),
\end{align}
where $\mathrm{Li}_2$ is the dilogarithm. The real part is needed only for $t > \mathcal{R}$, which occurs for $q > 2\,k_1$ in the arguments $1 + u$. The static polarisability is negative for every $q$, because $n_s (\bs{k}) - n_{s'} (\bs{k}')$ and $\varepsilon_s (\bs{k}) - \varepsilon_{s'} (\bs{k}')$ have opposite signs in every term of Eq.~\eqref{eq:pol}. The static dielectric function follows from Eq.~\eqref{eq:epsrpa} as
\begin{align}
\label{eq:kwstat-eps}
\epsilon^{\rm RPA} (q, 0) = 1 + \frac{\alpha}{4\,\pi\,c^3\,q^3}\,\mathcal{B} (c\,q)\,,
\end{align}
and it exceeds unity.

For $q \ll k_1$, we expand Eq.~\eqref{eq:kwstat-pi} in $c\,q$ and obtain
\begin{align}
\label{eq:kwstat-small}
\Pi (q, 0) = -\,\rho - \frac{g\,q^2}{12\,\pi^2\,v_0}\,\ln \!\left( \frac{\mathcal{R} + 1}{\mathcal{R} - 1} \right) + \frac{g\,q^2}{4\,\pi^2\,v_0} \left( \frac{1}{\mathcal{R}} - \frac{1}{9\,\mathcal{R}^3} \right) + \mathcal{O} (q^4),
\end{align}
with the DOS $\rho$ of Eq.~\eqref{eq:kwlin-dos}. The first term is the intraband screening of both FSs. The second term is the interband part, which reproduces $\epsilon_b$ of Eq.~\eqref{eq:kwlin-eb}. The third term is the curvature correction of the intraband response. The corresponding static dielectric function is
\begin{align}
\label{eq:kwstat-eps-small}
\epsilon^{\rm RPA} (q, 0) = \frac{q_{\rm TF}^2}{q^2} + \epsilon_b - \frac{\alpha}{\pi} \left( \frac{1}{\mathcal{R}} - \frac{1}{9\,\mathcal{R}^3} \right) + \mathcal{O} (q^2), \quad
q_{\rm TF}^2 = 4\,\pi\,e^2\,\rho = \frac{\alpha \left( \mathcal{R}^2 + 1 \right)}{\pi\,c^2\,\mathcal{R}}\,.
\end{align}
The Thomas-Fermi wavevector $q_{\rm TF}$ receives contributions from both FSs and is independent of $\Lambda$. For $\mu \to 0^+$ at fixed $q$, Eq.~\eqref{eq:kwstat-pi} stays finite, and $\rho$ tends to $g/(2\,\pi^2\,c^2\,v_0)$. The small-$q$ expansion then requires $q \ll k_1$, and its $q^2$ coefficient diverges logarithmically with $k_1 \to 0$.

The polarisability has a Kohn anomaly at $q = 2\,k_1$ for $\mathcal{R} < 2$, where $2\,k_1 < 1/c$ lies inside the range $c\,q < 1$. For $q \to 2\,k_1$, we find
\begin{align}
\label{eq:kwstat-kohn}
\Pi (q, 0) = \Pi_{\rm reg} (q) + \frac{g}{16\,\pi^2\,v_0\,\mathcal{R}} \left( q - 2\,k_1 \right)^2 \ln \left| q - 2\,k_1 \right| + \dots,
\end{align}
where $\Pi_{\rm reg}$ is smooth. The term linear in $q - 2\,k_1$ multiplying the logarithm, which is present for the 3d electron gas, cancels. This cancellation reflects the vanishing of the intraband overlap factor at backscattering. The second derivative of $\Pi (q, 0)$ diverges logarithmically at $q = 2\,k_1$. For $\mathcal{R} \geq 2$, the anomaly at $q = 2\,k_1 \geq 1/c$ lies outside the range $c\,q < 1$ of Eq.~\eqref{eq:kwstat-pi}. The same holds for the anomaly at $q = 2\,k_2 > 1/c$ of the band $s = -1$. The intermediate steps are given in Appendix~\ref{app:kwfull-static}.

The static response is shown in Fig.~\ref{fig:kw-plasmon-static}(b). The left panel displays the second derivative of $\Pi (q, 0)$ from Eq.~\eqref{eq:kwstat-pi} with respect to $c\,q$, in units of $g/(16\,\pi^2\,c^2\,v_0)$. It diverges logarithmically to $-\infty$ at $c\,q = \mathcal{R} - 1$, which is the Kohn momentum $q = 2\,k_F^{(1)}$ of Eq.~\eqref{eq:kwstat-kohn} and is marked by the dotted line. The right panel displays the static dielectric function of Eq.~\eqref{eq:kwstat-eps}. It approaches the Thomas-Fermi form of Eq.~\eqref{eq:kwstat-eps-small} at small $q$, which is drawn as a dashed line but is not separately visible. The Kohn momentum is again marked by a dotted line, and the curve remains smooth there, since only the second derivative of $\Pi (q, 0)$ is singular. All curves satisfy $c\,q < 1$.


\subsection{Friedel oscillations}
\label{seckwfried}

The non-analytic behaviour of $\Pi (q, 0)$ at $q = 2\,k_1$ follows from Eq.~\eqref{eq:kwstat-pi}. The only non-analytic ingredient of Eq.~\eqref{eq:kwstat-EX} at this wavevector is $\ln \left| \mathcal{R} - t \right|$ with $t = 1 + c\,q$. It appears in $\mathcal{E} (t)$ and, through the identity for $\Re\,\mathrm{Li}_2$ near unit argument, in $\Xi (t)$. The coefficient of $\ln \left| q - 2\,k_1 \right|$ is
\begin{align}
\label{eq:kwfried-sing}
\Pi (q, 0) = \Pi_{\rm reg} (q) + \frac{g}{16\,\pi^2\,v_0\,c^3\,q} \left[ \frac{c}{2} \left( q - 2\,k_1 \right) \left( 3 - \mathcal{R} - c\,q \right) - \left( 1 - c^2\,q^2 \right) \ln \!\left( \frac{1 + c\,q}{\mathcal{R}} \right) \right] \ln \left| q - 2\,k_1 \right|,
\end{align}
where $\Pi_{\rm reg}$ is analytic at $q = 2\,k_1$ and $\mathcal{R} < 2$. The bracket vanishes at $q = 2\,k_1$ together with its first derivative, and its expansion reproduces the leading term of Eq.~\eqref{eq:kwstat-kohn}. The static dielectric function at the anomaly is
\begin{align}
\label{eq:kwfried-eps}
& \epsilon_{2\,k_1} = \epsilon^{\rm RPA} (2\,k_1, 0) = 1 + \frac{\alpha}{4\,\pi \left( \mathcal{R} - 1 \right)^3}\,\mathcal{B}_{2\,k_1}\,, \nn
& \mathcal{B}_{2\,k_1} = 2\,\mathcal{R} \left( \mathcal{R} - 1 \right) + 4\,\mathcal{R}\,\ln \mathcal{R} - 2 \left( \mathcal{R} - 1 \right) \ln \left( \mathcal{R} - 1 \right) + \mathcal{R} \left( 2 - \mathcal{R} \right) \left[ 2\,\Xi (\mathcal{R} - 1) + \Xi (2 - \mathcal{R}) - \frac{\pi^2}{4} \right],
\end{align}
which is the value of Eq.~\eqref{eq:kwstat-eps} at $c\,q = \mathcal{R} - 1$.

We consider a static impurity with the potential energy $\mathcal{U} (q)$ of an electron. Within the RPA, the induced density is $\delta n (\bs{q}) = \Pi (q, 0)\,\mathcal{U} (q)/\epsilon^{\rm RPA} (q, 0)$. Its non-analytic part is $\Pi_{\rm sing}\,\mathcal{U}/\epsilon^2$, because $\epsilon^{\rm RPA}$ also depends on $\Pi$. We take the smooth factors at $q = 2\,k_1$. The model is isotropic, and the radial Fourier transform gives for $k_1\,r \gg 1$
\begin{align}
\label{eq:kwfried-dn}
\delta n (r) \simeq \frac{g\,k_1}{8\,\pi^3\,v_0\,\mathcal{R}}\,\frac{\mathcal{U} (2\,k_1)}{\epsilon_{2\,k_1}^2}\,\frac{\sin \!\left( 2\,k_1\,r \right)}{r^4}\,.
\end{align}
For a Coulomb impurity of charge $Z\,e$, with $\mathcal{U} (q) = -\,4\,\pi\,Z\,e^2/q^2$, we obtain
\begin{align}
\label{eq:kwfried-coul}
\delta n (r) \simeq -\,\frac{\alpha\,Z}{8\,\pi^2\,\mathcal{R}\,k_1\,\epsilon_{2\,k_1}^2}\,\frac{\sin \!\left( 2\,k_1\,r \right)}{r^4}\,.
\end{align}
The oscillations have the period $\pi/k_1$ and decay as $r^{-4}$, whereas the singularity $\left( q - 2\,k_F \right) \ln \left| q - 2\,k_F \right|$ of a parabolic band in 3d gives the decay $r^{-3}$. The weaker singularity originates from the overlap factor $\mathcal{F}_{1,1} = (1 + \cos\theta)/2$, which vanishes quadratically in the deviation from backscattering. The singular coefficient $g/(16\,\pi^2\,v_0\,\mathcal{R}) = \rho_1/(8\,k_1^2)$ is fixed by the DOS and the Fermi momentum of the band $s = 1$. For $\mathcal{R} \to 1$, the prefactor of Eq.~\eqref{eq:kwfried-coul} takes the form obtained for the untilted WSM, with $k_F$ replaced by $k_1$. The screening factor $\epsilon_{2\,k_1}^2$ differs from the WSM, because $\epsilon_{2\,k_1}$ contains the contribution of the FS of the band $s = -1$ and the interband part. The range $c\,q < 1$ of Eq.~\eqref{eq:kwstat-pi} contains the anomaly at $2\,k_1$ only for $\mathcal{R} < 2$. Eq.~\eqref{eq:kwstat-pi} does not cover $q \geq 1/c$, where the transitions from the filled states of the band $s = 1$ to the empty states of the band $s = -1$ open, and anomalies at such wavevectors are therefore not described. This applies in particular to the anomaly of the band $s = -1$ at $q = 2\,k_2 = \left( \mathcal{R} + 1 \right)/c$, which exceeds $1/c$ for every $\mu > 0$ and hence lies outside the range $c\,q < 1$. The density of Eq.~\eqref{eq:kwfried-coul} is plotted in the right panel of Fig.~\ref{fig:friedel-rsw-kw}, and the comparison with the other nodes is given in Sec.~\ref{seccompfried}. The intermediate steps are given in Appendix~\ref{app:kwfull-friedel}.


\begin{table}[t!]
\centering
\small
\renewcommand{\arraystretch}{2.0}
\begin{tabular}{|c|c|c|c|c|c|}
\hline
Quantity & RSW & KW & QTSM & TSM & WSM \\ \hline
\makecell{Chern number \\ of the node} & $\pm 4$ & $\pm 1$ & $\pm 2$ & $\pm 2$ & $\pm 1$ \\ \hline
Number of bands & four & two & three & three & two \\ \hline
\makecell{Interband pairs \\ at order $q^2$ \\ (doped regime)} & \makecell{$-1/2 \to 1/2$ and \\ $1/2 \leftrightarrow 3/2$, \\ both $v_0\,k$} & \makecell{$-1 \to 1$, \\ $2\,v_0\,k$} & \makecell{flat-band to upper \\ band, and $-1 \to 0$ \\ for $c > 0$, $v_0\,k$} & \makecell{flat-band to \\ upper band, $v_0\,k$} & \makecell{dispersive pair, \\ $2\,v_0\,k$} \\ \hline
Fermi momenta & $2\,k_\mu$, $2\,k_\mu/3$ & $\dfrac{\mathcal{R} \mp 1}{2\,c}$ & \makecell{$\dfrac{2\,k_\mu}{1 + \mathcal{R}}$, and \\ $\sqrt{k_\mu/c}$ for $c > 0$} & $k_\mu$ & $k_\mu$ \\ \hline
Fermi velocities & $v_0/2$, $3\,v_0/2$ & $\mathcal{R}\,v_0$ (both) & \makecell{$\mathcal{R}\,v_0$, and $2\,\sqrt{\xi}\,v_0$ \\ for $c > 0$} & $v_0$ & $v_0$ \\ \hline
$\omega_p^2$ & $\dfrac{16\,g\,e^2\,\mu^2}{9\,\pi\,v_0}$ & $\dfrac{g\,e^2\,v_0\,\mathcal{R} \left( \mathcal{R}^2 + 1 \right)}{3\,\pi\,c^2}$ & \makecell{$\dfrac{2\,g\,e^2\,\mathcal{R}\,\mu_1^2}{3\,\pi\,v_0}$, plus \\ $8\,\pi\,e^2\,c\,v_0\,\tilde n_0$ for $c > 0$} & $\dfrac{2\,g\,e^2\,\mu^2}{3\,\pi\,v_0}$ & $\dfrac{2\,g\,e^2\,\mu^2}{3\,\pi\,v_0}$ \\ \hline
$\omega_p^2$ for $\mu \to 0^+$ & $0$ & $\dfrac{2\,g\,e^2\,v_0}{3\,\pi\,c^2}$ & $0$ & $0$ & $0$ \\ \hline
\makecell{Scaling of $\omega_p$ \\ with density $\tilde n$} & $\tilde n^{1/3}$ & \makecell{$\tilde n^{1/2}$ for \\ $c\,\mu/v_0 \gg 1$} & \makecell{$\tilde n_1^{1/3}$ (top band), \\ $\tilde n_0^{1/2}$ (flat-band, \\ $c > 0$)} & $\tilde n^{1/3}$ & $\tilde n^{1/3}$ \\ \hline
$q_{\rm TF}^2/\omega_p^2$ & $\dfrac{28}{3\,v_0^2}$ & $\dfrac{3}{\mathcal{R}^2\,v_0^2}$ & \makecell{$\dfrac{3}{\mathcal{R}^2\,v_0^2}$ \\ for $c < 0$} & $\dfrac{3}{v_0^2}$ & $\dfrac{3}{v_0^2}$ \\ \hline
$\omega_{\rm th}$ & \makecell{$2\,\mu/3$, and \\ $2\,\mu$ for the \\ second step} & \makecell{$v_0 \left( \mathcal{R} - 1 \right)/c$, with \\ upper edge \\ $v_0 \left( \mathcal{R} + 1 \right)/c$} & $\mu_1$ & $\mu$ & $2\,\mu$ \\ \hline
$\epsilon_b - 1$ & \makecell{$\dfrac{8\,g\,e^2}{3\,\pi\,v_0}\,L_2$ \\ $+ \dfrac{2\,g\,e^2}{\pi\,v_0}\,\ln 3$} & $\dfrac{g\,e^2}{3\,\pi\,v_0}\,\ln \!\left( \dfrac{\mathcal{R} + 1}{\mathcal{R} - 1} \right)$ & $\dfrac{4\,g\,e^2}{3\,\pi\,v_0}\,L_1$ & $\dfrac{4\,g\,e^2}{3\,\pi\,v_0}\,L$ & $\dfrac{g\,e^2}{3\,\pi\,v_0}\,L$ \\ \hline
\makecell{$\Im \epsilon^{\rm RPA}$ above \\ all onsets} & $\dfrac{4\,g\,e^2}{3\,v_0}$ & \makecell{$\dfrac{g\,e^2}{6\,v_0}$ between the \\ edges, zero above} & $\dfrac{2\,g\,e^2}{3\,v_0}$ & $\dfrac{2\,g\,e^2}{3\,v_0}$ & $\dfrac{g\,e^2}{6\,v_0}$ \\ \hline
$\omega_p^2/\omega_{\rm th}^2$ & $\dfrac{4\,g\,e^2}{\pi\,v_0}$ & $\dfrac{g\,e^2\,\mathcal{R} \left( \mathcal{R}^2 + 1 \right)}{3\,\pi\,v_0 \left( \mathcal{R} - 1 \right)^2}$ & \makecell{$\dfrac{2\,g\,e^2\,\mathcal{R}}{3\,\pi\,v_0}$ \\ for $c < 0$} & $\dfrac{2\,g\,e^2}{3\,\pi\,v_0}$ & $\dfrac{g\,e^2}{6\,\pi\,v_0}$ \\ \hline
Undamped plasmon & \makecell{one, below \\ $2\,\mu/3$, with \\ $q_{\rm max} \leq k_\mu/3$} & \makecell{two, below $\omega_{\rm th}$ \\ with $q_{\rm max} \leq q_c$, \\ and above $\omega_{\rm max}$ \\ at long wavelengths} & \makecell{$q < q_{\rm max}$, with \\ $q_{\rm max} \leq q_1$ for $c > 0$, \\ $q_1 = k_F^{(0)} - k_F^{(1)}$} & \makecell{below $\mu$, ends \\ at $q_{\rm max} < k_F$} & \makecell{below $2\,\mu$, \\ exists for $q < k_F$} \\ \hline
Bohm-Gross velocity $V_{\rm BG}$ & $\sqrt{3}\,v_0/2$ & $\mathcal{R}\,v_0$ & \makecell{$\mathcal{R}\,v_0$ for $c < 0$} & $v_0$ & $v_0$ \\ \hline
\makecell{Momenta of the \\ Kohn singularities} & \makecell{$2\,k_F^{(3/2)}$, \\ $k_F^{(1/2)} + k_F^{(3/2)}$, \\ $2\,k_F^{(1/2)}$} & \makecell{$2\,k_F^{(1)}$ \\ (for $\mathcal{R} < 2$)} & \makecell{$2\,k_F^{(1)}$, and for $c > 0$ \\ $k_F^{(0)} + k_F^{(1)}$, $2\,k_F^{(0)}$} & $2\,k_F$ & $2\,k_F$ \\ \hline
Friedel decay & \makecell{$r^{-4}$, with \\ $r^{-5}$ and $r^{-6}$} & $r^{-4}$ & \makecell{$r^{-5}$ (top band), \\ $r^{-4}$ and $r^{-3}$ \\ for $c > 0$} & $r^{-5}$ & $r^{-4}$ \\ \hline
\makecell{Doped part of $\Pi (q, 0)$ \\ for $q \gg k_F$} & $-\,\dfrac{6272\,g\,k_\mu^4}{1215\,\pi^2\,v_0\,q^2}$ & \makecell{not applicable, since \\ $k_F^{(-1)} > 1/c > q$ \\ for $c\,q < 1$} & \makecell{$\propto 1/(c\,q^2)$, with \\ $\left( k_F^{(1)} \right)^3 + \Theta (c) \left( k_F^{(0)} \right)^3$} & \makecell{constant, \\ $\dfrac{g\,k_F^2}{6\,\pi^2\,v_0}$} & $\propto k_F^4/q^2$ \\ \hline
\makecell{Closed form of \\ $\Pi (q, 0)$ at all $q$} & \makecell{available, \\ elementary functions} & \makecell{available for \\ $c\,q < 1$, with \\ dilogarithms} & \makecell{not available, \\ integral of $\Im \Pi$} & available & available \\ \hline
\end{tabular}
\caption{\label{tab:comp-rsw-kw}
Comparison of the RSW node, the KW node, the QTSM, the TSM, and the WSM.}
\end{table}

\subsection{Plasmon dispersion and Landau damping}
\label{seckwplasmon}

We consider $q \to 0$ and frequencies in the window above the intraband edge $v_0\,\varpi_+ (q)$ of Eq.~\eqref{eq:kwim-edges} and below the onset $v_0\,\varpi_{\rm th} (q)$ of Eq.~\eqref{eq:kwim-th}. For $q \to 0$, the edge tends to $v_0\,\mathcal{R}\,q$ and the onset tends to $\omega_{\rm th}$ of Eq.~\eqref{eq:kwlin-onsets}. The window is therefore free of excitations at small $q$, and the expansion of $\Pi$ contains only even powers of $q$, $\Pi = q^2\,\Pi_2 + q^4\,\Pi_4 + \dots$. Equation~\eqref{eq:kwre-pi} gives
\begin{align}
\label{eq:kwplas-pi24}
& \Pi_2 = \frac{g}{24\,\pi^2\,v_0} \left[ \frac{2\,\mathcal{R} \left( \mathcal{R}^2 + 1 \right)}{c^2\,\varpi^2} - \ln \!\left( \frac{\left| \varpi^2 - 4\,k_2^2 \right|}{\left| \varpi^2 - 4\,k_1^2 \right|} \right) \right], \quad
\Pi_4 = \frac{g}{120\,\pi^2\,v_0}\,\tilde{\mathcal{P}} (\varpi), \nn
& \tilde{\mathcal{P}} (\varpi) = \frac{6\,\mathcal{R}^3 \left( \mathcal{R}^2 + 1 \right)}{c^2\,\varpi^4} - \frac{4\,\mathcal{R}}{\varpi^2} + \frac{8 \left( \mathcal{R} + 2 \right)^2 k_2^2}{\left( \varpi^2 - 4\,k_2^2 \right)^2} - \frac{8 \left( \mathcal{R} - 2 \right)^2 k_1^2}{\left( \varpi^2 - 4\,k_1^2 \right)^2} \nn
& \hspace{1.5cm} + \left( \mathcal{R}^2 - 2 \right) \left[ \frac{1}{\varpi^2 - 4\,k_1^2} - \frac{1}{\varpi^2 - 4\,k_2^2} \right] - c^2\,\ln \!\left( \frac{\left| \varpi^2 - 4\,k_2^2 \right|}{\left| \varpi^2 - 4\,k_1^2 \right|} \right).
\end{align}
The first term of $\Pi_2$ is the Drude term of Eq.~\eqref{eq:kwlin-wp}, and the logarithm is the interband screening of Eq.~\eqref{eq:kwlin-eps}. In $\Pi_4$, the first two terms originate from the intraband transitions of the two bands, and the remaining terms originate from the transitions $-1 \to 1$. The expansion requires $v_0\,\mathcal{R}\,q \ll \omega$ and a distance of $\omega$ from $\omega_{\rm th}$ and $\omega_{\rm max}$ that is large compared with $v_0\,q$. The intermediate steps are given in Appendix~\ref{app:kwfull-wdisp}.

Let $\epsilon_0 (\omega) = 1 - 4\,\pi\,e^2\,\Pi_2$ be the dielectric function of Eq.~\eqref{eq:kwlin-eps}, and let
\begin{align}
\mathcal{D} (\omega) = \partial_{\omega^2} \epsilon_0 = \frac{\omega_p^2}{\omega^4} + \frac{\alpha}{6\,\pi} \left[ \frac{1}{\omega^2 - \omega_{\rm max}^2} - \frac{1}{\omega^2 - \omega_{\rm th}^2} \right].
\end{align}
The zero of $\Re\epsilon^{\rm RPA} = \epsilon_0 - 4\,\pi\,e^2\,q^2\,\Pi_4$ obeys $\epsilon_0 (\omega) = 4\,\pi\,e^2\,q^2\,\Pi_4$. To order $q^2$, it is
\begin{align}
\label{eq:kwplas-wdisp}
\omega_*^2 (q) = \omega_*^2 (0) + \frac{\alpha\,q^2}{30\,\pi}\,\frac{\tilde{\mathcal{P}} (\omega_*/v_0)}{\mathcal{D} (\omega_*)} \Bigg|_{\omega_* = \omega_* (0)},
\end{align}
where $\omega_* (0)$ is the unique zero below the onset found in Sec.~\ref{seckwlinscreen}. Below the onset, the Drude term and the interband term of $\epsilon_0$ both increase with $\omega$, and hence $\mathcal{D} > 0$. The zero is therefore an analytic function of $q^2$ near $q = 0$, and it lies in the free window for small $q$. The plasmon is undamped.

For $\alpha \to 0$, we have $\omega_*^2 (0) \simeq \omega_p^2 \ll \omega_{\rm th}^2$ and $\mathcal{D} \to \omega_p^2/\omega_*^4$. The first term of $\tilde{\mathcal{P}}$ dominates, and Eq.~\eqref{eq:kwplas-wdisp} reduces to
\begin{align}
\label{eq:kwplas-bg}
\omega_*^2 (q) = \omega_p^2 + \frac{3}{5}\,V_F^2\,q^2\,, \quad V_F = v_0\,\mathcal{R}\,.
\end{align}
This is the Bohm-Gross dispersion~\cite{bohm-gross}. The velocity is the common Fermi speed of the two FSs, in contrast to the weighted average over different FS velocities that appears when the Fermi speeds differ. The corrections to Eq.~\eqref{eq:kwplas-bg} are of relative order $\alpha$.

For $\omega > \omega_{\rm max}$, Eq.~\eqref{eq:kwplas-pi24} applies in the same way, as long as $\omega$ is far from $\omega_{\rm max}$ on the scale of $v_0\,q$. The upper zero $\omega_*'$ of Sec.~\ref{seckwlinscreen} lies exponentially close to $\omega_{\rm max}$ for small $\alpha$, because $\Re\epsilon_0$ diverges logarithmically at the edge. The expansion in $q$ is controlled only for $q$ small compared with the distance of this zero from $\omega_{\rm max}/v_0$, and we do not use it for the upper mode.

For larger $q$, the lower plasmon leaves the free window through the intraband edge $v_0\,\varpi_+ (q)$ of Eq.~\eqref{eq:kwim-edges} or through the onset $v_0\,\varpi_{\rm th} (q)$ of Eq.~\eqref{eq:kwim-th}, whichever it reaches first. We denote the corresponding wavevector by $q_{\rm max}$. It is located by solving $\Re\epsilon^{\rm RPA} = 0$ on the relevant boundary with Eq.~\eqref{eq:kwre-pi}. The real part of Eq.~\eqref{eq:kwlin-eps} can be written as $\epsilon_0 (\omega) = 1 + \alpha\,\tilde{\mathcal{F}} (\omega)$, where $\tilde{\mathcal{F}} (\omega) = -\,\omega_p^2/(\alpha\,\omega^2) + \left[ 1/(6\,\pi) \right] \ln \!\left( |\omega^2 - \omega_{\rm max}^2|/|\omega^2 - \omega_{\rm th}^2| \right)$ does not depend on $\alpha$. By the monotonicity shown in Appendix~\ref{app:kwlin-screen}, $\tilde{\mathcal{F}}$ increases strictly from $-\,\infty$ to $+\,\infty$ on $(0, \omega_{\rm th})$ and has a unique zero $\omega_\infty < \omega_{\rm th}$. The condition $\tilde{\mathcal{F}} (\omega_* (0)) = -\,1/\alpha$ then shows that $\omega_* (0) < \omega_\infty$, and the plasmon therefore never starts at the onset. For $\alpha \to 0$, we have $\omega_* (0) \simeq \omega_p \ll \omega_{\rm th}$, because $\omega_p^2/\omega_{\rm th}^2 \propto \alpha$. Since the free window closes at $q = q_c$ of Eq.~\eqref{eq:kwim-qc}, we have $q_{\rm max} \leq q_c$. For $q > q_{\rm max}$, the zero of $\Re\epsilon^{\rm RPA}$ lies inside the continuum, where $\Im\epsilon^{\rm RPA} = -\left( 4\,\pi\,e^2/q^2 \right) \Im\Pi \neq 0$ with $\Im\Pi$ of Eq.~\eqref{eq:kwim-tot}. The mode is Landau damped with the rate $\gamma = \Im\epsilon^{\rm RPA}/\partial_\omega \Re\epsilon^{\rm RPA}$ at $\omega = \Re\,\omega_*$, which holds when $\gamma$ is small compared with the scales on which $\Im\epsilon^{\rm RPA}$ and $\Re\epsilon^{\rm RPA}$ vary. We do not evaluate $\gamma$ in closed form, because $\partial_\omega \Re\epsilon^{\rm RPA}$ at the zero requires $\Re \Pi$ at finite $q$ close to the edges of the continuum, where the expansion in $q$ is not controlled. The damping is due to the following transitions.
\begin{itemize}
\item The intraband transitions of both bands for $\omega < v_0\,\varpi_+ (q)$. The band $s = -1$ contributes for every $q$ in this range, because its lower edge is zero.
\item The interband transitions $-1 \to 1$ for $v_0\,\varpi_{\rm th} (q) < \omega < v_0\,\varpi_{\rm max} (q)$.
\end{itemize}
For $q > q_c$, the two continua overlap, and the whole range $0 < \omega < v_0\,\varpi_{\rm max} (q)$ is damped. The intermediate steps are given in Appendix~\ref{app:kwfull-wdisp}.

For the parameters of the figure, $\omega_p^2/\omega_{\rm th}^2 \simeq 3.4 \times 10^3$, and the logarithmic divergence of $\Re\epsilon^{\rm RPA}$ at the onset therefore places $\omega_* (0)$ exponentially close to $\omega_{\rm th}$. The expansion in $q$ is controlled only for $q$ small compared with the distance $\left( \omega_{\rm th} - \omega_* (0) \right)/v_0$ of this zero from the onset, and the undamped window is correspondingly narrow.


\section{Comparison across various systems}
\label{seccompare}

We compare the RSW and KW nodes with each other and with the untilted TSM, the QTSM, and the untilted WSM, using the results of Ref.~\cite{ips-tsm-plasmon} for the last three. The main results are collected in Table~\ref{tab:comp-rsw-kw}. All five systems have a BC-monopole and an isotropic response. The RSW node, the TSM, and the WSM have the Hamiltonian $v_0\,\bs{\tilde{k}} \cdot \bs{\mathcal{S}}$ with the pseudospin $3/2$, $1$, and $1/2$, respectively. The KW node has the eigenstates of the WSM and a spectrum with a quadratic term, whereas the QTSM has the eigenstates of the TSM and the same quadratic term $c\,v_0\,k^2$, which is proportional to the identity. The Friedel oscillations are compared in Sec.~\ref{seccompfried}. All five systems have a unique undamped plasmon below their lowest onset at long wavelengths, because the monotonicity argument of Secs.~\ref{secrswlinscreen} and~\ref{seckwlinscreen} applies to each of them. We write $k_\mu = \mu/v_0$ and $\alpha = g\,e^2/v_0$ throughout. For the QTSM, we write $\xi = c\,\mu/v_0$ and $\mathcal{R} = \sqrt{1 + 4\,\xi}$, which is the quantity denoted $\beta$ in Ref.~\cite{ips-tsm-plasmon}, and $\mu_1 = 2\,\mu/\left( 1 + \mathcal{R} \right)$. The Fermi momenta of the top band and of the flat-band are $k_F^{(1)} = \mu_1/v_0$ and $k_F^{(0)} = \sqrt{\mu/(c\,v_0)}$. The QTSM entries refer to $c < 0$ unless $c > 0$ is indicated, and $\tilde{n}_0$ denotes the density of the flat-band. The cut-off dependence of the table enters through $L_2 = \ln \!\left( v_0\,\Lambda/(2\,\mu) \right)$, $L_1 = \ln \!\left( v_0\,\Lambda/\mu_1 \right)$, and $L = \ln \!\left( v_0\,\Lambda/\mu \right)$, and the constants of $\epsilon_b$ that depend on the cut-off scheme are omitted. The Chern number of a node is the magnitude of the sum of the Chern numbers of the bands below the node. The Bohm-Gross velocity is defined by $\omega_*^2 = \omega_p^2 + (3/5)\,V_{\rm BG}^2\,q^2$ for $\omega_p \ll \omega_{\rm th}$. The entry of the KW node for $\Im \epsilon^{\rm RPA}$ holds between $\omega_{\rm th}$ and $\omega_{\rm max}$.

\subsection{RSW versus KW}
\label{seccomprswkw}

Both nodes have two concentric FSs in the doped regime and an isotropic response. In the RSW node, the two FSs have the different Fermi speeds $v_0/2$ and $3\,v_0/2$, and both shrink to the node for $\mu \to 0^+$. In the KW node, the two FSs have the common Fermi speed $\mathcal{R}\,v_0$, and the FS of the band $s = -1$ stays at $k = 1/c$ for $\mu \to 0^+$. The Drude weight of the RSW node therefore vanishes as $\mu^2$, with the weights of the two bands in the ratio $3 : 1$, whereas that of the KW node tends to a finite value, with the weights in the ratio $\left( \mathcal{R} - 1 \right)^2 : \left( \mathcal{R} + 1 \right)^2$. The intrinsic RSW node has no undamped plasmon, in contrast to the KW node. For $c\,\mu/v_0 \gg 1$, the KW plasmon becomes that of a parabolic band with $\omega_p \propto \tilde n^{1/2}$, whereas $\omega_p \propto \tilde n^{1/3}$ for the RSW node at all doping. The Thomas-Fermi wavevector obeys $q_{\rm TF}^2 = 3\,\omega_p^2/V_F^2$ for the KW node, as for a single isotropic band. For the RSW node, the ratio $q_{\rm TF}^2/\omega_p^2 = 3 \sum_s \rho_s/\sum_s \rho_s\,V_s^2$ takes the value $28/(3\,v_0^2)$, because the two bands have different Fermi speeds.

The interband responses differ more strongly. In the RSW node, three interband pairs carry weight at order $q^2$ in the intrinsic limit, all with the transition energy $v_0\,k$, and the transitions $-1/2 \to 1/2$ and $1/2 \leftrightarrow 3/2$ remain in the doped regime. The continuum starts at $2\,\mu/3$, has a step at $2\,\mu$, and extends up to the cut-off. The background $\epsilon_b$ therefore depends logarithmically on the cut-off. In the KW node, a single pair carries weight, with the transition energy $2\,v_0\,k$ independent of $c$, and the continuum occupies the finite window $\left( \omega_{\rm th}, \omega_{\rm max} \right)$ of the width $2\,v_0/c$ with a constant absorption. The background is therefore independent of the cut-off, and it diverges logarithmically for $\mu \to 0^+$. The ratio $\omega_p^2/\omega_{\rm th}^2$ is independent of $\mu$ for the RSW node and depends on $c\,\mu/v_0$ for the KW node. Since the KW continuum has an upper edge, the KW node supports a second undamped plasmon above it at long wavelengths, whereas the absorption of the RSW node is nonzero at every $\omega > 2\,\mu/3$ and supports only the lower plasmon.

The lower plasmons obey the Bohm-Gross dispersion for $\omega_p \ll \omega_{\rm th}$, with $V_{\rm BG}^2 = 3\,v_0^2/4$ for the RSW node and $V_{\rm BG} = V_F = \mathcal{R}\,v_0$ for the KW node. The former is the average of $V_s^2$ over the FSs weighted with the Drude weights, and the latter reflects the common Fermi speed. The undamped lower plasmon of the RSW node ends at $q_{\rm max} \leq k_\mu/3$, which is one sixth of the larger Fermi momentum, whereas that of the KW node ends at $q_{\rm max} \leq q_c$ of Eq.~\eqref{eq:kwim-qc}. The static polarisability of the RSW node has a closed form in terms of elementary functions, with three Kohn singularities, whereas that of the KW node requires the dilogarithm and has a single Kohn singularity at $2\,k_F^{(1)}$ for $\mathcal{R} < 2$, within the range $c\,q < 1$ of its derivation. Within this range, the Fermi momentum $k_F^{(-1)} > 1/c$ exceeds $q$, and the doped part of the KW polarisability for $q \gg k_F$ is therefore not accessible. The corresponding part of the RSW polarisability decays as $q^{-2}$. The leading Friedel oscillations of both nodes decay as $r^{-4}$, and the RSW node has the two additional terms of Sec.~\ref{seccompfried}.

\subsection{RSW versus TSM}
\label{seccomprswtsm}

The RSW node and the TSM have the same structure of the Hamiltonian and differ in the spin representation. The TSM has a flat-band at zero energy, which has no FS in the linear model, and only its upper band has an FS. The RSW node has no flat-band and has two FSs. At the same $\mu$, the DOS of the RSW node exceeds that of the TSM by the factor $224/27$, and the Drude weight, and hence $\omega_p^2$, exceeds it by the factor $8/3$. The ratio $q_{\rm TF}^2/\omega_p^2$ is $28/(3\,v_0^2)$ instead of $3/v_0^2$, the Bohm-Gross velocity is $\sqrt{3}\,v_0/2$ instead of $v_0$, and the Chern number of the node is $\pm 4$ instead of $\pm 2$.

The interband part follows from the quantum metric. For the TSM, the transitions between the flat-band and the upper band have the metric $\left( \delta_{ij} - \hat{k}_i\,\hat{k}_j \right)/(2\,k^2)$ and the transition energy $v_0\,k$. For the RSW node, the transitions $-1/2 \to 1/2$ have the metric $\left( \delta_{ij} - \hat{k}_i\,\hat{k}_j \right)/k^2$, which is twice as large, and the same transition energy. The intrinsic interband polarisability of the RSW node is therefore twice that of the TSM. The coefficient of the logarithm in $\epsilon_b - 1$ and the absorption above the highest onset are also twice the TSM values, namely $8\,g\,e^2/(3\,\pi\,v_0)$ and $4\,g\,e^2/(3\,v_0)$. The RSW node has in addition the transitions $1/2 \leftrightarrow 3/2$, which lower the onset to $2\,\mu/3$, below the onset $\mu$ of the TSM, and they add the finite term $2\,g\,e^2\,\ln 3/(\pi\,v_0)$ to $\epsilon_b - 1$. The ratio $\omega_p^2/\omega_{\rm th}^2$ is six times larger for the RSW node.

At finite wavevectors, the free region of the TSM closes at $q = k_F$, whereas that of the RSW node closes at $q = k_\mu/3$, since the intraband edge of the faster band and the lowest interband onset approach each other more quickly. The doped part of the polarisability at large $q$ tends to a constant for the TSM, whereas it decays as $q^{-2}$ for the RSW node. The leading Friedel oscillations of the RSW node decay as $r^{-4}$, which is slower than the $r^{-5}$ of the TSM, despite the higher pseudospin. The decay is set by the power with which the intraband overlap factor vanishes at backscattering and not by the pseudospin, as explained in Sec.~\ref{seccompfried}.

\subsection{RSW versus QTSM}
\label{seccomprswqtsm}

The QTSM is the TSM with the quadratic term $c\,v_0\,k^2$. Its eigenstates and overlap factors are those of the TSM and do not depend on $c$. For $c < 0$, only the top band has an FS, with the Fermi speed $\mathcal{R}\,v_0$. For $c > 0$, the flat-band disperses as $c\,v_0\,k^2$ and has an FS that encloses the one of the top band, and the QTSM then has two concentric FSs, as the RSW node has. In the RSW node, both FSs belong to linear bands with the speeds $v_0/2$ and $3\,v_0/2$, both scale with $\mu$, and the larger FS belongs to the slower band. In the QTSM, the larger FS belongs to a parabolic band with the effective mass $1/(2\,c\,v_0)$. The top band of the QTSM contributes the TSM value multiplied by $4\,\mathcal{R}/(1 + \mathcal{R})^2$ to the Drude weight, and the flat-band adds the term $8\,\pi\,e^2\,c\,v_0\,\tilde n_0$, which is the plasma frequency of a parabolic band and gives $\omega_p \propto \tilde n_0^{1/2}$. At the same $\mu$, the bare plasma frequency of the RSW node exceeds that of the QTSM with $c < 0$ by the factor $2 \left( 1 + \mathcal{R} \right)^2/(3\,\mathcal{R})$, which is at least $8/3$.

The interband response of the QTSM does not depend on the sign of $c$. For $c < 0$, it has the form of the TSM with $\mu$ replaced by $\mu_1$. It has a single onset at $\mu_1$, the coefficient $4\,g\,e^2/(3\,\pi\,v_0)$ of the logarithm in $\epsilon_b - 1$, and the absorption $2\,g\,e^2/(3\,v_0)$ above the onset. For $c > 0$, the transitions $-1 \to 0$ take over from the transitions $0 \to 1$ beyond $k_F^{(0)}$. Both groups have the same metric and the same transition energy $v_0\,k$, and the absorption therefore has no step at the second FS. The RSW node has twice the coefficient and twice the absorption of the QTSM, as for the TSM. Its absorption has a step at $2\,\mu$, because the pairs $1/2 \leftrightarrow 3/2$ and $-1/2 \to 1/2$ have different weights. The lowest onset $2\,\mu/3$ of the RSW node lies below $\mu_1$ for $\mathcal{R} < 2$, that is for $4\,c\,\mu/v_0 < 3$. The ratio $\omega_p^2/\omega_{\rm th}^2$ is independent of $\mu$ for the RSW node, whereas for the QTSM it is $2\,g\,e^2\,\mathcal{R}/(3\,\pi\,v_0)$ for $c < 0$. For $c > 0$, it acquires the flat-band term $\left( 1 + \mathcal{R} \right)^2/(2\,\sqrt{\xi})$ inside the bracket of Ref.~\cite{ips-tsm-plasmon}, which grows without bound for $c \to 0^+$.

At finite wavevectors, the undamped plasmon of both nodes ends at a critical wavevector. For the RSW node, the free region closes at $q = k_\mu/3$, and hence $q_{\rm max} \leq k_\mu/3$. For the QTSM with $c > 0$, the transitions between the two FSs have zero energy for $q_1 \leq q \leq q_2$, with $q_1 = k_F^{(0)} - k_F^{(1)}$ and $q_2 = k_F^{(0)} + k_F^{(1)}$, and hence $q_{\rm max} \leq q_1$. The analogous range of the RSW node starts at $k_F^{(1/2)} - k_F^{(3/2)} = 4\,k_\mu/3$, which exceeds $k_\mu/3$, and the bound of the RSW node is therefore set by the closing of the free region. For $\omega_p \ll \omega_{\rm th}$, both plasmons have the Bohm-Gross dispersion. The velocity is the average of $V^2$ over the FSs weighted with the Drude weights, namely $\sqrt{3}\,v_0/2$ for the RSW node and $\mathcal{R}\,v_0$ for the QTSM with $c < 0$. The static polarisability of the RSW node has a closed form in terms of elementary functions, with the doped part decaying as $q^{-2}$ at large $q$ and depending on no parameter other than $k_\mu$. For the QTSM, only the forms at small and large $q$ are available, and the value at arbitrary $q$ follows from the integral of the imaginary part. The large-$q$ correction is proportional to $1/(c\,q^2)$ with $\left( k_F^{(1)} \right)^3 + \Theta (c) \left( k_F^{(0)} \right)^3$. It is controlled by the quadratic term, and the limits $c \to 0$ and $q \to \infty$ do not commute. For $c > 0$, the Friedel oscillations of the two nodes follow the same hierarchy. The larger FS gives the slowest decay, the transitions between the FSs the next one, and the smaller FS the fastest one, namely $r^{-4}$, $r^{-5}$, and $r^{-6}$ for the RSW node and $r^{-3}$, $r^{-4}$, and $r^{-5}$ for the QTSM. The exponents of the QTSM are lower by one, because the flat-band has unit overlap at backscattering, whereas the factor $4\,a$ of the larger FS of the RSW node vanishes linearly, as explained in Sec.~\ref{seccompfried}.

\subsection{RSW versus WSM}
\label{seccomprswwsm}

The WSM has two bands, and its interband pair has the metric $\left( \delta_{ij} - \hat{k}_i\,\hat{k}_j \right)/(4\,k^2)$ and the transition energy $2\,v_0\,k$. The pair $-1/2 \to 1/2$ of the RSW node has four times this metric and half this energy. Since the radial integral contributes a factor $1/\upsilon$ for a transition energy $\upsilon\,v_0\,k$, the intrinsic interband polarisability of the RSW node is eight times that of the WSM. The coefficient of the logarithm in $\epsilon_b - 1$ and the absorption above the highest onset are also eight times the WSM values, whereas the corresponding factor for the TSM is four. The Drude weight of the RSW node is $8/3$ times that of the WSM at the same $\mu$, and its lowest onset $2\,\mu/3$ is one third of the WSM onset $2\,\mu$. The ratio $\omega_p^2/\omega_{\rm th}^2$ is therefore $24$ times larger, and the ratio $q_{\rm TF}^2/\omega_p^2$ is $28/(3\,v_0^2)$ instead of $3/v_0^2$.

The Bohm-Gross velocity is $\sqrt{3}\,v_0/2$ for the RSW node and $v_0$ for the WSM. The WSM plasmon exists for $q < k_F$, whereas that of the RSW node ends at $q_{\rm max} \leq k_\mu/3$. The doped parts of both static polarisabilities fall off as $q^{-2}$ for $q \gg k_F$, in contrast to the constant of the TSM. The leading Friedel oscillations of both nodes decay as $r^{-4}$, since the overlap factor of the outer band of the RSW node vanishes linearly in $a = (1 + \cos\theta)/2$ at backscattering, as does that of the WSM. The RSW node has the additional terms that decay as $r^{-5}$ and $r^{-6}$.

\subsection{KW versus QTSM}
\label{seccompkwqtsm}

The KW node and the QTSM have the same energies for the two dispersive bands, $\varepsilon_{\pm 1} = \pm\,v_0\,k + c\,v_0\,k^2$. For $\mu > 0$, the band $s = 1$ therefore has the same Fermi momentum $k_F^{(1)} = \left( \mathcal{R} - 1 \right)/(2\,c)$, the same Fermi speed $\mathcal{R}\,v_0$, the same DOS, and the same Drude weight in both systems. The systems differ in three respects. First, the eigenstates are the spin-$1/2$ coherent states for the KW node and the spin-$1$ states for the QTSM, and the intraband overlap factor of the band $s = 1$ is $a$ for the KW node and $a^2$ for the QTSM, with $a = (1 + \cos\theta)/2$. Second, the QTSM has a flat-band, which acquires an FS for $c > 0$. Third, the linearised description of the QTSM is restricted to $|c|\,\Lambda < 1$, and its band $s = -1$ is filled throughout this range, since its FS lies at $\left( 1 + \mathcal{R} \right)/(2\,c) > 1/c$. But the KW node has this lower dispersive band folded up to give a second FS, which has the same Fermi speed $\mathcal{R}\,v_0$. The second FS of the QTSM with $c > 0$ belongs to the flat-band and has the speed $2\,\sqrt{\xi}\,v_0$.

The band $s = 1$ contributes the same amount to $\omega_p^2$ in both systems. The second FS adds the contribution of the band $s = -1$ for the KW node and the parabolic flat-band term for the QTSM with $c > 0$, as listed in Table~\ref{tab:comp-rsw-kw}. Since $k_F^{(-1)} \to 1/c$ for $\mu \to 0^+$, the KW node retains a finite Drude weight in this limit, whereas that of the QTSM vanishes. For $c\,\mu/v_0 \gg 1$, both systems have $\omega_p \propto \tilde n^{1/2}$. The relation $q_{\rm TF}^2 = 3\,\omega_p^2/V_F^2$ with $V_F = \mathcal{R}\,v_0$ holds for the KW node and for the QTSM with $c < 0$. For the QTSM with $c > 0$, the ratio involves the different speeds of the two FSs.

The KW node has the interband pair of the WSM and the QTSM has those of the TSM, with unchanged transition energies, because the quadratic term is proportional to the identity. The coefficient of the logarithm in $\epsilon_b - 1$ and the absorption are therefore four times larger in the QTSM, as for the TSM and the WSM. At the same $k_F^{(1)}$, the onset of the KW node is twice that of the QTSM. The logarithm of the QTSM depends on the cut-off through $\Lambda/k_F^{(1)}$, whereas that of the KW node contains the finite ratio $k_F^{(-1)}/k_F^{(1)}$, because the transitions end at the Fermi momentum of the band $s = -1$. The absorption of the QTSM extends to the cut-off, whereas that of the KW node ends at $\omega_{\rm max}$. At long wavelengths, the KW node therefore has a second undamped plasmon above $\omega_{\rm max}$, which the QTSM does not have.

At finite wavevectors, the lower plasmon of the KW node ends at $q_{\rm max} \leq q_c$, whereas that of the QTSM with $c > 0$ ends at $q_{\rm max} \leq q_1 = k_F^{(0)} - k_F^{(1)}$, because the transitions between its two FSs damp the mode at all frequencies for $q_1 \leq q \leq q_2$. In the KW node, the intraband continuum of the band $s = -1$ reaches $\omega \to 0$ for every $q$, and both bands share the upper edge $q \left( \mathcal{R} + c\,q \right)$. For $\omega_p \ll \omega_{\rm th}$, the Bohm-Gross velocity is $V_{\rm BG} = \mathcal{R}\,v_0$ for the KW node and an average over the two Fermi speeds for the QTSM with $c > 0$. The static polarisability of the KW node has a closed form with dilogarithms for $c\,q < 1$, whereas only its limits at small and large $q$ and the integral of the imaginary part are available for the QTSM. The doped part of the QTSM at large $q$ is proportional to $1/(c\,q^2)$, and the corresponding part of the KW node is not accessible, since $k_F^{(-1)} > 1/c > q$. The Kohn singularity of the band $s = 1$ is $\left( q - 2\,k_F^{(1)} \right)^2 \ln |q - 2\,k_F^{(1)}|$ for the KW node, with the coefficient $g/(16\,\pi^2\,v_0\,\mathcal{R})$, and $\left( q - 2\,k_F^{(1)} \right)^3 \ln |q - 2\,k_F^{(1)}|$ for the QTSM, with the coefficient $-\,g/(24\,\pi^2\,\mathcal{R}\,v_0\,k_F^{(1)})$. Both coefficients are inversely proportional to the Fermi speed $\mathcal{R}\,v_0$. The Friedel oscillations of this band therefore decay as $r^{-4}$ for the KW node and as $r^{-5}$ for the QTSM, which reflects the overlap factors $a$ and $a^2$, since the energies of the band are identical. The QTSM with $c > 0$ has in addition the oscillations of the flat-band and of the transitions between the two FSs, which decay as $r^{-3}$ and $r^{-4}$ and have no counterpart in the KW node within the range $c\,q < 1$.

\subsection{KW versus WSM}
\label{seccompkwwsm}

The KW node has the eigenstates and the overlap factors of the WSM, and the two nodes differ only in the energies through the quadratic term. For the band $s = 1$, the interband response is that of the WSM, with the same metric, the same transition energy $2\,v_0\,k$, and the same absorption $g\,e^2/(6\,v_0)$ above the onset $\omega_{\rm th} = 2\,v_0\,k_F^{(1)}$, with $k_F^{(1)}$ in place of $k_F$. The difference lies in the occupations. In the WSM, the transitions extend to the cut-off, whereas in the KW node, they end at the Fermi momentum $k_F^{(-1)}$ of the band $s = -1$. The coefficient of the logarithm in $\epsilon_b - 1$ is the same, namely $g\,e^2/(3\,\pi\,v_0)$, with $\Lambda$ replaced by $k_F^{(-1)}$ and $k_F$ by $k_F^{(1)}$. The cut-off dependence of the WSM is thus replaced by the finite ratio $k_F^{(-1)}/k_F^{(1)}$.

The Drude weights differ. The WSM has a single FS and $\omega_p^2 \propto \mu^2$, whereas the KW node has two FSs with the common Fermi speed $\mathcal{R}\,v_0$, and the FS of the band $s = -1$ dominates the Drude weight. The limit $c \to 0$ at fixed $\mu$ is not smooth, as discussed in Sec.~\ref{seckwlindrude}, and the contribution of the band $s = 1$ to $\omega_p^2$ tends to the WSM value in this limit. The ratio $q_{\rm TF}^2/\omega_p^2$ equals $3/V_F^2$ with $V_F = \mathcal{R}\,v_0$, instead of $3/v_0^2$.

The WSM has one undamped plasmon below $2\,\mu$, which exists for $q < k_F$. The KW node has two undamped plasmons, one below $\omega_{\rm th}$, which ends at $q_{\rm max} \leq q_c$, and one above $\omega_{\rm max}$ at long wavelengths, which is possible because the continuum has an upper edge. The Bohm-Gross velocity of the KW node is $\mathcal{R}\,v_0$, compared with $v_0$ for the WSM. The Kohn singularity of the KW node is quadratic at $2\,k_F^{(1)}$, as for the WSM at $2\,k_F$, because the overlap factor $\mathcal{F}_{1,1} = (1 + \cos\theta)/2$ is the same, and the leading Friedel oscillations of both nodes decay as $r^{-4}$. The screening factor at the singularity differs, since $\epsilon_{2\,k_1}$ of the KW node contains the contribution of the FS of the band $s = -1$.

\subsection{Friedel oscillations}
\label{seccompfried}

\begin{figure}[t!]
\centering
\includegraphics[width=0.8 \linewidth]{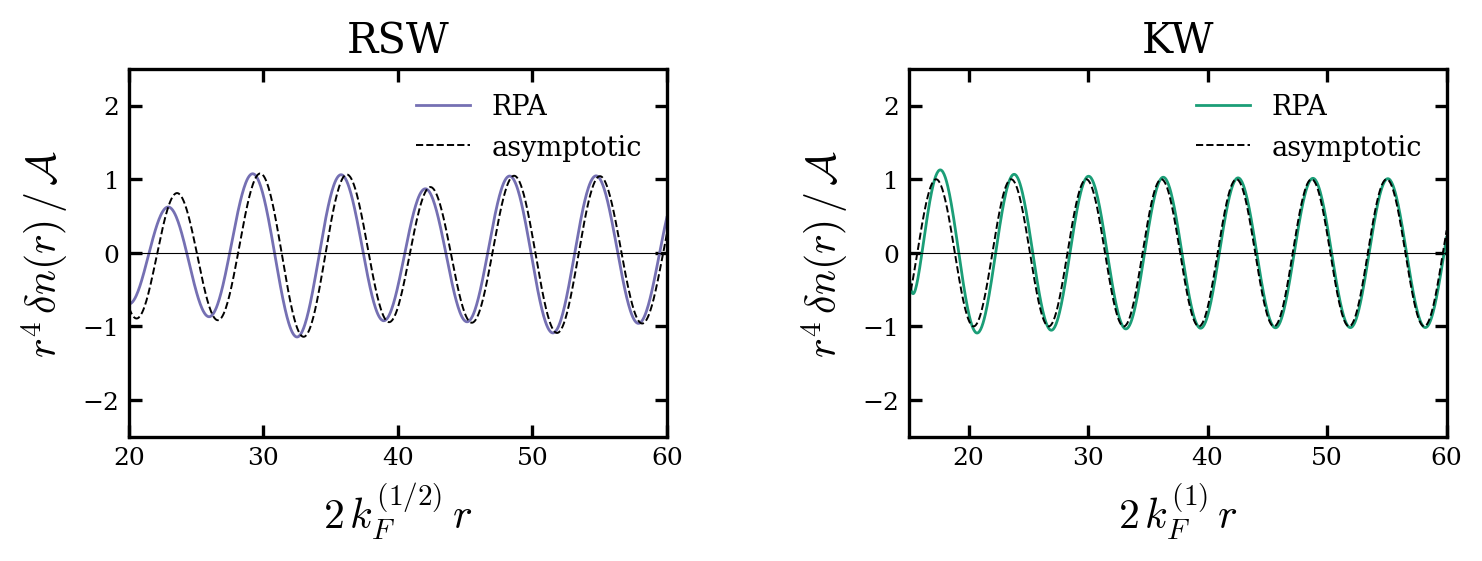}
\caption{\label{fig:friedel-rsw-kw}
Friedel oscillations for the RSW node (left) and the KW node (right).}
\end{figure}

The decay of the Friedel oscillations is fixed by the power $p$ with which the overlap factor of the responsible channel vanishes at backscattering, $\bs{k}' = -\,\bs{k}$, in terms of $a = (1 + \cos\theta)/2$. The channel gives a singularity $(q - q_0)^n \ln |q - q_0|$ of $\Pi (q, 0)$ with $n = p + 1$ at its momentum $q_0$, which is twice the Fermi momentum for an intraband channel and the sum of the Fermi momenta for an interband channel. The oscillation of $\delta n (r)$ at the wavenumber $q_0$ then decays as $r^{-n-2}$, as derived for the RSW node in Appendix~\ref{app:rswfull-friedel}. For a parabolic band in three dimensions, the overlap factor is unity, $p = 0$, and the decay is $r^{-3}$.

For the WSM and the KW node, the intraband factor is $a$, which gives $p = 1$ and $r^{-4}$. For the TSM, the factor of the top band is $a^2$, which gives $p = 2$ and $r^{-5}$, and the top band of the QTSM has the same factor. For $c > 0$, the flat-band factor $\mathcal{F}_{0 0} = \cos^2\theta = (2\,a - 1)^2$ equals unity at backscattering, which gives $p = 0$ and $r^{-3}$, and the interband channel between the two FSs has $\mathcal{F}_{0 1} = 2\,a\,b$ with $b = (1 - \cos\theta)/2$, which gives $p = 1$ and $r^{-4}$. For the RSW node, the factor is $4\,a$ for the band $s = 1/2$, $3\,a^2$ for the interband channel $1/2 \leftrightarrow 3/2$, and $a^3$ for the band $s = 3/2$, which gives $p = 1$, $2$, and $3$, and hence $r^{-4}$, $r^{-5}$, and $r^{-6}$. The leading decay of the RSW node is therefore the same as for the WSM and the KW node, and slower than for the TSM and the QTSM with $c < 0$, although its pseudospin is the largest. The QTSM with $c > 0$ decays more slowly still, since its flat-band has unit overlap at backscattering, and its three terms follow the same hierarchy as those of the RSW node, with every exponent lowered by one. The pseudospin does not fix the decay, which is fixed by the suppression of backscattering within the band that has the largest Fermi momentum. For the KW node, the anomaly of the band $s = -1$ lies outside the range $c\,q < 1$ and is therefore not included, as explained in Sec.~\ref{seckwfried}.

The induced charge density $\delta n(r)$ around a Coulomb impurity of charge $Z\,e$ is shown in Fig.~\ref{fig:friedel-rsw-kw}. Each panel displays $r^4\,\delta n(r)$ normalised by the amplitude of the leading term, which makes the leading envelope flat. The horizontal axis is scaled by $2\,k_F^{(1/2)}$ for the RSW node and by $2\,k_F^{(1)}$ for the KW node. The solid lines are the numerical Fourier transform of Eq.~\eqref{eq:rswfried-dnq} with a Gaussian regulator in $q$, which removes the short-distance part. The regulator is divided out at the leading Kohn momentum. For the RSW node, the static $\epsilon^{\rm RPA}$ has an unphysical zero just above $\Lambda$, where the linearised model is not valid. The simple pole of $1/\epsilon^{\rm RPA}$ at this zero is subtracted before the transform. The dashed lines are the asymptotic forms of Eqs.~\eqref{eq:rswfried-dn} and~\eqref{eq:kwfried-coul}. In the RSW panel, the two subleading terms, which decay as $r^{-5}$ and $r^{-6}$, make the asymptotic curve deviate from a flat envelope at moderate distances. At large distances, the numerical and the asymptotic curves agree in both panels, and the single intraband singularity at $q = 2\,k_F^{(1)}$ dominates the response of the KW node. The parameters are $\mu = 0.01$\,eV and $Z = 1$, together with $c = 0.005$ for the KW node. They give $k_\mu = 25$\,eV, $k_F^{(1/2)} = 50$\,eV, and $k_F^{(3/2)} = 50/3$\,eV for the RSW node, and $\mathcal{R} = \sqrt{3/2}$ and $k_F^{(1)} \simeq 22.5$\,eV for the KW node. The regulator widths are $100$\,eV for the RSW node and $30$\,eV for the KW node.

\section{Summary and outlook}
\label{secsum}

We have studied the collective charge excitations of the RSW and KW nodes within the RPA, as a companion to the corresponding analysis of the TSM and the WSM in Ref.~\cite{ips-tsm-plasmon}. For the TSM and the WSM, only the upper dispersive band crosses the Fermi level at a positive chemical potential. In the two nodes considered here, more than one band contributes to the intraband response, as shown in Fig.~\ref{fig:rsw-kw-bands}. The RSW node has two concentric FSs, which arise from its two upper bands, whereas both bands of the KW node cross the Fermi level because of the quadratic term.

For the RSW node, we have obtained the DOS, the Drude weights of the two FSs, and the interband screening at small momentum-transfers. The plasmon frequency is set by the sum of the two Drude weights, and the interband transitions renormalise it through a background dielectric constant, which depends on the cut-off only through a logarithm. At arbitrary momentum-transfers, we have derived the imaginary part of the polarisability channel by channel, identified the edges of the particle-hole continua, and obtained the real part through the Kramers-Kronig relation. The undamped plasmon is confined to a window below the lowest interband onset, which closes at a finite wavevector, and the Landau damping beyond it arises from the particle-hole continuum. The static polarisability and the screening are available in terms of elementary functions, and its non-analytic points give the Friedel oscillations, whose decay is fixed by the suppression of backscattering within the responsible channel.

For the KW node, the Drude weight remains finite as the chemical potential approaches the node, and the interband transitions are confined to a finite window of frequencies. The background dielectric constant is therefore independent of the cut-off. Because the continuum has an upper edge, the KW node supports a second undamped plasmon above it at long wavelengths, in addition to the one below the onset. The expressions at arbitrary momentum-transfers involve dilogarithms and are valid for momentum-transfers smaller than the inverse of the coefficient of the quadratic term. In this range, the Friedel oscillations are obtained only for the band with the smaller Fermi momentum, because the anomaly of the other band lies outside it.

The comparison with the TSM, the QTSM, and the WSM shows that the intraband response is fixed by the Fermi momenta and the band-velocities, whereas the interband response is governed by the quantum metric of the transitions and by their energies. The differences in the screening, the plasmon branches, and the range of the undamped plasmons can therefore be traced separately to the spectrum and to the eigenstates.

Several extensions are natural. The effects of rotation-symmetry-breaking terms in the RSW node can be incorporated by the same channel decomposition, at the cost of the isotropy of the response. A finite temperature smears the Fermi edges and modifies the continua, and disorder broadens the plasmons beyond the Landau damping computed here. The RPA neglects vertex corrections and local-field effects, and going beyond it would test the role of the multiple FSs in the screening. A magnetic field couples the plasmons to the BC and to the OMM of the bands. Finally, nodes of higher pseudospin and lattice models, in which the cut-off is provided by the bandstructure, would allow a quantitative assessment of the regime of validity of the linearised description.

\section*{Statements and Declarations}

\begin{itemize}

\item Funding: Not applicable.

\item Availability of data and material: This manuscript has no associated data.

\item Competing interests: The author declares that she has no known competing financial interests or personal relationships that could have appeared to influence the work reported in this paper.

\item Authors Contributions: Not applicable.

\end{itemize}

\bibliography{ref_pl}

\newpage

\appendix

\setcounter{page}{1}
\section*{Supplemental Material for ``Plasmon modes in three-dimensional multifold semimetals. II. Rarita-Schwinger-Weyl and Kramers-Weyl nodes''}

In this supplemental material, we provide the details of our derivations and arguments presented in the main text.

\section{RSW at small momentum-transfers}
\label{app:rswlin}

This appendix gives the intermediate steps behind Sec.~\ref{secrswlin}, where $\chi = 1$. Throughout, $\varepsilon_s (\bs{k}) = s\,v_0\,k$ denotes the energy of the band $s$, and $k_F^{(s)} = \mu/(s\,v_0)$ denotes the Fermi momentum of the band $s \in \{1/2, 3/2\}$ for $\mu > 0$.

\subsection{Eigenstates, overlap factors, and quantum metric}
\label{app:rsw-states}

The spin-$3/2$ matrices are $\bs{\mathcal{J}} = \sum_{n = 1}^{3} \bs{\sigma}^{(n)}/2$, with the Pauli matrices $\bs{\sigma}^{(n)}$ acting on the $n$th of three spin-$1/2$ factors. The eigenstates along a unit vector $\bs{n}$ are the symmetric states of the three spin-$1/2$ eigenstates $|{\uparrow_{\bs{n}}}\rangle$ and $|{\downarrow_{\bs{n}}}\rangle$ of $\bs{n} \cdot \bs{\sigma}$,
\begin{align}
|3/2, \bs{n} \rangle = |{\uparrow\uparrow\uparrow}\rangle\,, \quad
|1/2, \bs{n} \rangle = \frac{|{\uparrow\uparrow\downarrow}\rangle + |{\uparrow\downarrow\uparrow}\rangle + |{\downarrow\uparrow\uparrow}\rangle}{\sqrt{3}}\,, \quad
|-1/2, \bs{n} \rangle = \frac{|{\uparrow\downarrow\downarrow}\rangle + |{\downarrow\uparrow\downarrow}\rangle + |{\downarrow\downarrow\uparrow}\rangle}{\sqrt{3}}\,, \quad
|-3/2, \bs{n} \rangle = |{\downarrow\downarrow\downarrow}\rangle\,,
\end{align}
where all the spin-$1/2$ states refer to $\bs{n}$. They satisfy $\bs{n} \cdot \bs{\mathcal{J}}\,|s, \bs{n} \rangle = s\,|s, \bs{n} \rangle$. For $\mathcal{H}_{\rm RSW, \chi}$, we have $\bs{n} = \hat{\bs{\tilde k}}$, and $\bs{\tilde k} \cdot \bs{\tilde k}' = \bs{k} \cdot \bs{k}'$ because $\chi^2 = 1$. The overlap factors therefore depend only on $\cos\theta = \hat{\bs{k}} \cdot \hat{\bs{k}}'$.

Let $\bs{n}$ and $\bs{n}'$ enclose the angle $\theta$. As for the spin-$1$ case of Ref.~\cite{ips-tsm-plasmon}, the two spin-$1/2$ bases are related by an SU(2) matrix, which fixes $\langle {\uparrow'} | {\uparrow} \rangle = \langle {\downarrow'} | {\downarrow} \rangle^* = \mathfrak{u}$ and $\langle {\downarrow'} | {\uparrow} \rangle = -\,\langle {\uparrow'} | {\downarrow} \rangle^* = \mathfrak{w}$, with $|\mathfrak{u}|^2 = \cos^2 \!\left( \theta/2 \right) = a$ and $|\mathfrak{w}|^2 = \sin^2 \!\left( \theta/2 \right) = b$. The amplitudes between the spin-$3/2$ states are the sums of products of three spin-$1/2$ overlaps. For the band $s = 3/2$,
\begin{align}
\langle 3/2' | 3/2 \rangle = \mathfrak{u}^3\,, \quad
\langle 1/2' | 3/2 \rangle = \sqrt{3}\,\mathfrak{u}^2\,\mathfrak{w}\,, \quad
\langle -1/2' | 3/2 \rangle = \sqrt{3}\,\mathfrak{u}\,\mathfrak{w}^2\,, \quad
\langle -3/2' | 3/2 \rangle = \mathfrak{w}^3\,.
\end{align}
For the band $s = 1/2$, the three terms in which the position of the down spin is unchanged contribute $|\mathfrak{u}|^2\,\mathfrak{u}$ each to $\langle 1/2' | 1/2 \rangle$, and the six terms in which it changes contribute $-\,|\mathfrak{w}|^2\,\mathfrak{u}$ each. Hence
\begin{align}
\langle 1/2' | 1/2 \rangle = \mathfrak{u}\,(a - 2\,b)\,, \quad
\langle -1/2' | 1/2 \rangle = \mathfrak{w}\,(2\,a - b)\,,
\end{align}
where the second amplitude follows in the same manner. With $a - 2\,b = (3\,\cos\theta - 1)/2$ and $2\,a - b = (3\,\cos\theta + 1)/2$, the squared moduli give the factors of Eq.~\eqref{eq:rsw-ff}. The remaining factors follow from $\mathcal{F}_{s's} = \mathcal{F}_{ss'}$ and from the symmetry $\mathcal{F}_{-s,-s'} = \mathcal{F}_{ss'}$, which exchanges the roles of the up and down spins. As a check, the sum rule for the row $s = 3/2$ is $a^3 + 3\,a^2\,b + 3\,a\,b^2 + b^3 = (a + b)^3 = 1$, and the sum rule for the row $s = 1/2$ is $3\,a^2\,b + a\,(a - 2\,b)^2 + b\,(2\,a - b)^2 + 3\,a\,b^2 = (a + b)^3 = 1$.

For $\bs{k}' = \bs{k} + \bs{q}$, the sine of the angle is $\sin\theta = |\bs{k} \times \bs{q}|/(k\,k')$, and for $q \ll k$,
\begin{align}
\sin^2\theta \simeq \frac{q^2 - \left( \bs{q} \cdot \hat{\bs{k}} \right)^2}{k^2} = \frac{q_i\,q_j \left( \delta_{ij} - \hat k_i\,\hat k_j \right)}{k^2}\,.
\end{align}
With $b \simeq \sin^2\theta/4$, $a \simeq 1$, and $3\,\cos\theta + 1 \simeq 4$, we obtain $\mathcal{F}_{3/2, 1/2} \simeq 3\,\sin^2\theta/4$ and $\mathcal{F}_{1/2, -1/2} \simeq \sin^2\theta$, which give $Q^{(3/2, 1/2)}_{ij}$ and $Q^{(1/2, -1/2)}_{ij}$ of Eq.~\eqref{eq:rsw-qm-ff}. The remaining pairs are $\mathcal{F}_{3/2, -1/2} \simeq 3\,\sin^4\theta/16$ of order $q^4$ and $\mathcal{F}_{3/2, -3/2} = b^3$ of order $q^6$. The quantum metric follows from Eq.~\eqref{eq:qmetric}. For the band $s = 3/2$, only the pair $(3/2, 1/2)$ contributes at order $q^2$, and for the band $s = 1/2$, the pairs $(1/2, 3/2)$ and $(1/2, -1/2)$ contribute, which gives $Q^{(1/2)}_{ij} = (3/4 + 1)\,(\delta_{ij} - \hat k_i\,\hat k_j)/k^2$. These are the coefficients $\tilde{\mathcal{G}}_s$ of Eq.~\eqref{eq:rsw-qmetric}.

The angular average follows from $\int d\tilde\Omega\,\hat k_i\,\hat k_j = 4\,\pi\,\delta_{ij}/3$ and $\int d\tilde\Omega = 4\,\pi$. For a pair with the coefficient $\kappa$ in $Q^{(ss')}_{ij} = \kappa\,(\delta_{ij} - \hat k_i\,\hat k_j)/k^2$,
\begin{align}
\label{eq:rswlin-ang}
\int d\tilde\Omega\;k^2\,Q^{(ss')}_{ij} = \kappa \left( 4\,\pi - \frac{4\,\pi}{3} \right) \delta_{ij} = \frac{8\,\pi\,\kappa}{3}\,\delta_{ij}\,.
\end{align}
The values $\kappa = 1$ and $\kappa = 3/4$ give $8\,\pi/3$ and $2\,\pi$ for the pairs $(1/2, -1/2)$ and $(3/2, 1/2)$, respectively.

\subsection{Intrinsic limit}
\label{app:rswlin-intr}

We give the steps behind Eqs.~\eqref{eq:rswlin-pi0} to~\eqref{eq:rswlin-eps0}. For $\mu \to 0^+$ at $T = 0$, the occupations are $f_0 (\varepsilon_{-3/2}) = f_0 (\varepsilon_{-1/2}) = 1$ and $f_0 (\varepsilon_{1/2}) = f_0 (\varepsilon_{3/2}) = 0$. The occupation factor in Eq.~\eqref{eq:pol} then vanishes for $s = s'$ and for the pairs $(1/2, 3/2)$ and $(-1/2, -3/2)$ in either order. The pairs $(-3/2, 1/2)$, $(-1/2, 3/2)$, and $(-3/2, 3/2)$ carry the factors of order $q^4$ and $q^6$ and are dropped. The pair $(-1/2, 1/2)$ has the occupation factor $+1$ and the denominator $\omega - v_0\,k'$ at the leading order, while the pair $(1/2, -1/2)$ has the factor $-1$ and the denominator $\omega + v_0\,k$. Since $\mathcal{F}_{1/2, -1/2}$ is already of order $q^2$, we set $k' = k$ in the denominators, and the sum becomes
\begin{align}
\mathcal{F}_{1/2, -1/2} \left( \frac{1}{\omega - v_0\,k + i\,0^+} - \frac{1}{\omega + v_0\,k + i\,0^+} \right) = \mathcal{F}_{1/2, -1/2}\,\frac{2\,v_0\,k}{(\omega + i\,0^+)^2 - v_0^2\,k^2}\,.
\end{align}
As for the TSM~\cite{ips-tsm-plasmon}, this replacement requires $v_0\,q \ll \omega$.

With the measure $d^3 \bs{k} = k^2\,dk\,d\tilde\Omega$, the angular integral of Eq.~\eqref{eq:rswlin-ang} with $\kappa = 1$ yields the prefactor $g\,q_i\,q_j\,(8\,\pi/3)\,\delta_{ij}/(2\,\pi)^3 = g\,q^2/(3\,\pi^2)$ of Eq.~\eqref{eq:rswlin-pi0}. The principal value of the radial integral is
\begin{align}
\Re \int_0^{\Lambda} dk\, \frac{2\,v_0\,k}{\omega^2 - v_0^2\,k^2} = -\,\frac{1}{v_0} \left[ \ln \!\left( \left| \omega^2 - v_0^2\,k^2 \right| \right) \right]_0^{\Lambda} \simeq -\,\frac{1}{v_0}\,\ln \!\left( \frac{v_0^2\,\Lambda^2}{\omega^2} \right),
\end{align}
where the last step uses $\omega \ll v_0\,\Lambda$. The imaginary part is $-\,\pi \int_0^{\Lambda} dk\, \delta (\omega - v_0\,k) = -\,\pi/v_0$ for $\omega > 0$. Multiplying these results by $g\,q^2/(3\,\pi^2)$ gives Eq.~\eqref{eq:rswlin-pi0b}. Because $\Pi_{\rm intra} = 0$ at the intrinsic point, Eq.~\eqref{eq:rswlin-eps0} follows from $\Re \epsilon^{\rm RPA} = 1 - 4\,\pi\,e^2\,\Re \Pi/q^2$ and $\Im \epsilon^{\rm RPA} = -\,4\,\pi\,e^2\,\Im \Pi/q^2$, where the second relation gives $4\,\pi\,e^2\,g/(3\,\pi\,v_0) = 4\,g\,e^2/(3\,v_0)$.

\subsection{Doped regime: DOS and Drude weights}
\label{app:rswlin-drude}

We give the steps behind Eqs.~\eqref{eq:rswlin-dos} and~\eqref{eq:rswlin-wp}. For $\mu > 0$, the bands $s = -3/2$ and $s = -1/2$ are filled, and the bands $s = 1/2$ and $s = 3/2$ have the FSs with $\delta (\varepsilon_s - \mu) = \delta (k - k_F^{(s)})/(s\,v_0)$. The DOS of the band $s$, $\rho_s = g \int d^3 \bs{k}/(2\,\pi)^3\,\delta (\varepsilon_s - \mu)$, is
\begin{align}
\rho_s (\mu) = \frac{g}{(2\,\pi)^3}\,\frac{4\,\pi \left( k_F^{(s)} \right)^2}{s\,v_0} = \frac{g\,\mu^2}{2\,\pi^2\,s^3\,v_0^3}\,.
\end{align}
The sum over the two bands uses $\sum_s s^{-3} = 8 + 8/27 = 224/27$ and gives $\rho = 112\,g\,\mu^2/(27\,\pi^2\,v_0^3)$. The density measured from $\mu = 0$ is $\tilde n = \int_0^\mu d\mu'\,\rho (\mu')$, which equals the number of states inside the two Fermi spheres, $g \sum_s 4\,\pi \left( k_F^{(s)} \right)^3/[3\,(2\,\pi)^3]$. With $\bs{V}^{(s)} = s\,v_0\,\hat{\bs{k}}$ and $\langle \hat k_i\,\hat k_j \rangle = \delta_{ij}/3$, Eq.~\eqref{eq:drude} gives
\begin{align}
D_{ij} = \pi\,e^2 \sum_s \rho_s\,s^2\,v_0^2\,\frac{\delta_{ij}}{3} = \frac{g\,e^2\,\mu^2}{6\,\pi\,v_0} \sum_s \frac{1}{s}\,\delta_{ij} = \frac{g\,e^2\,\mu^2}{6\,\pi\,v_0} \left( 2 + \frac{2}{3} \right) \delta_{ij}\,.
\end{align}
The contributions of the bands $s = 1/2$ and $s = 3/2$ are in the ratio $3 : 1$.

We now derive Eq.~\eqref{eq:intra}. The filled bands do not contribute, and $\mathcal{F}_{ss} = 1 - \mathcal{O} \left( q^2 \right)$ for the partially occupied bands $s = 1/2$ and $s = 3/2$. Expanding in $q$ at fixed $\omega$ and $T = 0$ gives
\begin{align}
f_0 (\varepsilon_s (\bs{k})) - f_0 (\varepsilon_s (\bs{k}')) \simeq q_i\,V_i^{(s)}\,\delta (\varepsilon_s - \mu)\,, \quad
\omega + \varepsilon_s (\bs{k}) - \varepsilon_s (\bs{k}') \simeq \omega - q_j\,V_j^{(s)}\,.
\end{align}
Expanding $1/(\omega - q_j\,V_j^{(s)})$ in powers of $q/\omega$ gives
\begin{align}
\Pi_{\rm intra} (\bs{q}, \omega) \simeq g \sum_s \int \frac{d^3 \bs{k}}{(2\,\pi)^3}\,\delta (\varepsilon_s - \mu) \left[ \frac{q_i\,V_i^{(s)}}{\omega} + \frac{q_i\,q_j\,V_i^{(s)}\,V_j^{(s)}}{\omega^2} \right].
\end{align}
The first term vanishes, because it is odd under $\hat{\bs{k}} \to -\,\hat{\bs{k}}$. The second term equals $q_i\,q_j\,D_{ij}/(\pi\,e^2\,\omega^2)$, which is Eq.~\eqref{eq:intra}. Since $D_{ij}$ is proportional to $\delta_{ij}$, the zero of $1 - 4\,D_{ii}/\omega^2$ gives $\omega_p^2 = 4\,D_{ii} = 16\,g\,e^2\,\mu^2/(9\,\pi\,v_0)$.

\subsection{Interband screening and plasmons}
\label{app:rswlin-screen}

We provide the steps behind Eqs.~\eqref{eq:rswlin-eps} and~\eqref{eq:rswlin-eb} and behind the bound on $\omega_*$ in Sec.~\ref{secrswlinscreen}, assuming $\mu > 0$, $v_0\,q \ll \omega$, and $q \ll k_F^{(3/2)}$ for the expansion of the overlap factors. The occupations are $f_0 (\varepsilon_{-3/2}) = f_0 (\varepsilon_{-1/2}) = 1$ and $f_0 (\varepsilon_s) = \Theta (k_F^{(s)} - k)$ for $s = 1/2, 3/2$. Both Fermi momenta satisfy $k_F^{(3/2)} = k_F^{(1/2)}/3$. The pairs $(-3/2, -1/2)$ remain blocked at every $k$, and the pairs of order $q^4$ and $q^6$ are dropped as in Appendix~\ref{app:rswlin-intr}.

For the pairs $(-1/2, 1/2)$ and $(1/2, -1/2)$, the occupation factors are $1 - \Theta (k_F^{(1/2)} - k) = \Theta (k - k_F^{(1/2)})$ and its negative. The result of Appendix~\ref{app:rswlin-intr} therefore holds with the lower limit of the radial integral replaced by $k_F^{(1/2)}$. The real part becomes
\begin{align}
\Re \int_{k_F^{(1/2)}}^{\Lambda} dk\, \frac{2\,v_0\,k}{\omega^2 - v_0^2\,k^2} = -\,\frac{1}{v_0}\,\ln \!\left( \frac{v_0^2\,\Lambda^2 - \omega^2}{|4\,\mu^2 - \omega^2|} \right),
\end{align}
and the imaginary part is $-\,\pi\,\Theta (\omega - 2\,\mu)/v_0$, because the delta function fixes $k = \omega/v_0$, which lies above $k_F^{(1/2)}$ if and only if $\omega > 2\,\mu$.

For the pairs $(1/2, 3/2)$ and $(3/2, 1/2)$, the occupation factor is $f_0 (\varepsilon_{1/2}) - f_0 (\varepsilon_{3/2}) = \Theta (k_F^{(1/2)} - k) - \Theta (k_F^{(3/2)} - k)$, which equals $1$ for $k_F^{(3/2)} < k < k_F^{(1/2)}$ and vanishes otherwise. The energy difference is $\varepsilon_{3/2} - \varepsilon_{1/2} = v_0\,k$, and the angular average of Eq.~\eqref{eq:rswlin-ang} with $\kappa = 3/4$ gives the prefactor $g\,q_i\,q_j\,2\,\pi\,\delta_{ij}/(2\,\pi)^3 = g\,q^2/(4\,\pi^2)$. Hence
\begin{align}
\Pi_{1/2 \leftrightarrow 3/2} (\bs{q}, \omega) = \frac{g\,q^2}{4\,\pi^2} \int_{k_F^{(3/2)}}^{k_F^{(1/2)}} dk\, \frac{2\,v_0\,k}{(\omega + i\,0^+)^2 - v_0^2\,k^2}\,,
\end{align}
with the real part $-\,[g\,q^2/(4\,\pi^2\,v_0)]\,\ln \!\left( |4\,\mu^2 - \omega^2|/|4\,\mu^2/9 - \omega^2| \right)$, where $v_0\,k_F^{(1/2)} = 2\,\mu$ and $v_0\,k_F^{(3/2)} = 2\,\mu/3$ have been used. The imaginary part is $-\,g\,q^2/(4\,\pi\,v_0)$ for $2\,\mu/3 < \omega < 2\,\mu$ and zero otherwise.

After adding the three contributions and $\Pi_{\rm intra} = q^2\,D_{ii}/(\pi\,e^2\,\omega^2)$ from Appendix~\ref{app:rswlin-drude}, Eq.~\eqref{eq:epsrpa} gives Eq.~\eqref{eq:rswlin-eps}, where the coefficients follow from $4\,\pi\,e^2\,g/(3\,\pi^2\,v_0) = 4\,g\,e^2/(3\,\pi\,v_0)$ and $4\,\pi\,e^2\,g/(4\,\pi^2\,v_0) = g\,e^2/(\pi\,v_0)$. In the static limit $\omega \to 0$, the two logarithms tend to $2\,\ln \!\left( v_0\,\Lambda/(2\,\mu) \right)$ and $\ln 9$, which gives Eq.~\eqref{eq:rswlin-eb}.

For the zero of the real part, we write $\Re \epsilon^{\rm RPA} = \epsilon_{\rm inter} (\omega) - \omega_p^2/\omega^2$ for $0 < \omega < 2\,\mu/3$, with
\begin{align}
\epsilon_{\rm inter} (\omega) = 1 + \frac{4\,g\,e^2}{3\,\pi\,v_0}\,\ln \!\left( \frac{v_0^2\,\Lambda^2 - \omega^2}{4\,\mu^2 - \omega^2} \right) + \frac{g\,e^2}{\pi\,v_0}\,\ln \!\left( \frac{4\,\mu^2 - \omega^2}{4\,\mu^2/9 - \omega^2} \right).
\end{align}
The derivative of the first logarithm is $2\,\omega\left[ 1/(4\,\mu^2 - \omega^2) - 1/(v_0^2\,\Lambda^2 - \omega^2) \right]$, which is positive for $4\,\mu^2 < v_0^2\,\Lambda^2$. The derivative of the second logarithm is $2\,\omega \left[ 1/(4\,\mu^2/9 - \omega^2) - 1/(4\,\mu^2 - \omega^2) \right]$, which is positive as well. The derivative of the Drude term, $2\,\omega_p^2/\omega^3$, is also positive. The real part is therefore continuous and strictly increasing on $(0, 2\,\mu/3)$. It tends to $-\,\infty$ for $\omega \to 0$ because the Drude term diverges whereas $\epsilon_{\rm inter} (\omega)$ stays finite, and it tends to $+\,\infty$ for $\omega \to (2\,\mu/3)^-$ because the second logarithm diverges. There is hence exactly one zero $\omega_*$ in $(0, 2\,\mu/3)$. Since $\epsilon_{\rm inter} (\omega) \geq \epsilon_b$, we have
\begin{align}
0 = \Re \epsilon^{\rm RPA} (\omega_*) \geq \epsilon_b - \frac{\omega_p^2}{\omega_*^2} \quad \Rightarrow \quad \omega_*^2 \leq \frac{\omega_p^2}{\epsilon_b}\,,
\end{align}
which is the bound stated in Sec.~\ref{secrswlinscreen}. The imaginary part of $\epsilon^{\rm RPA}$ vanishes below $\omega_{\rm th} = 2\,\mu/3$, and the mode at $\omega_*$ is therefore undamped.

\section{RSW at arbitrary momentum-transfers}
\label{app:rswfull}

We give the steps behind Sec.~\ref{secrswim}. Throughout, $k_\mu = \mu/v_0 > 0$, $\varpi > 0$, and $k' = |\bs{k} + \bs{q}|$.

\subsection{Imaginary part of polarisability: Particle-hole continua and their edges}
\label{app:rswfull-im}

\subsubsection*{Reduction to a single integral}

For $\varpi > 0$, the imaginary part of Eq.~\eqref{eq:pol} is
\begin{align}
\Im \Pi (\bs{q}, \omega) = -\,\pi\,g \int \frac{d^3 \bs{k}}{(2\,\pi)^3} \sum_{s, s'} \mathcal{F}_{ss'} \left[ f_0 (\varepsilon_s (\bs{k})) - f_0 (\varepsilon_{s'} (\bs{k}')) \right] \delta \!\left( \omega + \varepsilon_s (\bs{k}) - \varepsilon_{s'} (\bs{k}') \right).
\end{align}
The delta function is nonzero only if $\varepsilon_{s'} (\bs{k}') > \varepsilon_s (\bs{k})$. At $T = 0$, the occupation difference then equals $1$ if the state $(s, \bs{k})$ is occupied and the state $(s', \bs{k}')$ is empty, and it vanishes otherwise, since an occupied final state of higher energy implies an occupied initial state. The pairs within the filled bands $s = -3/2$ and $s = -1/2$ are blocked, and the pairs with $\varepsilon_{s'} (\bs{k}') < \varepsilon_s (\bs{k})$ do not contribute. The eight channels $\lambda\,s \to s'$ of Sec.~\ref{secrswim} remain. The measure is $d^3 \bs{k} = 2\,\pi\,k\,k'\,dk\,dk'/q$ on the region $|k - q| \leq k' \leq k + q$, and the delta function gives $\int dk'\,\delta (\omega - v_0 \left( s'\,k' - \lambda\,s\,k \right)) = 1/(v_0\,s')$ at $k' = (\varpi + \lambda\,s\,k)/s'$. Hence
\begin{align}
\Im \Pi^{(\lambda\,s \to s')} = -\,\frac{g}{4\,\pi\,v_0\,q\,s'} \int dk\; k\,k'\,\mathcal{F}_{\lambda\,s,\,s'}\,.
\end{align}
By the law of cosines, $1 + \cos\theta = A/(2\,k\,k')$ and $1 - \cos\theta = B/(2\,k\,k')$, which gives $a = A/(4\,k\,k')$ and $b = B/(4\,k\,k')$. Inserting these into Eq.~\eqref{eq:rsw-ff}, and using $\mathcal{F}_{-s,-s'} = \mathcal{F}_{ss'}$ for the valence channels, we obtain the polynomials of Eq.~\eqref{eq:rswim-P}. For example, $\mathcal{F}_{1/2, 1/2} = a \left( a - 2\,b \right)^2 = A \left( A - 2\,B \right)^2/[64\,(k\,k')^3]$, since $\left( 3\,\cos\theta - 1 \right)/2 = a - 2\,b$. This yields Eq.~\eqref{eq:rswim-int}.

\subsubsection*{Limits of integration}

The conditions on $k$ are linear. The triangle condition consists of $k' \leq k + q$, $k' \geq k - q$, and $k' \geq q - k$. With $d = s' - \lambda\,s$, they read
\begin{align}
d\,k \geq \varpi - s'\,q\,, \quad d\,k \leq \varpi + s'\,q\,, \quad \left( \lambda\,s + s' \right) k \geq s'\,q - \varpi\,.
\end{align}
The remaining conditions are $k' > 0$ and the occupation constraints. The initial state is occupied if $s\,k < k_\mu$ for $\lambda = +$, and always for $\lambda = -$. The final state is empty if $s'\,k' > k_\mu$, which reads $\varpi + \lambda\,s\,k > k_\mu$. The interval of $k$ is the intersection of these bounds, and comparing every lower bound with every upper bound gives the conditions on $(q, \varpi)$ for a nonempty interval. After discarding the redundant bounds, we obtain Table~\ref{tab:rswim-ranges}. For $d = 0$, which occurs for the intraband channels with $\lambda = +$, the first two inequalities do not involve $k$ and reduce to $\varpi \leq s'\,q$ and $\varpi \geq -\,s'\,q$. For $\left( \lambda\,s + s' \right) = 0$, which occurs for the channels $-1/2 \to 1/2$ and $-3/2 \to 3/2$, the third inequality reduces to $\varpi \geq s'\,q$.

\subsubsection*{Closed form of the channels}

Under $\varpi \to -\,\varpi$ in a conduction channel $s \to s'$, the final momentum changes sign, $k' \to -\,k'$, and hence $A \to -\,B$ and $B \to -\,A$. The polynomials satisfy $\mathcal{P}_{s,s'} (-\,B, -\,A) = -\,\mathcal{P}_{-s,s'} (A, B)$, because $\mathcal{F}_{-s,s'} = \mathcal{F}_{s,-s'}$ exchanges $a$ and $b$, and the denominator $(k\,k')^2$ is unchanged. The integrand of the valence channel $-s \to s'$ is therefore minus that of the conduction channel $s \to s'$ evaluated at $-\,\varpi$. We therefore need only the four conduction integrands. In terms of $k$ and $k'$, the weight $W = \mathcal{P}_{s,s'}/[64\,(k\,k')^2]$ has the partial-fraction form
\begin{align}
\label{eq:rswim-pf}
W_{ss'} (k) = \beta_2\,k^2 + \beta_1\,k + \beta_0 + \frac{c_2}{k^2} + \frac{c_1}{k} + \frac{d_2}{k'^2} + \frac{d_1}{k'}\,.
\end{align}
The coefficients are given in Tables~\ref{tab:rswim-pf1} and~\ref{tab:rswim-pf2}, in terms of
\begin{align}
\mathcal{Z}_{1/2} = \frac{\left( q^2 - 4\,\varpi^2 \right)^3}{\varpi^2}\,, \quad
\mathcal{Y}_{1/2} = \frac{\left( q^2 - 4\,\varpi^2 \right)^2 \left( 3\,q^2 + 28\,\varpi^2 \right)}{\varpi^3}\,, \quad
\mathcal{Z}_{3/2} = \frac{\left( 9\,q^2 - 4\,\varpi^2 \right)^3}{\varpi^2}\,, \quad
\mathcal{Y}_{3/2} = \frac{\left( 9\,q^2 - 4\,\varpi^2 \right)^2 \left( 9\,q^2 + 20\,\varpi^2 \right)}{\varpi^3}\,.
\end{align}
The poles at $k = 0$ carry the coefficients of the band $s'$, and the poles at $k' = 0$ carry those of the band $s$.

\begin{table}[t!]
\centering
\caption{\label{tab:rswim-pf1} Polynomial coefficients of Eq.~\eqref{eq:rswim-pf}.}
\small
\renewcommand{\arraystretch}{1.4}
\begin{tabular}{|c|c|c|c|}
\hline
$s \to s'$ & $\beta_2$ & $\beta_1$ & $\beta_0$ \\
\hline
$\frac{1}{2} \to \frac{1}{2}$ & $1$ & $2\,\varpi$ & $7\,\varpi^2 - \frac{7\,q^2}{4}$ \\
\hline
$\frac{3}{2} \to \frac{3}{2}$ & $1$ & $\frac{2\,\varpi}{3}$ & $\frac{\varpi^2}{3} - \frac{3\,q^2}{4}$ \\
\hline
$\frac{1}{2} \to \frac{3}{2}$ & $-\frac{16}{27}$ & $\frac{64\,\varpi}{27}$ & $2\,q^2 - \frac{56\,\varpi^2}{9}$ \\
\hline
$\frac{3}{2} \to \frac{1}{2}$ & $-\frac{16}{3}$ & $-\frac{128\,\varpi}{9}$ & $2\,q^2 - \frac{40\,\varpi^2}{3}$ \\
\hline
\end{tabular}
\end{table}

\begin{table}[t!]
\centering
\caption{\label{tab:rswim-pf2} Pole coefficients of Eq.~\eqref{eq:rswim-pf}.}
\small
\renewcommand{\arraystretch}{1.4}
\begin{tabular}{|c|c|c|c|c|}
\hline
$s \to s'$ & $c_2$ & $c_1$ & $d_2$ & $d_1$ \\
\hline
$\frac{1}{2} \to \frac{1}{2}$ & $-\frac{9}{256}\,\mathcal{Z}_{1/2}$ & $\frac{3}{256}\,\mathcal{Y}_{1/2}$ & $-\frac{9}{256}\,\mathcal{Z}_{1/2}$ & $-\frac{3}{256}\,\mathcal{Y}_{1/2}$ \\
\hline
$\frac{3}{2} \to \frac{3}{2}$ & $-\frac{1}{20736}\,\mathcal{Z}_{3/2}$ & $\frac{1}{6912}\,\mathcal{Y}_{3/2}$ & $-\frac{1}{20736}\,\mathcal{Z}_{3/2}$ & $-\frac{1}{6912}\,\mathcal{Y}_{3/2}$ \\
\hline
$\frac{1}{2} \to \frac{3}{2}$ & $\frac{1}{6912}\,\mathcal{Z}_{3/2}$ & $-\frac{1}{6912}\,\mathcal{Y}_{3/2}$ & $\frac{3}{256}\,\mathcal{Z}_{1/2}$ & $\frac{3}{256}\,\mathcal{Y}_{1/2}$ \\
\hline
$\frac{3}{2} \to \frac{1}{2}$ & $\frac{3}{256}\,\mathcal{Z}_{1/2}$ & $-\frac{3}{256}\,\mathcal{Y}_{1/2}$ & $\frac{1}{6912}\,\mathcal{Z}_{3/2}$ & $\frac{1}{6912}\,\mathcal{Y}_{3/2}$ \\
\hline
\end{tabular}
\end{table}

Since $dk' = (s/s')\,dk$ for $\lambda = +$, the antiderivative of Eq.~\eqref{eq:rswim-pf} is
\begin{align}
\label{eq:rswim-G}
\mathcal{I}^{+}_{ss'} (p) = \frac{\beta_2\,p^3}{3} + \frac{\beta_1\,p^2}{2} + \beta_0\,p - \frac{c_2}{p} + c_1\,\ln p + \frac{s'}{s} \left[ -\,\frac{d_2}{p'} + d_1\,\ln \!\left( |p'| \right) \right], \quad p' = \frac{\varpi + s\,p}{s'}\,,
\end{align}
and the valence channels follow from $\mathcal{I}^{-}_{ss'} (p; \varpi) = -\,\mathcal{I}^{+}_{ss'} (p; -\,\varpi)$. The imaginary part of each channel is then
\begin{align}
\Im \Pi^{(\lambda\,s \to s')} = -\,\frac{g}{4\,\pi\,v_0\,q\,s'} \left[ \mathcal{I}^{\lambda}_{ss'} (k_+) - \mathcal{I}^{\lambda}_{ss'} (k_-) \right].
\end{align}
At a channel onset, $k_-$ may tend to zero or $k'$ may tend to zero at a limit. The individual pole terms then diverge, and their sum has a finite limit.

\subsubsection*{Intrinsic limit}

For $k_\mu \to 0$, the occupation constraints disappear, and only the valence channels remain, because the conduction channels require an occupied initial state in the empty bands. For $\varpi > 3\,q/2$, the triangle condition fixes the limits of the four channels as $k_\pm = \varpi \pm q/2$ for $-1/2 \to 1/2$, $k_\pm = (\varpi \pm 3\,q/2)/2$ for $-1/2 \to 3/2$, $k_\pm = (\varpi \pm q/2)/2$ for $-3/2 \to 1/2$, and $k_\pm = (\varpi \pm 3\,q/2)/3$ for $-3/2 \to 3/2$. Inserting them into Eq.~\eqref{eq:rswim-G}, the logarithms cancel in the sum of the four channels, although not in any pair of them, and the remaining terms add up to the constant of Eq.~\eqref{eq:rswim-intr}.

For $q/2 < \varpi < 3\,q/2$, the channel $-3/2 \to 3/2$ is absent, and the limits of the other channels change to $k_- = 3\,q/2 - \varpi$ for $-1/2 \to 3/2$ and $k_+ = \varpi - q/2$ for $-3/2 \to 1/2$. The logarithms no longer cancel. Collecting the terms with the same argument, we obtain Eq.~\eqref{eq:rswim-intr2}. At the lower end, $\mathfrak{f} (1/2^+) = 18/72 = 1/4$, and the logarithmic terms vanish because their prefactors contain $\left( 4\,v^2 - 1 \right)^2$. At the upper end, $\mathfrak{f} (3/2^-) = 594/648 = 11/12$, and the logarithmic terms vanish because their prefactors contain $\left( 9 - 4\,v^2 \right)^2$ and the argument of the first logarithm equals $1$. At $\varpi = q/2$, the channel $-1/2 \to 1/2$ has $k_- = 0$ and $k_+ = q$ with $k' = q - k$, which corresponds to antiparallel $\bs{k}$ and $\bs{k}'$. The overlap factor equals $1$ there, and the integral gives $-\,\left( g/(2\,\pi\,v_0\,q) \right) \int_0^q dk\,k \left( q - k \right) = -\,g\,q^2/(12\,\pi\,v_0)$. The step at $\varpi = 3\,q/2$ follows in the same manner from the channel $-3/2 \to 3/2$, with the prefactor $g/(6\,\pi\,v_0\,q)$ and the same integral, which gives $-\,g\,q^2/(36\,\pi\,v_0)$.

\subsubsection*{Doped regime: edges and free regions}

For $\varpi > 3\,q/2 + 4\,k_\mu$, the occupation constraints are inactive in the four valence channels, because the upper limits $2\,(\varpi - k_\mu)$, $2\,(\varpi - k_\mu)$, $2\,(\varpi - k_\mu)/3$, and $2\,(\varpi - k_\mu)/3$ exceed the triangle limits $\varpi + q/2$, $(\varpi + 3\,q/2)/2$, $(\varpi + q/2)/2$, and $(\varpi + 3\,q/2)/3$ once $\varpi \geq q/2 + 2\,k_\mu$, $q/2 + 4\,k_\mu/3$, $3\,q/2 + 4\,k_\mu$, and $3\,q/2 + 2\,k_\mu$, respectively. The conduction channels vanish for $\varpi > 2\,k_\mu + 3\,q/2$. The sum therefore equals the intrinsic one.

The edges follow from Table~\ref{tab:rswim-ranges}. A point of the $(q, \varpi)$ plane is free of single-particle excitations if it lies outside the ranges of all eight channels. For $q < k_\mu/3$, the lowest lower edge among the interband channels is $2\,k_\mu/3 - q/2$ of the channel $1/2 \to 3/2$, because the lower edges of the other interband channels are at least $2\,k_\mu/3 + q/2$, $2\,k_\mu - 3\,q/2$, $2\,k_\mu - q/2$, and $4\,k_\mu - 3\,q/2$, and the channels $1/2 \to 1/2$, $3/2 \to 3/2$, and $3/2 \to 1/2$ end at $3\,q/2$ at the latest. This gives the first region of Eq.~\eqref{eq:rswim-free}, which is nonempty if $3\,q/2 < 2\,k_\mu/3 - q/2$. For $q > 4\,k_\mu$, the lowest lower edge among all channels is $q/2 - 2\,k_\mu$ of the intraband channel $1/2 \to 1/2$. The lower edges of the other channels are $3\,q/2 - 2\,k_\mu$, $3\,q/2 - 4\,k_\mu$, $q/2 - 4\,k_\mu/3$, and at least $q/2$, which are all larger for $q > 4\,k_\mu$. This gives the second region. For the remaining values of $q$, the ranges of Table~\ref{tab:rswim-ranges} overlap and cover the whole half-line $\varpi > 0$.


\subsection{Real part of polarisability and RPA dielectric function}
\label{app:rswfull-re}

We write $\Pi = \Pi^{(0)} + \Pi_\mu$, where $\Pi^{(0)}$ is the polarisability of the undoped system with the same cut-off, and $\Pi_\mu$ is the change caused by filling the bands $s = 1/2$ and $s = 3/2$ up to $\mu$. Only the terms of Eq.~\eqref{eq:pol} containing the occupation of a conduction state contribute to $\Pi_\mu$, and they involve states with $k \leq k_F^{(s)}$ only. The cut-off therefore does not enter $\Pi_\mu$. The retarded function $\Pi_\mu$ is analytic in the upper half of the complex $\omega$-plane. It decays as $1/\omega^2$ for $|\omega| \to \infty$, and its imaginary part is odd in $\omega$. Hence
\begin{align}
\Re \Pi_\mu (q, \omega) = \frac{2}{\pi}\,\mathrm{P} \!\! \int_0^\infty d\omega'\, \frac{\omega'\,\Im \Pi_\mu (q, \omega')}{\omega'^2 - \omega^2}\,,
\end{align}
with $\Im \Pi_\mu = \Im \Pi - \Im \Pi^{(0)}$. By Sec.~\ref{secrswim}, $\Im \Pi_\mu$ vanishes for $\omega' > \omega_{\rm w} = v_0 \left( 3\,q/2 + 4\,k_\mu \right)$, since every channel has reached its intrinsic value there. The integral is therefore restricted to the window $0 < \omega' < \omega_{\rm w}$, and it does not depend on $\Lambda$. The intrinsic part obeys the same relation with the upper limit $v_0\,\Lambda$ on the transition energy, and for $v_0\,\Lambda \gg \omega, v_0\,q$ we split $\Im \Pi^{(0)} = -\,g\,q^2\,\mathfrak{f}_0 (\omega'/(v_0\,q))/(3\,\pi\,v_0)$ into the constant $\Theta (\omega' - 3\,v_0\,q/2)$ of Eq.~\eqref{eq:rswim-intr} and the remainder $\mathfrak{f}_0 - \Theta$. The constant contributes
\begin{align}
-\,\frac{2\,g\,q^2}{3\,\pi^2\,v_0}\,\mathrm{P} \!\! \int_{3\,v_0\,q/2}^{v_0\,\Lambda} d\omega'\, \frac{\omega'}{\omega'^2 - \omega^2} = -\,\frac{g\,q^2}{3\,\pi^2\,v_0}\,\ln \!\left( \frac{v_0^2\,\Lambda^2}{\left| \omega^2 - 9\,v_0^2\,q^2/4 \right|} \right),
\end{align}
where we have neglected $\omega^2$ against $v_0^2\,\Lambda^2$ in the numerator of the argument. The remainder is nonzero only for $1/2 < v < 3/2$ with $v = \omega'/(v_0\,q)$, where it equals $\mathfrak{f} (v)$, and it contributes
\begin{align}
-\,\frac{2\,g\,q^2}{3\,\pi^2\,v_0}\,\mathrm{P} \!\! \int_{1/2}^{3/2} dv\, \frac{v\,\mathfrak{f} (v)}{v^2 - W^2} = -\,\frac{g\,q^2}{3\,\pi^2\,v_0}\,\mathcal{K} (W)\,,
\end{align}
which gives Eqs.~\eqref{eq:rswre-kk} and~\eqref{eq:rswre-K}. The dielectric function of Eq.~\eqref{eq:rswre-eps} follows from Eq.~\eqref{eq:epsrpa} with the prefactors $4\,\pi\,e^2 \cdot g/(3\,\pi^2\,v_0) = 4\,g\,e^2/(3\,\pi\,v_0)$ and $4\,\pi\,e^2 \cdot 2/\pi = 8\,e^2$.

For $W > 3/2$, the denominator of Eq.~\eqref{eq:rswre-K} does not vanish, and the integrand is a finite combination of elementary functions with the logarithms of Eq.~\eqref{eq:rswim-intr2}. For $W \to \infty$, we expand $1/(v^2 - W^2) = -\,W^{-2} \left( 1 + v^2/W^2 + \dots \right)$ and use the moments
\begin{align}
\int_{1/2}^{3/2} dv\, v\,\mathfrak{f} (v) = 1 - \frac{3 \ln 2}{5}\,, \quad
\int_{1/2}^{3/2} dv\, \frac{\mathfrak{f} (v)}{v} = \frac{31 \ln 3}{8} - \frac{10 \ln 2}{3} - \frac{31}{18}\,,
\end{align}
which are elementary because the logarithms of Eq.~\eqref{eq:rswim-intr2} have arguments of the form $(2\,v - 1)/(2\,v + 1)$ and $(2\,v + 3)/(3 - 2\,v)$. The first moment gives the decay of $\mathcal{K}$ stated in Sec.~\ref{secrswre}, and the second gives $\mathcal{K} (0)$.


\subsection{Static polarisability and screening}
\label{app:rswfull-static}

We evaluate $\Pi_\mu (q, 0)$ of Appendix~\ref{app:rswfull-re} directly from Eq.~\eqref{eq:pol}. Under the exchange of $(s, \bs{k})$ and $(s', \bs{k}')$, the overlap factor is invariant, $\mathcal{F}_{s's} (\bs{k}', \bs{k}) = \mathcal{F}_{ss'} (\bs{k}, \bs{k}')$, and the denominator changes sign. At $\omega = 0$, the terms of Eq.~\eqref{eq:pol} with the occupation of the final state are therefore equal to those with the occupation of the initial state, and
\begin{align}
\Pi (q, 0) = 2\,g \sum_{s, s'} \mathrm{P} \!\! \int \frac{d^3 \bs{k}}{(2\,\pi)^3}\, f_0 (\varepsilon_s (\bs{k}))\, \frac{\mathcal{F}_{ss'} (\bs{k}, \bs{k}')}{\varepsilon_s (\bs{k}) - \varepsilon_{s'} (\bs{k}')}\,.
\end{align}
The terms with $s < 0$ do not depend on $\mu$ and belong to $\Pi^{(0)}$. The terms with $s = 1/2$ and $s = 3/2$ give $\Pi_\mu$. With $d^3 \bs{k} = 2\,\pi\,k\,k'\,dk\,dk'/q$ as in Appendix~\ref{app:rswfull-im}, we obtain
\begin{align}
\label{eq:rswstat-app-pi}
& \Pi_\mu (q, 0) = -\,\frac{g}{2\,\pi^2\,v_0\,q} \sum_{s = 1/2, 3/2} \int_0^{k_F^{(s)}} dk\; J_s (k)\,, \nn
& J_s (k) = \sum_{s'} \mathrm{P} \!\! \int_{|k - q|}^{k + q} dk'\, \frac{k\,k'\,\mathcal{F}_{ss'}}{s'\,k' - s\,k}\,,
\end{align}
where $s'$ runs over all four bands. The principal value is needed only for $s' > 0$, where $s'\,k' = s\,k$ can occur inside the interval.

By Eq.~\eqref{eq:rswim-P}, the integrand of $J_s$ equals $\mathcal{P}_{ss'}/[64\,k^2\,k'^2\,(s'\,k' - s\,k)]$, which has the partial-fraction form
\begin{align}
\frac{k\,k'\,\mathcal{F}_{ss'}}{s'\,k' - s\,k} = Q_{ss'} (k') + \frac{\alpha_{ss'}}{k' - s\,k/s'} + \frac{\beta_{ss'}}{k'} + \frac{\gamma_{ss'}}{k'^2}\,,
\end{align}
with a cubic polynomial $Q_{ss'}$ in $k'$. A direct computation of the coefficients gives $\alpha_{s, -s'} = \alpha_{ss'}$, $\beta_{s, -s'} = \beta_{ss'}$, and $\gamma_{s, -s'} = -\,\gamma_{ss'}$, and $Q_{ss'} + Q_{s, -s'}$ is odd in $k'$. In the sum over $\pm s'$, the double poles therefore cancel, the antiderivative of the polynomial is even in $k'$, and the simple poles combine into $\ln |k'^2 - s^2\,k^2/s'^2|$. Every term is then an even function of $k'$, and the lower limit $|k - q|$ can be replaced by $k - q$. The coefficients $\beta_{s, 1/2}$ and $\beta_{s, 3/2}$ are equal and opposite, and the logarithms of $k'$ cancel in the sum over the two final bands. For the shell functions, we obtain
\begin{align}
\label{eq:rswstat-J}
J_{1/2} (k) = {} & \frac{2\,q \left( 17\,k^2 - 9\,q^2 \right)}{9\,k} + \frac{\left( q^2 - 4\,k^2 \right) \left( 3\,q^2 - 4\,k^2 \right)^2}{32\,k^4} \ln \left| \frac{q - 2\,k}{q + 2\,k} \right| \nn
& + \frac{\left( 9\,q^2 - 16\,k^2 \right)^2 \left( 9\,q^2 - 4\,k^2 \right)}{2592\,k^4} \ln \left| \frac{\left( 3\,q + 2\,k \right) \left( 3\,q + 4\,k \right)}{\left( 3\,q - 2\,k \right) \left( 3\,q - 4\,k \right)} \right|, \nn
J_{3/2} (k) = {} & \frac{2\,q \left( q^2 - 7\,k^2 \right)}{3\,k} + \frac{\left( q^2 - 4\,k^2 \right) \left( q^4 - 20\,q^2\,k^2 + 136\,k^4 \right)}{48\,k^4} \ln \left| \frac{q - 2\,k}{q + 2\,k} \right| \nn
& + \frac{\left( q^2 - 16\,k^2 \right)^2 \left( q^2 - 4\,k^2 \right)}{96\,k^4} \ln \left| \frac{q + 4\,k}{q - 4\,k} \right|.
\end{align}
They are odd in $k$ and regular at $k \to 0$, where the poles in $k$ of the individual terms cancel.

The prefactors of the logarithms in Eq.~\eqref{eq:rswstat-J} are polynomials in $k^2$ divided by $k^4$. Each logarithm splits into terms $\ln |q + c\,k|$, which multiply only even powers of $k$, and the integral $\int dk\; k^n \ln |q + c\,k|$ is elementary for every $n \neq -1$. No dilogarithm therefore appears, in contrast to the KW node. Integration by parts gives the antiderivatives, which are finite for $k \to 0$ after the sum of all terms. We denote them by $\Phi_s (k)$. Their limits at the origin are
\begin{align}
\Phi_{1/2} (0^+) = q^3 \left( -\,\frac{8}{3} + 9 \ln 3 - \frac{5}{9} \ln 2 - \frac{1}{3} \ln q \right), \quad
\Phi_{3/2} (0^+) = q^3 \left( \frac{8}{9} - \frac{29}{9} \ln 2 - \ln q \right).
\end{align}
At the Fermi momenta $k_F^{(1/2)} = 2\,k_\mu$ and $k_F^{(3/2)} = 2\,k_\mu/3$, the arguments of all the logarithms reduce, up to constant factors, to $4\,k_\mu \pm q$, $4\,k_\mu \pm 3\,q$, and $8\,k_\mu \pm 3\,q$. The constant factors give $\ln 2$ and $\ln 3$. The coefficients of the logarithms sum to $-\,4\,q^3/3$, and the resulting $\ln q$ cancels against that of $\Phi_s (0^+)$, which leaves $\ln x$. In terms of $x = q/(4\,k_\mu)$, the doping part is
\begin{align}
\Pi_\mu (q, 0) = -\,\frac{g\,q^2}{2\,\pi^2\,v_0}\,\mathfrak{s} (x)\,, \quad
\mathfrak{s} (x) = \frac{2}{3}\,\mathfrak{p} (x) + \frac{4}{3} \ln x + \frac{16}{9} + \frac{80}{9} \ln 2 - 8 \ln 3\,,
\end{align}
with $\mathfrak{p}$ of Eq.~\eqref{eq:rswstat-p}. Adding Eq.~\eqref{eq:rswstat-pi0}, and using $\ln \left( 4\,\Lambda^2/(9\,q^2) \right) = 2 \ln \left( \Lambda/(2\,k_\mu) \right) - 2 \ln 3 - 2 \ln x$ together with the value of $\mathcal{K} (0)$, we find that the terms in $\ln x$ cancel and that the terms in $\ln 2$ cancel against those of $\mathcal{K} (0)$. The remaining constants combine to $3 \ln 3/2 - 38/9$, which gives Eq.~\eqref{eq:rswstat-pi}.

The logarithms of $\mathfrak{p}$ come in the pairs $\ln |1 \pm a\,x|$ with $a = 1$ and $3$, and $\ln |2 \pm 3\,x|$. The coefficient of the member with the lower sign is that of the member with the upper sign evaluated at $-\,x$. The part that is even in $x$ multiplies $\ln |1 - a^2\,x^2|$ or $\ln |1 - 9\,x^2/4|$, and the part that is odd in $x$ gives the functions $\mathfrak{a}$, $\mathfrak{b}$, and $\mathfrak{c}$. The coefficients of the lower-sign members contain the factors $(x - 1)^2$, $(3\,x - 1)^4$, and $(3\,x - 2)^3$, which gives the exponents $n = 2$, $4$, and $3$ stated in Sec.~\ref{secrswstat}. For $x \to 0$, the series of the logarithms gives $\mathfrak{p} (x) = 7/(9\,x^2) + 4\,x^2/5 + \mathcal{O} (x^4)$, whose first term reproduces the DOS of Eq.~\eqref{eq:rswlin-dos} through $q_{\rm TF}^2$. For $x \to \infty$, we find $\mathfrak{s} (x) = 49/(1215\,x^4) + 4546/(688905\,x^6) + \mathcal{O} (x^{-8})$, which gives Eqs.~\eqref{eq:rswstat-large} and~\eqref{eq:rswstat-eps-large}.

\subsection{Kohn singularity and Friedel oscillations}
\label{app:rswfull-friedel}

We first extract the singular parts of Eq.~\eqref{eq:rswstat-p}. The logarithm $\ln |1 - 9\,x^2/4|$ splits into $\ln |2 - 3\,x| + \ln |2 + 3\,x| - \ln 4$. The coefficients of the three logarithms with vanishing arguments are therefore $-\,7 - \mathfrak{a} (x)$ for $\ln |1 - x|$, $-\,2/3 - \mathfrak{b} (x)$ for $\ln |1 - 3\,x|$, and $20/3 - \mathfrak{c} (x)$ for $\ln |2 - 3\,x|$. They factorise as
\begin{align}
& -\,7 - \mathfrak{a} (x) = -\,\frac{\left( x - 1 \right)^2 \left( 9\,x^4 + 18\,x^3 - 18\,x^2 + 2\,x + 1 \right)}{8\,x^3}\,, \nn
& -\,\frac{2}{3} - \mathfrak{b} (x) = -\,\frac{\left( 3\,x - 1 \right)^4 \left( 9\,x^2 + 12\,x + 1 \right)}{648\,x^3}\,, \nn
& \frac{20}{3} - \mathfrak{c} (x) = \frac{\left( 3\,x - 2 \right)^3 \left( 27\,x^3 + 54\,x^2 - 9\,x - 2 \right)}{324\,x^3}\,.
\end{align}
The remaining logarithms are regular at the respective points. Dividing each coefficient by $(x - x_i)^{\ell_i}$ and setting $x = x_i$ gives the amplitudes $A_i$ of Eq.~\eqref{eq:rswfried-sing}. The rescaling $\ln |1 - 3\,x| = \ln 3 + \ln |x - 1/3|$ and the analogous one for $\ln |2 - 3\,x|$ only add analytic terms. Setting $x = x_i$ in the rest of Eq.~\eqref{eq:rswstat-p}, where the coefficient of the singular logarithm vanishes, gives the values of Eq.~\eqref{eq:rswfried-pvals}.

For the dielectric function, Eq.~\eqref{eq:epsrpa} gives $\delta \epsilon^{\rm RPA} = -\,4\,\pi\,e^2\,\delta \Pi/q^2$, and hence $\delta \left( 1/\epsilon^{\rm RPA} \right) = 4\,\pi\,e^2\,\delta \Pi/(q^2\,\epsilon^2)$. The singular part of Eq.~\eqref{eq:rswfried-dnq} is then $\delta n_{\rm sing} (q) = -\,4\,\pi\,Z\,e^2\,\Pi_{\rm sing}/(q_i^2\,\epsilon_i^2)$ to leading order in $q - q_i$. By Eq.~\eqref{eq:rswstat-pi}, $\Pi_{\rm sing} (q, 0) = -\,g\,q_i^2\,A_i\,(q - q_i)^{\ell_i} \ln |q - q_i|/\left[ 3\,\pi^2\,v_0\,(4\,k_\mu)^{\ell_i} \right]$, and the factors of $q_i^2$ cancel. This gives Eq.~\eqref{eq:rswfried-C}, with $\alpha = g\,e^2/v_0$.

The Fourier transform of $\ln |t|$ for $\kappa > 0$ is
\begin{align}
\int_{-\infty}^\infty dt\, e^{i\,\kappa\,t} \ln |t| = -\,\frac{\pi}{\kappa}\,,
\end{align}
and the multiplication by $t^{\ell}$ corresponds to the operator $\left( -\,i\,\partial_\kappa \right)^{\ell}$. Therefore
\begin{align}
\int_{-\infty}^\infty dt\, e^{i\,\kappa\,t}\,t^{\ell} \ln |t| = -\,\frac{\pi\,\ell!\,i^{\ell}}{\kappa^{\ell+1}}\,.
\end{align}
We write Eq.~\eqref{eq:rswfried-ft} as $\delta n (r) = \Im \int_0^\infty dq\, q\,e^{i\,q\,r}\,\delta n (q)/(2\,\pi^2\,r)$ and evaluate the factor $q$ at $q_i$ in the neighbourhood of each singularity. The corrections are smaller by $1/(q_i\,r)$. Each singular point then contributes
\begin{align}
\delta n_i (r) \simeq -\,\frac{C_i\,q_i\,\ell_i!}{2\,\pi\,r^{\ell_i + 2}}\,\Im \left[ i^{\,\ell_i}\,e^{i\,q_i\,r} \right].
\end{align}
Inserting $C_i$ of Eq.~\eqref{eq:rswfried-C}, $q_i = 4\,k_\mu\,x_i$, and the values of $(x_i, \ell_i, A_i)$ gives the three terms of Eq.~\eqref{eq:rswfried-dn}. For $i = 1$, for example, we have $\Im \left[ i^2\,e^{i\,q_1\,r} \right] = -\,\sin (q_1\,r)$ and $C_1 = -\,Z\,\alpha/(8\,\pi\,\epsilon_1^2\,k_\mu^2)$, which yields the first term.


\subsection{Plasmon dispersion and Landau damping}
\label{app:rswfull-wdisp}

We give the steps behind Eqs.~\eqref{eq:rswplas-pi4}--\eqref{eq:rswplas-imeps}. Throughout, $k_\mu = \mu/v_0 > 0$, $q < k_\mu/3$, and $0 < \varpi < 2\,k_\mu/3 - q/2$ with $\varpi > 3\,q/2$, so that $\varpi$ lies in the first free region of Eq.~\eqref{eq:rswim-free}.

\subsubsection*{Reduction to channels}

At $T = 0$, the occupation difference in Eq.~\eqref{eq:pol} is $f_0 (\varepsilon_s (\bs{k})) - f_0 (\varepsilon_{s'} (\bs{k}')) = f_0 (\varepsilon_s (\bs{k})) \left[ 1 - f_0 (\varepsilon_{s'} (\bs{k}')) \right] - f_0 (\varepsilon_{s'} (\bs{k}')) \left[ 1 - f_0 (\varepsilon_s (\bs{k})) \right]$. In the second term, we exchange $(s, \bs{k}) \leftrightarrow (s', \bs{k}')$ and use $\bs{k} \to -\,\bs{k} - \bs{q}$, which leaves the measure and the overlap factor invariant. This gives
\begin{align}
\Pi (\bs{q}, \omega) = g \int \frac{d^3 \bs{k}}{(2\,\pi)^3} \sum_{s, s'} \mathcal{F}_{ss'}\,f_0 (\varepsilon_s (\bs{k})) \left[ 1 - f_0 (\varepsilon_{s'} (\bs{k}')) \right] \left[ \frac{1}{\omega + \varepsilon_s (\bs{k}) - \varepsilon_{s'} (\bs{k}')} - \frac{1}{\omega - \varepsilon_s (\bs{k}) + \varepsilon_{s'} (\bs{k}')} \right].
\end{align}
Only the terms with an empty final state survive. The final band is therefore $s' \in \lbrace 1/2, 3/2 \rbrace$ with $k' > k_F^{(s')}$. The initial band is any of the four bands, and for $s \in \lbrace 1/2, 3/2 \rbrace$ it is restricted to $k < k_F^{(s)}$. The channel $3/2 \to 1/2$ requires $k < 2\,k_\mu/3$ and $k' > 2\,k_\mu$, and it is absent for $q < 4\,k_\mu/3$, in agreement with Table~\ref{tab:rswim-ranges}. In this form, no Pauli-blocked region of the filled bands appears, and no region close to the node contributes to the doped system.

We use $\Sigma = k + k'$ and $\check{D} = k' - k$ with $|\check{D}| \leq q \leq \Sigma$. The measure is $d^3 \bs{k} = \pi \left( \Sigma^2 - \check{D}^2 \right) d\Sigma\,d\check{D}/(4\,q)$, and Eq.~\eqref{eq:rswim-P} gives $\mathcal{F}_{ss'} = \mathcal{P}_{ss'}/\left( \Sigma^2 - \check{D}^2 \right)^3$ with $A = \Sigma^2 - q^2$ and $B = q^2 - \check{D}^2$. Writing $\varepsilon_s (\bs{k}) - \varepsilon_{s'} (\bs{k}') = v_0\,\Delta$, with $\Delta = \left[ \left( s - s' \right) \Sigma - \left( s + s' \right) \check{D} \right]/2$, we obtain
\begin{align}
\label{eq:rswplas-app-pi}
\Pi (q, \omega) = \frac{g}{32\,\pi^2\,v_0\,q} \sum_{s \to s'} \iint d\Sigma\,d\check{D}\; \frac{\mathcal{P}_{ss'}}{\left( \Sigma^2 - \check{D}^2 \right)^2} \left[ \frac{1}{\varpi + \Delta} - \frac{1}{\varpi - \Delta} \right],
\end{align}
where the domain of each channel follows from the conditions above. For the valence channels $s \in \lbrace -3/2, -1/2 \rbrace$, it is $\Sigma + \check{D} > 2\,k_F^{(s')}$ with $\Sigma$ up to the cut-off. For the conduction channels, it is $\Sigma - \check{D} < 2\,k_F^{(s)}$ and $\Sigma + \check{D} > 2\,k_F^{(s')}$. The denominators $\varpi \pm \Delta$ do not vanish on any domain. For the valence channels, $|\Delta| \geq \left( |s| + s' \right) k_F^{(s')}$ at the lower limit of $\Sigma$ up to terms of order $q$, which exceeds $2\,k_\mu/3$ for all four pairs. For the channel $1/2 \to 3/2$, $|\Delta| \geq 2\,k_\mu/3 - q/2$ on the domain, and $\varpi < 2\,k_\mu/3 - q/2$ is the condition for the free region.

\subsubsection*{Expansion in $q$}

We write $\check{D} = q\,x$ with $-1 \leq x \leq 1$ and denote the integrand of Eq.~\eqref{eq:rswplas-app-pi} by $G_{ss'} (\Sigma, x, q)$. For the valence channels, the lower limit of $\Sigma$ is $2\,k_F^{(s')} - q\,x$, which exceeds $q$ for small $q$. At fixed $\Sigma$ and $x$, the expansion of $G_{ss'}$ in $q$ starts at order $q^2$, and its coefficients are rational functions of $\Sigma$ with poles at $\Sigma = 0$ and at $\Sigma = 2\,\varpi/\left( |s| + s' \right)$. The latter lies below the lower limit. We write the integral over $\Sigma$ as the integral from $2\,k_F^{(s')}$ to the cut-off plus the integral from $2\,k_F^{(s')} - q\,x$ to $2\,k_F^{(s')}$, and we expand the second one in $q\,x$ with the Taylor series of the primitive. The primitives are elementary, and the integral over $x$ is a polynomial integral. The primitive of the coefficient of $q^4$ has a finite limit for $\Sigma \to \infty$, and $\Pi_4$ is therefore independent of the cut-off. The cut-off enters only the coefficient of $q^2$, where it gives the logarithm of Eq.~\eqref{eq:rswlin-eps}.

For the intraband channels $s \to s$, the conditions are $\Sigma - \check{D} < 2\,k_F^{(s)} < \Sigma + \check{D}$, and we write $\Sigma = 2\,k_F^{(s)} + q\,y$. The domain is $0 < x \leq 1$ and $|y| < x$, and $\Delta = -\,s\,q\,x$. We expand $G_{ss}$ in $q$ at fixed $x$ and $y$ and integrate over the triangle. For the channel $1/2 \to 3/2$, the domain is $4\,k_\mu/3 - \check{D} < \Sigma < 4\,k_\mu + \check{D}$, and both limits are treated as above.

\subsubsection*{Results of the channels}

The coefficient of $q^2$ of the intraband channels is $g\,k_\mu^2/(3\,\pi^2\,v_0\,\varpi^2)$ for $s = 1/2$ and $g\,k_\mu^2/(9\,\pi^2\,v_0\,\varpi^2)$ for $s = 3/2$. They are in the ratio $3 : 1$ of the Drude weights of Sec.~\ref{secrswlindrude}, and their sum is the Drude term of Eq.~\eqref{eq:rswlin-eps}. The remaining channels give the logarithms of that equation. The coefficients of $q^4$ are
\begin{align}
& \Pi_4^{(1/2 \to 1/2)} = \frac{g}{\pi^2\,v_0} \left[ \frac{k_\mu^2}{20\,\varpi^4} - \frac{1}{15\,\varpi^2} \right], \quad
\Pi_4^{(3/2 \to 3/2)} = \frac{g}{\pi^2\,v_0} \left[ \frac{3\,k_\mu^2}{20\,\varpi^4} - \frac{1}{10\,\varpi^2} \right], \nn
& \Pi_4^{(-3/2 \to 1/2)} = \frac{g}{10\,\pi^2\,v_0\,\varpi^2}\,\ln \!\left( 1 - \frac{\varpi^2}{16\,k_\mu^2} \right), \quad
\Pi_4^{(-1/2 \to 3/2)} = \frac{g}{10\,\pi^2\,v_0\,\varpi^2}\,\ln \!\left( 1 - \frac{9\,\varpi^2}{16\,k_\mu^2} \right), \quad
\Pi_4^{(-3/2 \to 3/2)} = 0\,, \nn
& \Pi_4^{(-1/2 \to 1/2)} = \frac{g}{\pi^2\,v_0} \left[ -\,\frac{1}{5\,\varpi^2}\,\ln \!\left( 1 - \frac{\varpi^2}{4\,k_\mu^2} \right) - \frac{1}{60 \left( \varpi^2 - 4\,k_\mu^2 \right)} - \frac{2\,k_\mu^2}{15 \left( \varpi^2 - 4\,k_\mu^2 \right)^2} \right], \nn
& \Pi_4^{(1/2 \to 3/2)} = \frac{g}{\pi^2\,v_0} \Bigg[ \frac{1}{10\,\varpi^2}\,\ln \!\left( \frac{4\,k_\mu^2 - \varpi^2}{4\,k_\mu^2 - 9\,\varpi^2} \right) + \frac{9}{80 \left( \varpi^2 - 4\,k_\mu^2 \right)} + \frac{9\,k_\mu^2}{10 \left( \varpi^2 - 4\,k_\mu^2 \right)^2} \nn
& \hspace{2.5cm} - \frac{9}{80 \left( 9\,\varpi^2 - 4\,k_\mu^2 \right)} - \frac{9\,k_\mu^2}{10 \left( 9\,\varpi^2 - 4\,k_\mu^2 \right)^2} \Bigg].
\end{align}
The channel $-3/2 \to 3/2$ does not contribute at this order, since its overlap factor $b^3$ is of order $q^6$ at small $q$. The sum of the seven contributions is Eq.~\eqref{eq:rswplas-pi4} with Eq.~\eqref{eq:rswplas-P}. The two intraband channels give the first two terms of $\tilde{\mathcal{P}}$, and the five interband channels give the rest. The odd powers of $q$ vanish in every channel.

\subsubsection*{Dispersion and Bohm-Gross limit}

Since $\Pi = q^2\,\Pi_2 + q^4\,\Pi_4 + \mathcal{O} \left( q^6 \right)$ and $\epsilon^{\rm RPA} = 1 - 4\,\pi\,e^2\,\Pi/q^2$, we have $\Re\epsilon^{\rm RPA} = \epsilon_0 - 4\,\pi\,e^2\,q^2\,\Pi_4 + \mathcal{O} \left( q^4 \right)$, and $4\,\pi\,e^2\,g/(240\,\pi^2\,v_0) = \alpha/(60\,\pi)$. The function $\mathcal{D}$ of Eq.~\eqref{eq:rswplas-D} follows by differentiating the real part of Eq.~\eqref{eq:rswlin-eps} with respect to $\omega^2$ and neglecting $\omega^2$ against $v_0^2\,\Lambda^2$, as in that equation. The derivatives of the two interband terms are $\left[ 4\,\alpha/(3\,\pi) \right] /\left( 4\,\mu^2 - \omega^2 \right)$ and $\left( \alpha/\pi \right) \left[ 1/\left( 4\,\mu^2/9 - \omega^2 \right) - 1/\left( 4\,\mu^2 - \omega^2 \right) \right]$, which add up to the last two terms of Eq.~\eqref{eq:rswplas-D}. Expanding $\epsilon_0$ about $\omega_*^2 (0)$ with $\epsilon_0 (\omega_*^2 (0)) = 0$ in the condition $\epsilon_0 = \alpha\,q^2\,\tilde{\mathcal{P}}/(60\,\pi)$ gives Eq.~\eqref{eq:rswplas-wdisp}. Because $\mathcal{D} > 0$, the implicit-function theorem shows that $\omega_*^2$ is analytic in $q^2$.

For $\alpha \to 0$, we use $\tilde{\mathcal{P}} \simeq 48\,k_\mu^2\,v_0^4/\omega_*^4$ and $\mathcal{D} \simeq \omega_p^2/\omega_*^4$ with $\omega_p^2 = 16\,\alpha\,v_0^2\,k_\mu^2/(9\,\pi)$ from Eq.~\eqref{eq:rswlin-wp}. Then
\begin{align}
\omega_*^2 (q) - \omega_p^2 = \frac{\alpha\,q^2}{60\,\pi}\,\frac{48\,k_\mu^2\,v_0^4}{\omega_p^2} = \frac{9}{20}\,v_0^2\,q^2\,,
\end{align}
which is Eq.~\eqref{eq:rswplas-bg}. The ratio of the sums quoted below that equation follows from $\sum_s s^{-3}\,s^4 = \sum_s s = 2$ and $\sum_s s^{-3}\,s^2 = \sum_s 1/s = 8/3$ over $s = 1/2, 3/2$, which gives $3\,v_0^2/4$.

\subsubsection*{Edges and damping}

The free region closes when $3\,q/2 = 2\,k_\mu/3 - q/2$, that is, at $q = k_\mu/3$, which gives the bound $q_{\rm max} \leq k_\mu/3$. For $q < k_\mu/3$ and $\varpi < 3\,q/2$, Table~\ref{tab:rswim-ranges} shows that the channels $3/2 \to 3/2$ and $1/2 \to 1/2$ are the only ones that can be nonzero. The remaining channels begin at $\varpi = 2\,k_\mu/3 - q/2$ or higher, which exceeds $3\,q/2$, and the channel $3/2 \to 1/2$ is absent. The channel $1/2 \to 1/2$ ends at $\varpi = q/2$. For $k_\mu/3 < q < 4\,k_\mu$, the ranges of Table~\ref{tab:rswim-ranges} overlap, as shown in Appendix~\ref{app:rswfull-im}.

For Eq.~\eqref{eq:rswplas-imeps}, we use Eq.~\eqref{eq:rswim-int} for the channel $3/2 \to 3/2$ with $q \ll k_\mu$ and $q/2 < \varpi < 3\,q/2$. The limits of Table~\ref{tab:rswim-ranges} are $k_- = 2 \left( k_\mu - \varpi \right)/3$ and $k_+ = 2\,k_\mu/3$, since the first entry of the maximum is smaller for small $q$. Over this interval of width $2\,\varpi/3$, we have $k' = k + \mathcal{O} (q)$ and $\mathcal{P}_{3/2, 3/2}/\left( k\,k' \right)^2 = A^3/\left( k\,k' \right)^2 \to 64\,k^2$ with $k = 2\,k_\mu/3$. Hence
\begin{align}
\Im \Pi = -\,\frac{g}{256\,\pi\,v_0\,q}\,\frac{2}{3}\,\frac{2\,\varpi}{3}\,\frac{256\,k_\mu^2}{9} = -\,\frac{4\,g\,k_\mu^2\,\varpi}{81\,\pi\,v_0\,q}\,,
\end{align}
and $\Im\epsilon^{\rm RPA} = -\left( 4\,\pi\,e^2/q^2 \right) \Im \Pi$ gives Eq.~\eqref{eq:rswplas-imeps}. The same result follows from the Landau damping $-\,\pi\,\rho_{3/2}\,\omega/\left( 2\,V\,q \right)$ of a 3d Fermi gas with $V = 3\,v_0/2$ and $\rho_{3/2} = 4\,g\,k_\mu^2/\left( 27\,\pi^2\,v_0 \right)$.

\section{KW at small momentum-transfers}
\label{app:kwlin}

This appendix gives the intermediate steps behind Sec.~\ref{seckwlin}, where $c > 0$ and $\chi = 1$. Throughout, $\varepsilon_s (\bs{k}) = s\,v_0\,k + c\,v_0\,k^2$ denotes the energy of the band $s = \pm 1$, and $\mathcal{R}$ and $k_F^{(s)}$ are defined in Eq.~\eqref{eq:kwlin-R}.

\subsection{Eigenstates, overlap factors, and quantum metric}
\label{app:kw-states}

The eigenstates of Eq.~\eqref{eq:kw-states} are the spin-$1/2$ eigenstates of $\hat{\bs{\tilde k}} \cdot \bs{\sigma}$. For $\mathcal{H}_{\rm KW, \chi}$, we have $\bs{\tilde k} \cdot \bs{\tilde k}' = \bs{k} \cdot \bs{k}'$ because $\chi^2 = 1$, and the overlap factors depend only on $\cos\theta = \hat{\bs{k}} \cdot \hat{\bs{k}}'$. The two spin-$1/2$ bases enclosing the angle $\theta$ are related by an SU(2) matrix with the elements $\mathfrak{u}$ and $\mathfrak{w}$ of Appendix~\ref{app:rsw-states}, which gives $|\langle s | s \rangle'|^2 = |\mathfrak{u}|^2 = \cos^2 \!\left( \theta/2 \right)$ and $|\langle -s | s \rangle'|^2 = |\mathfrak{w}|^2 = \sin^2 \!\left( \theta/2 \right)$. These are the factors of Eq.~\eqref{eq:kw-ff}. For $\bs{k}' = \bs{k} + \bs{q}$ with $q \ll k$, we have $\mathcal{F}_{+-} = \sin^2 \!\left( \theta/2 \right) \simeq \sin^2\theta/4$ with
\begin{align}
\sin^2\theta \simeq \frac{q_i\,q_j \left( \delta_{ij} - \hat k_i\,\hat k_j \right)}{k^2}\,,
\end{align}
which gives $Q^{(+-)}_{ij}$ of Eq.~\eqref{eq:kw-qmetric}. Since the pair $(+, -)$ is the only interband pair, the quantum metric of each band equals $Q^{(+-)}_{ij}$. The angular average follows from $\int d\tilde\Omega\,\hat k_i\,\hat k_j = 4\,\pi\,\delta_{ij}/3$ and $\int d\tilde\Omega = 4\,\pi$,
\begin{align}
\label{eq:kwlin-ang}
\int d\tilde\Omega\;k^2\,Q^{(+-)}_{ij} = \frac{1}{4} \left( 4\,\pi - \frac{4\,\pi}{3} \right) \delta_{ij} = \frac{2\,\pi}{3}\,\delta_{ij}\,.
\end{align}
The quadratic term does not enter the overlap factors, and the angular average is therefore independent of $c$.

\subsection{Intrinsic limit}
\label{app:kwlin-intr}

We give the steps behind Eqs.~\eqref{eq:kwlin-pi0} to~\eqref{eq:kwlin-eps0}. The band $s = -1$ has the energy $\varepsilon_{-1} (k) = v_0\,k\,(c\,k - 1)$, which is below $\mu = 0^+$ for $k < 1/c$ and above it for $k > 1/c$. The occupations are therefore $f_0 (\varepsilon_{-1}) = \Theta (1/c - k)$ and $f_0 (\varepsilon_1) = 0$ for $\mu \to 0^+$. The occupation factor in Eq.~\eqref{eq:pol} vanishes for $s = s'$ at the order considered here. For the pair $(-1, 1)$, it equals $\Theta (1/c - k)$, and the energy denominator is $\omega - 2\,v_0\,k$ at the leading order, since the quadratic term cancels in $\varepsilon_1 - \varepsilon_{-1} = 2\,v_0\,k$. For the pair $(1, -1)$, the factor is $-\,\Theta (1/c - k)$ and the denominator is $\omega + 2\,v_0\,k$. Since $\mathcal{F}_{+-}$ is of order $q^2$, we set $k' = k$ in the denominators, which requires $v_0\,q \ll \omega$ as for the RSW node.

The angular integral of Eq.~\eqref{eq:kwlin-ang} yields the prefactor $g\,q_i\,q_j\,(2\,\pi/3)\,\delta_{ij}/(2\,\pi)^3 = g\,q^2/(12\,\pi^2)$ of Eq.~\eqref{eq:kwlin-pi0}. The principal value of the radial integral is
\begin{align}
\Re \int_{0}^{1/c} dk\, \frac{4\,v_0\,k}{\omega^2 - 4\,v_0^2\,k^2} = -\,\frac{1}{2\,v_0} \left[ \ln \!\left( \left| \omega^2 - 4\,v_0^2\,k^2 \right| \right) \right]_0^{1/c} = -\,\frac{1}{2\,v_0}\,\ln \!\left( \frac{\left| 4\,v_0^2/c^2 - \omega^2 \right|}{\omega^2} \right),
\end{align}
and the imaginary part is $-\,\pi \int_0^{1/c} dk\, 4\,v_0\,k\,\delta (\omega^2 - 4\,v_0^2\,k^2) = -\,\pi/(2\,v_0)$ for $0 < \omega < 2\,v_0/c$, where the delta function fixes $k = \omega/(2\,v_0)$. Multiplying these results by $g\,q^2/(12\,\pi^2)$ gives Eq.~\eqref{eq:kwlin-pi0b}. The relations $\Re \epsilon^{\rm RPA}_{\rm inter} = 1 - 4\,\pi\,e^2\,\Re \Pi_{\rm inter}/q^2$ and $\Im \epsilon^{\rm RPA}_{\rm inter} = -\,4\,\pi\,e^2\,\Im \Pi_{\rm inter}/q^2$ then give Eq.~\eqref{eq:kwlin-eps0}, with $4\,\pi\,e^2\,g/(24\,\pi^2\,v_0) = g\,e^2/(6\,\pi\,v_0)$.

\subsection{Doped regime: DOS and Drude weights}
\label{app:kwlin-drude}

We give the steps behind Eqs.~\eqref{eq:kwlin-R}, \eqref{eq:kwlin-dos}, and~\eqref{eq:kwlin-wp}. The Fermi momenta solve $c\,k^2 + s\,k - \mu/v_0 = 0$, whose roots are $k_F^{(s)} = \left( -\,s + \mathcal{R} \right)/(2\,c)$, which is Eq.~\eqref{eq:kw-kf}. The difference of the two roots is $1/c$. The Fermi speeds are $v_s (k_F^{(s)}) = v_0 \left( s + 2\,c\,k_F^{(s)} \right) = v_0\,\mathcal{R}$ for both bands, and $\delta (\varepsilon_s - \mu) = \delta (k - k_F^{(s)})/(v_0\,\mathcal{R})$. The DOS of the band $s$ is
\begin{align}
\rho_s (\mu) = \frac{g}{(2\,\pi)^3}\,\frac{4\,\pi \left( k_F^{(s)} \right)^2}{v_0\,\mathcal{R}} = \frac{g \left( k_F^{(s)} \right)^2}{2\,\pi^2\,v_0\,\mathcal{R}}\,.
\end{align}
The sum over the two bands uses $\left( k_F^{(1)} \right)^2 + \left( k_F^{(-1)} \right)^2 = \left( \mathcal{R}^2 + 1 \right)/(2\,c^2)$ and gives $\rho$ of Eq.~\eqref{eq:kwlin-dos}. The density measured from $\mu = 0$ is the number of states in the two Fermi spheres minus the number of states of the band $s = -1$ below $k = 1/c$,
\begin{align}
\tilde n = \frac{g}{6\,\pi^2} \left[ \left( k_F^{(1)} \right)^3 + \left( k_F^{(-1)} \right)^3 - \frac{1}{c^3} \right] = \frac{g}{6\,\pi^2\,c^3} \left[ \frac{\mathcal{R} \left( \mathcal{R}^2 + 3 \right)}{4} - 1 \right],
\end{align}
which agrees with $\tilde n = \int_0^\mu d\mu'\,\rho (\mu')$, as follows from $d\mathcal{R}/d\mu = 2\,c/(v_0\,\mathcal{R})$. With $\bs{V}^{(s)} = v_0\,\mathcal{R}\,\hat{\bs{k}}$ and $\langle \hat k_i\,\hat k_j \rangle = \delta_{ij}/3$, Eq.~\eqref{eq:drude} gives
\begin{align}
D_{ij} = \pi\,e^2 \sum_s \rho_s\,v_0^2\,\mathcal{R}^2\,\frac{\delta_{ij}}{3} = \frac{g\,e^2\,v_0\,\mathcal{R}}{6\,\pi} \sum_s \left( k_F^{(s)} \right)^2 \delta_{ij} = \frac{g\,e^2\,v_0\,\mathcal{R} \left( \mathcal{R}^2 + 1 \right)}{12\,\pi\,c^2}\,\delta_{ij}\,.
\end{align}
The expansion of $\Pi_{\rm intra}$ in $q/\omega$ proceeds as in Appendix~\ref{app:rswlin-drude}, with the sum over the two bands $s = \pm 1$, and gives $\omega_p^2 = 4\,D_{ii}$ of Eq.~\eqref{eq:kwlin-wp}. For $c\,\mu/v_0 \gg 1$, we have $\mathcal{R} \simeq 2\sqrt{c\,\mu/v_0}$, which gives $\tilde n \simeq g\,\mathcal{R}^3/(24\,\pi^2\,c^3) = g\,\mu^{3/2}/[3\,\pi^2\,(c\,v_0)^{3/2}]$ and $\omega_p^2 \simeq g\,e^2\,v_0\,\mathcal{R}^3/(3\,\pi\,c^2) = 8\,\pi\,e^2\,c\,v_0\,\tilde n$.

\subsection{Interband screening and plasmons}
\label{app:kwlin-screen}

We provide the steps behind Eqs.~\eqref{eq:kwlin-eps} and~\eqref{eq:kwlin-eb} and behind the two zeros in Sec.~\ref{seckwlinscreen}, assuming $\mu > 0$, $v_0\,q \ll \omega$, and $q \ll k_F^{(1)}$. The occupations are $f_0 (\varepsilon_s) = \Theta (k_F^{(s)} - k)$ for both bands, with $k_F^{(1)} < k_F^{(-1)}$. The occupation factor of the pair $(-1, 1)$ is $f_0 (\varepsilon_{-1}) - f_0 (\varepsilon_1) = \Theta (k_F^{(-1)} - k) - \Theta (k_F^{(1)} - k)$, which equals $1$ for $k_F^{(1)} < k < k_F^{(-1)}$ and vanishes otherwise. The result of Appendix~\ref{app:kwlin-intr} therefore holds with the limits of the radial integral replaced by $k_F^{(1)}$ and $k_F^{(-1)}$. The real part becomes
\begin{align}
\Re \int_{k_F^{(1)}}^{k_F^{(-1)}} dk\, \frac{4\,v_0\,k}{\omega^2 - 4\,v_0^2\,k^2} = -\,\frac{1}{2\,v_0}\,\ln \!\left( \frac{\left| \omega^2 - \omega_{\rm max}^2 \right|}{\left| \omega^2 - \omega_{\rm th}^2 \right|} \right),
\end{align}
where $2\,v_0\,k_F^{(\pm 1)}$ have been identified as in Eq.~\eqref{eq:kwlin-onsets}. The imaginary part is $-\,\pi/(2\,v_0)$ if $k = \omega/(2\,v_0)$ lies in the window, that is, for $\omega_{\rm th} < \omega < \omega_{\rm max}$, and zero otherwise. After adding $\Pi_{\rm intra} = q^2\,D_{ii}/(\pi\,e^2\,\omega^2)$ from Appendix~\ref{app:kwlin-drude}, Eq.~\eqref{eq:epsrpa} gives Eq.~\eqref{eq:kwlin-eps}. In the static limit $\omega \to 0$, the logarithm tends to $2\,\ln \!\left( k_F^{(-1)}/k_F^{(1)} \right) = 2\,\ln \!\left( (\mathcal{R} + 1)/(\mathcal{R} - 1) \right)$, which gives Eq.~\eqref{eq:kwlin-eb}.

For the zeros of the real part, we write $\Re \epsilon^{\rm RPA} = \epsilon_{\rm inter} (\omega) - \omega_p^2/\omega^2$ with
\begin{align}
\epsilon_{\rm inter} (\omega) = 1 + \frac{g\,e^2}{6\,\pi\,v_0}\,\ln \!\left( \frac{\left| \omega^2 - \omega_{\rm max}^2 \right|}{\left| \omega^2 - \omega_{\rm th}^2 \right|} \right).
\end{align}
For $0 < \omega < \omega_{\rm th}$, the derivative of $\epsilon_{\rm inter}$ is $2\,\omega\,\left[ 1/(\omega_{\rm th}^2 - \omega^2) - 1/(\omega_{\rm max}^2 - \omega^2) \right]\,g\,e^2/(6\,\pi\,v_0)$, which is positive because $\omega_{\rm max} > \omega_{\rm th}$. The derivative of the Drude term, $2\,\omega_p^2/\omega^3$, is also positive. The real part is therefore continuous and strictly increasing on $(0, \omega_{\rm th})$. It tends to $-\,\infty$ for $\omega \to 0$ and to $+\,\infty$ for $\omega \to \omega_{\rm th}^-$, and there is exactly one zero $\omega_*$ in this interval. Since $\epsilon_{\rm inter} (\omega) \geq \epsilon_b$, we have $0 = \Re \epsilon^{\rm RPA} (\omega_*) \geq \epsilon_b - \omega_p^2/\omega_*^2$, which gives $\omega_*^2 \leq \omega_p^2/\epsilon_b$.

For $\omega > \omega_{\rm max}$, the derivative of $\epsilon_{\rm inter}$ is $2\,\omega\,\left[ 1/(\omega^2 - \omega_{\rm max}^2) - 1/(\omega^2 - \omega_{\rm th}^2) \right]\,g\,e^2/(6\,\pi\,v_0)$, which is again positive. The real part is continuous and strictly increasing on $(\omega_{\rm max}, \infty)$. It tends to $-\,\infty$ for $\omega \to \omega_{\rm max}^+$ because the logarithm diverges, and it tends to $1$ for $\omega \to \infty$. There is hence exactly one zero $\omega_*'$ in this interval. In this range, $\epsilon_{\rm inter} (\omega) < 1$, and the condition $\Re \epsilon^{\rm RPA} (\omega_*') = 0$ gives $\omega_*'^2 = \omega_p^2/\epsilon_{\rm inter} (\omega_*') > \omega_p^2$. The imaginary part of $\epsilon^{\rm RPA}$ vanishes for $\omega < \omega_{\rm th}$ and for $\omega > \omega_{\rm max}$, and both modes are therefore undamped.

\section{KW at arbitrary momentum-transfers}
\label{app:kwfull}

\subsection{Imaginary part of polarisability: Particle-hole continua and their edges}
\label{app:kwfull-im}

We give the steps behind Eqs.~\eqref{eq:kwim-edges}--\eqref{eq:kwim-plateau}. The symmetrised kernels, the variables $\Sigma$ and $\check{D}$, and the integration regions are those of Appendix~\ref{app:kwfull-re}, and we assume $2\,c\,q < 1$ and $\varpi > 0$.

For the intraband channel of the band $s$, the kernel contains $x/(\varpi^2 - x^2)$ with $x = \check{D}\,u$ and $u = s + c\,\Sigma > 0$. Its imaginary part is $-\,\pi\,x\,\delta (\varpi^2 - x^2)$, and the integral over $\check{D}$ gives $\int d\check{D}\,\check{D}\,u\,\delta (\varpi^2 - \check{D}^2 u^2) = 1/(2\,u)$ at $\check{D} = \varpi/u$. With the prefactor $g/(16\,\pi^2\,v_0\,q)$ of the intraband channels, we obtain
\begin{align}
\Im \Pi_{s,s} = -\,\frac{g}{32\,\pi\,v_0\,q} \int d\Sigma\; \frac{\Sigma^2 - q^2}{s + c\,\Sigma}\,,
\end{align}
where the integral runs over the values of $\Sigma$ for which $\check{D} = \varpi/(s + c\,\Sigma)$ lies in the allowed range $|\Sigma - 2\,k_s| < \check{D} \leq q$ with $\Sigma \geq q$. For the transitions $-1 \to 1$, the integral over $\Sigma$ at fixed $\check{D}$ gives $1/(2\,w)$ with $w = 1 + c\,\check{D}$ at $\Sigma = \varpi/w$, and we obtain
\begin{align}
\Im \Pi_{-1,1} = -\,\frac{g}{32\,\pi\,v_0\,q} \int d\check{D}\; \frac{q^2 - \check{D}^2}{1 + c\,\check{D}}\,,
\end{align}
over the values of $\check{D} \in (-q, q)$ for which $\max \lbrace q, 2\,k_1 - \check{D} \rbrace < \varpi/(1 + c\,\check{D}) < 2\,k_2 + \check{D}$.

The conditions on the integration variables are solved as follows. For the band $s = 1$, the condition $|\Sigma - 2\,k_1| < \varpi/(1 + c\,\Sigma)$ is $g_1 (\Sigma) < \varpi$ with $g_1 = |\Sigma - 2\,k_1| \left( 1 + c\,\Sigma \right)$. For $\Sigma > 2\,k_1$, the function $g_1$ increases, and the condition holds for $\Sigma < \Sigma_+^{(1)}$, where $\Sigma_+^{(1)}$ is the positive root of $c\,\Sigma^2 + (2 - \mathcal{R})\,\Sigma - 2\,k_1 - \varpi = 0$. Its discriminant is $\mathcal{R}^2 + 4\,c\,\varpi$, which gives the expression in Eq.~\eqref{eq:kwim-intra}. We used $1 - 2\,c\,k_1 = 2 - \mathcal{R}$. For $\Sigma < 2\,k_1$, the function $g_1$ is a concave parabola with the vertex at $\Sigma = (\mathcal{R} - 2)/(2\,c)$, and its roots at the level $\varpi$ are $\left( \mathcal{R} - 2 \pm \mathfrak{r}_- \right)/(2\,c)$, since $(\mathcal{R} - 2)^2 + 4\,(\mathcal{R} - 1) - 4\,c\,\varpi = \mathcal{R}^2 - 4\,c\,\varpi$. The lower end $\Sigma_a$ of the window lies to the right of the vertex. For $q \leq k_1$, we have $\Sigma_a = 2\,k_1 - q \geq k_1$, and for $q > k_1$, we have $\Sigma_a = q > k_1$, while the vertex $\left( \mathcal{R} - 2 \right)/(2\,c)$ lies below $k_1$. The function $g_1$ is hence decreasing on the window, and the condition is $\Sigma > \left( \mathcal{R} - 2 + \mathfrak{r}_- \right)/(2\,c)$. For $\varpi > \mathcal{R}^2/(4\,c)$, the parabola lies below the level $\varpi$ everywhere, and no condition arises. The remaining conditions are $\Sigma \geq q$, $\Sigma > 2\,k_1 - q$ from $\check{D} \leq q$, and $\check{D} \leq q$ itself, which reads $\Sigma \geq (\varpi - q)/(c\,q)$. Together they give $\Sigma_-^{(1)}$ of Eq.~\eqref{eq:kwim-intra}. The bound $\Sigma < 2\,k_1 + q$ is equivalent to $g_1 (2\,k_1 + q) = q \left( \mathcal{R} + c\,q \right) > \varpi$, which gives the upper edge $\varpi_+$.

For the band $s = -1$, we have $g_{-1} = |\Sigma - 2\,k_2| \left( c\,\Sigma - 1 \right)$. The right root satisfies $c\,\Sigma^2 - (\mathcal{R} + 2)\,\Sigma + 2\,k_2 - \varpi = 0$ with the discriminant $\mathcal{R}^2 + 4\,c\,\varpi$, where we used $1 + 2\,c\,k_2 = \mathcal{R} + 2$ and $(\mathcal{R} + 2)^2 - 4\,(\mathcal{R} + 1) = \mathcal{R}^2$. The left parabola has the vertex at $(\mathcal{R} + 2)/(2\,c)$, which lies to the left of the window since $c\,q < \mathcal{R}/2$. The condition on the window is then $\Sigma > \left( \mathcal{R} + 2 + \mathfrak{r}_- \right)/(2\,c)$, and $\check{D} \leq q$ gives $\Sigma \geq (q + \varpi)/(c\,q)$. The bound $\Sigma \geq q$ is automatic for $\Sigma > 2\,k_2 - q$. We obtain $\Sigma_\pm^{(-1)}$ of Eq.~\eqref{eq:kwim-intra}.

For the transitions $-1 \to 1$, the condition $\varpi < \left( 2\,k_2 + \check{D} \right) \left( 1 + c\,\check{D} \right) = c\,\check{D}^2 + (\mathcal{R} + 2)\,\check{D} + 2\,k_2$ gives $\check{D} > \left( -\,\mathcal{R} - 2 + \mathfrak{r}_+ \right)/(2\,c)$, because the right-hand side increases with $\check{D}$ for $\check{D} > -\,q$. The condition $\varpi > q \left( 1 + c\,\check{D} \right)$ gives $\check{D} < (\varpi - q)/(c\,q)$. The condition $\varpi > \left( 2\,k_1 - \check{D} \right) \left( 1 + c\,\check{D} \right)$ is equivalent to $c\,\check{D}^2 - (\mathcal{R} - 2)\,\check{D} + \varpi - 2\,k_1 > 0$, and it holds outside the interval $(\check{D}_-, \check{D}_+)$ with the roots $\check{D}_\mp$ of Eq.~\eqref{eq:kwim-inter}. For $\varpi > \mathcal{R}^2/(4\,c)$, it holds for every $\check{D}$. The vertex $\check{D} = (\mathcal{R} - 2)/(2\,c)$ can lie inside $(-q, q)$ for $\mathcal{R} < 2$, and the allowed set then consists of two segments, which gives the two terms in Eq.~\eqref{eq:kwim-inter}. The condition $\varpi > \left( 2\,k_1 - \check{D} \right) \left( 1 + c\,\check{D} \right)$ is the statement $k' > k_1$, and the excluded interval is therefore the Pauli-blocked region of the band $s = 1$.

The primitives of Eq.~\eqref{eq:kwim-G} follow from the partial fractions
\begin{align}
\frac{\Sigma^2 - q^2}{1 + c\,\Sigma} = \frac{\Sigma}{c} - \frac{1}{c^2} + \frac{1 - c^2\,q^2}{c^2 \left( 1 + c\,\Sigma \right)}\,, \quad
\frac{\Sigma^2 - q^2}{c\,\Sigma - 1} = \frac{\Sigma}{c} + \frac{1}{c^2} + \frac{1 - c^2\,q^2}{c^2 \left( c\,\Sigma - 1 \right)}\,, \quad
\frac{q^2 - \check{D}^2}{1 + c\,\check{D}} = -\,\frac{\check{D}}{c} + \frac{1}{c^2} - \frac{1 - c^2\,q^2}{c^2 \left( 1 + c\,\check{D} \right)}\,.
\end{align}
Evaluating $\mathcal{G}_0$ at $\check{D} = \pm\,q$ gives Eq.~\eqref{eq:kwim-plateau}, since the integrand is positive and the full range gives the largest absorption.

The edges follow from the extremal transition energies. For the band $s = -1$, the energy $v_0\,\check{D} \left( c\,\Sigma - 1 \right)$ increases with $k'$ at fixed $k$, because $\partial_{k'} \left[ \left( k' - k \right) \left( c\,k + c\,k' - 1 \right) \right] = 2\,c\,k' - 1 > 0$ for $k' > k_2 > 1/(2\,c)$. The maximum is then at $k' = k + q$ and $k = k_2$, which gives $q \left( \mathcal{R} + c\,q \right)$, and the minimum is zero at $k = k' = k_2$. For the band $s = 1$, the maximum is at the same configuration of $k_1$. The minimum is zero for $q \leq 2\,k_1$ and equals the energy of $k = k_1$ and $k' = q - k_1$ for $q > 2\,k_1$, which is $\left( q - 2\,k_1 \right) \left( 1 + c\,q \right)$. For the transitions $-1 \to 1$, the minimum of $\Sigma_l (\check{D}) \left( 1 + c\,\check{D} \right)$ with $\Sigma_l = \max \lbrace q, 2\,k_1 - \check{D} \rbrace$ is attained at an endpoint. The function $\left( 2\,k_1 - \check{D} \right) \left( 1 + c\,\check{D} \right)$ is concave, and its values at $\check{D} = \pm\,q$ differ by $2\,q \left( 2 - \mathcal{R} \right)$. For $q > k_1$, the branch $q \left( 1 + c\,\check{D} \right)$ increases with $\check{D}$, and the minimum is $q \left( \mathcal{R} - c\,q \right)$ at $\check{D} = 2\,k_1 - q$. The latter is smaller than $\left( 2\,k_1 + q \right) \left( 1 - c\,q \right)$ by $\left( \mathcal{R} - 1 \right) \left( 1/c - 2\,q \right) > 0$. The upper edge is the value at $\Sigma = 2\,k_2 + \check{D}$ with $\check{D} = q$.

The wavevector $q_c$ follows from $\varpi_+ = \varpi_{\rm th}$. For $\mathcal{R} \leq 2$ and $q \leq k_1$, the equation $q \left( \mathcal{R} + c\,q \right) = \left( 2\,k_1 - q \right) \left( 1 + c\,q \right)$ reduces to $c\,q^2 + q - k_1 = 0$ and has the root $q_c = \left( \sqrt{2\,\mathcal{R} - 1} - 1 \right)/(2\,c)$, which obeys $q_c \leq k_1$. For $\mathcal{R} \geq 2$, the equation $q \left( \mathcal{R} + c\,q \right) = \left( 2\,k_1 + q \right) \left( 1 - c\,q \right)$ reduces to $c\,q^2 + (\mathcal{R} - 1)\,q - k_1 = 0$ and gives the second line of Eq.~\eqref{eq:kwim-qc}. Both expressions agree at $\mathcal{R} = 2$.

Finally, for $q \to 0$ at fixed $\varpi/q < \mathcal{R}$, we have $\Sigma_\pm^{(1)} \to 2\,k_1 \pm \varpi/\mathcal{R}$, since $\mathfrak{r}_\pm \to \mathcal{R} \pm 2\,c\,\varpi/\mathcal{R}$. The integrand tends to $4\,k_1^2/\mathcal{R}$, and the width of the integration range is $2\,\varpi/\mathcal{R}$. The band $s = -1$ follows with $k_1 \to k_2$, and this gives Eq.~\eqref{eq:kwim-landau}. For the transitions $-1 \to 1$ and $2\,k_1 < \varpi < 2\,k_2$, none of the constraints restricts $\check{D}$ at $q \to 0$, and $\int_{-q}^{q} d\check{D} \left( q^2 - \check{D}^2 \right) = 4\,q^3/3$ gives $-\,g\,q^2/(24\,\pi\,v_0)$.

\subsection{Real part of polarisability and RPA dielectric function}
\label{app:kwfull-re}

We give the steps behind Eqs.~\eqref{eq:kwre-pi}--\eqref{eq:kwre-eps}. The occupations at $T = 0$ are $n_{1} (\bs{k}) = \Theta (k_1 - k)$ and $n_{-1} (\bs{k}) = \Theta (k_2 - k)$. For the band $s = -1$, the energy $\varepsilon_{-1} = v_0\,k \left( c\,k - 1 \right)$ is below $\mu$ for $k < k_2$ and above $\mu$ for $k > k_2$, since $k_2 > 1/(2\,c)$ and the band increases beyond this momentum. The substitution $\bs{k} \to -\,\bs{k} - \bs{q}$ exchanges $k$ and $k'$ and leaves the overlap factors invariant. For the pair $(-1, 1)$, the two time orderings of Eq.~\eqref{eq:pol} then combine exactly into
\begin{align}
\Pi_{-1,1} = g \int \frac{d^3 \bs{k}}{(2\,\pi)^3}\,\mathcal{F}_{-1,1} \left[ n_{-1} (\bs{k}) - n_1 (\bs{k}') \right] \frac{2\,X}{(\omega + i\,0^+)^2 - X^2}\,, \quad X = \varepsilon_1 (k') - \varepsilon_{-1} (k) = v_0\,\Sigma \left( 1 + c\,\check{D} \right).
\end{align}
The occupation difference is $n_{-1} (\bs{k}) - n_1 (\bs{k}') = \Theta (k_2 - k)\,\Theta (k' - k_1) - \Theta (k - k_2)\,\Theta (k_1 - k')$. The second term requires $k - k' > k_2 - k_1 = 1/c$ and vanishes for $q < 1/c$. For $q \geq 1/c$, the transitions from the filled states of the band $s = 1$ to the empty states of the band $s = -1$ would open, and Eq.~\eqref{eq:kwre-pi} would acquire a fourth channel. We exclude this range. For the intraband pairs, the same substitution gives
\begin{align}
\Pi_{ss} = g \int \frac{d^3 \bs{k}}{(2\,\pi)^3}\,\mathcal{F}_{ss} \left[ n_s (\bs{k}) - n_s (\bs{k}') \right] \frac{Y_s}{(\omega + i\,0^+)^2 - Y_s^2}\,, \quad Y_s = \varepsilon_s (k') - \varepsilon_s (k) = v_0\,\check{D} \left( s + c\,\Sigma \right).
\end{align}
The difference $n_s (\bs{k}) - n_s (\bs{k}')$ contains the two terms $\Theta (k_s - k)\,\Theta (k' - k_s)$ and $-\,\Theta (k - k_s)\,\Theta (k_s - k')$, where $k_s$ is the Fermi momentum of the band $s$. They are exchanged by $k \leftrightarrow k'$, and $Y_s$ is odd under this exchange. The two terms therefore give equal contributions, and we obtain $\Pi_{ss} = 2\,g \int d^3 \bs{k}/(2\,\pi)^3\,\mathcal{F}_{ss}\,\Theta (k_s - k)\,\Theta (k' - k_s)\,Y_s/(\omega^2 - Y_s^2)$. For $s = -1$, the transitions $k < k_2 < k'$ have $c \left( k + k' \right) > c\,k_2 > 1$, since $c\,k_2 = \left( \mathcal{R} + 1 \right)/2$, and $Y_{-1} > 0$ accordingly.

The measure follows from $d^3 \bs{k} = 2\,\pi\,k\,k'\,dk\,dk'/q$ on the region $|k - q| \leq k' \leq k + q$, together with $dk\,dk' = d\Sigma\,d\check{D}/2$ and $k\,k' = \left( \Sigma^2 - \check{D}^2 \right)/4$. The law of cosines gives $1 + \cos\theta = \left( \Sigma^2 - q^2 \right)/(2\,k\,k')$ and $1 - \cos\theta = \left( q^2 - \check{D}^2 \right)/(2\,k\,k')$, which yields the overlap factors of Sec.~\ref{seckwre}. Hence
\begin{align}
\frac{d^3 \bs{k}}{(2\,\pi)^3}\,\mathcal{F}_{ss} = \frac{\left( \Sigma^2 - q^2 \right) d\Sigma\,d\check{D}}{32\,\pi^2\,q}\,, \quad
\frac{d^3 \bs{k}}{(2\,\pi)^3}\,\mathcal{F}_{-1,1} = \frac{\left( q^2 - \check{D}^2 \right) d\Sigma\,d\check{D}}{32\,\pi^2\,q}\,.
\end{align}
The region is $|\check{D}| \leq q \leq \Sigma$ in every channel. The conditions on $k = (\Sigma - \check{D})/2$ and $k' = (\Sigma + \check{D})/2$ then give the following limits.
\begin{itemize}
\item For the intraband channel of the band $s = 1$, the conditions $k < k_1 < k'$ give $\check{D} > 0$ and $2\,k_1 - \check{D} < \Sigma < 2\,k_1 + \check{D}$. At fixed $\Sigma$, the range of $\check{D}$ is $|\Sigma - 2\,k_1| < \check{D} \leq q$. This requires $\max \lbrace q, 2\,k_1 - q \rbrace < \Sigma < 2\,k_1 + q$.
\item For the band $s = -1$, the same conditions with $k_2$ give $|\Sigma - 2\,k_2| < \check{D} \leq q$ and $2\,k_2 - q < \Sigma < 2\,k_2 + q$. The bound $\Sigma \geq q$ is automatic, because $k_2 > 1/c > q$.
\item For the transitions $-1 \to 1$, the conditions $k < k_2$ and $k' > k_1$ give $\max \lbrace q, 2\,k_1 - \check{D} \rbrace < \Sigma < 2\,k_2 + \check{D}$ at fixed $\check{D} \in (-q, q)$.
\end{itemize}
For the intraband channels, the integral over $\check{D}$ at fixed $\Sigma$ is
\begin{align}
\int d\check{D}\; \frac{\check{D} \left( s + c\,\Sigma \right)}{\varpi^2 - \check{D}^2 \left( s + c\,\Sigma \right)^2} = -\,\frac{1}{2 \left( s + c\,\Sigma \right)}\,\ln \!\left( \left| \varpi^2 - \check{D}^2 \left( s + c\,\Sigma \right)^2 \right| \right),
\end{align}
and the principal value is the logarithm of the absolute value. For the interband channel, the integral over $\Sigma$ at fixed $\check{D}$ is
\begin{align}
\int d\Sigma\; \frac{\Sigma \left( 1 + c\,\check{D} \right)}{\varpi^2 - \Sigma^2 \left( 1 + c\,\check{D} \right)^2} = -\,\frac{1}{2 \left( 1 + c\,\check{D} \right)}\,\ln \!\left( \left| \varpi^2 - \Sigma^2 \left( 1 + c\,\check{D} \right)^2 \right| \right).
\end{align}
The denominators $1 + c\,\Sigma$, $c\,\Sigma - 1$, and $1 + c\,\check{D}$ are positive on the ranges of integration. The first is positive since $\Sigma > 0$. For the second, $c\,\Sigma - 1 > c \left( 2\,k_2 - q \right) - 1 = \mathcal{R} - c\,q > 0$ follows from $c\,q < 1 < \mathcal{R}$. The third obeys $1 + c\,\check{D} > 1 - c\,q > 0$. Evaluating the primitives at the limits and collecting the prefactors $g/(32\,\pi^2\,v_0\,q)$, with the factor $2$ in front of each channel cancelling the factor $1/2$ of the primitives, gives Eqs.~\eqref{eq:kwre-pi} and~\eqref{eq:kwre-N}. The expression for $\Re\epsilon^{\rm RPA}$ follows from $\mathcal{V} (q)\,g/(32\,\pi^2\,v_0\,q) = \alpha/(8\,\pi\,q^3)$.

For the limit $q \to 0$, we take $q \ll k_1$ and write $\Sigma = 2\,k_1 + x$ with $|x| \leq q$ in the first channel. The prefactor tends to $4\,k_1^2/\mathcal{R}$, since $1 + 2\,c\,k_1 = \mathcal{R}$. The logarithm tends to $\left( x^2 - q^2 \right) \mathcal{R}^2/\varpi^2$. With $\int_{-q}^{q} dx \left( x^2 - q^2 \right) = -\,4\,q^3/3$, we obtain $\mathcal{N}_{1,1} \to -\,16\,k_1^2\,\mathcal{R}\,q^3/(3\,\varpi^2)$, which gives the first Drude term quoted in Sec.~\ref{seckwre}. The band $s = -1$ follows with $k_1 \to k_2$, since $2\,c\,k_2 - 1 = \mathcal{R}$. The sum is $\left( k_1^2 + k_2^2 \right) = \left( \mathcal{R}^2 + 1 \right)/(2\,c^2)$ times the common factor, in agreement with $D_{ii}$ of Eq.~\eqref{eq:kwlin-dos}. In the interband channel, the logarithm tends to $\ln \!\left( |\varpi^2 - 4\,k_2^2|/|\varpi^2 - 4\,k_1^2| \right)$ and $\int_{-q}^{q} d\check{D} \left( q^2 - \check{D}^2 \right) = 4\,q^3/3$, which gives the interband term quoted in Sec.~\ref{seckwre}. Here $\varpi^2 - 4\,k_1^2 = \left( \omega^2 - \omega_{\rm th}^2 \right)/v_0^2$ and $\varpi^2 - 4\,k_2^2 = \left( \omega^2 - \omega_{\rm max}^2 \right)/v_0^2$ with $\omega_{\rm th}$ and $\omega_{\rm max}$ of Eq.~\eqref{eq:kwlin-onsets}.

\subsection{Static polarisability and screening}
\label{app:kwfull-static}

We give the steps behind Eqs.~\eqref{eq:kwstat-pi}--\eqref{eq:kwstat-kohn}. At $\varpi = 0$, Eq.~\eqref{eq:kwre-pi} becomes
\begin{align}
\Pi (q, 0) = -\,\frac{g}{16\,\pi^2\,v_0\,q} \left( \tilde{\mathcal{J}}_{1} + \tilde{\mathcal{J}}_{-1} + \tilde{\mathcal{J}}_{0} \right),
\end{align}
with
\begin{align}
& \tilde{\mathcal{J}}_{1} = \int_{\Sigma_a}^{2\,k_1 + q} d\Sigma\; A_+ (\Sigma)\,\ln \!\left( \frac{q}{\left| \Sigma - 2\,k_1 \right|} \right), \quad
\tilde{\mathcal{J}}_{-1} = \int_{2\,k_2 - q}^{2\,k_2 + q} d\Sigma\; A_- (\Sigma)\,\ln \!\left( \frac{q}{\left| \Sigma - 2\,k_2 \right|} \right), \nn
& \tilde{\mathcal{J}}_{0} = \int_{-q}^{q} d\check{D}\; B (\check{D})\,\ln \!\left( \frac{2\,k_2 + \check{D}}{\Sigma_l} \right),
\end{align}
where $A_+ = \left( \Sigma^2 - q^2 \right)/\left( 1 + c\,\Sigma \right)$, $A_- = \left( \Sigma^2 - q^2 \right)/\left( c\,\Sigma - 1 \right)$, and $B = \left( q^2 - \check{D}^2 \right)/\left( 1 + c\,\check{D} \right)$. The factors $1 + c\,\Sigma$, $c\,\Sigma - 1$, and $1 + c\,\check{D}$ have cancelled inside the logarithms.

We first take $q \leq k_1$, so that $\Sigma_a = 2\,k_1 - q$ and $\Sigma_l = 2\,k_1 - \check{D}$. With $\Sigma = 2\,k_s + x$ and $y = \mathcal{R} + c\,x$, we have $1 + c\,\Sigma = y$ for $s = 1$ and $c\,\Sigma - 1 = y$ for $s = -1$. The numerators are $\left( 2\,k_1 + x \right)^2 - q^2 = \left[ (y - 1)^2 - c^2\,q^2 \right]/c^2$ and $\left( 2\,k_2 + x \right)^2 - q^2 = \left[ (y + 1)^2 - c^2\,q^2 \right]/c^2$. The two intraband integrals therefore combine into
\begin{align}
\tilde{\mathcal{J}}_{1} + \tilde{\mathcal{J}}_{-1} = \int_{-q}^{q} dx\,\ln \!\left( \frac{q}{|x|} \right) \left[ \frac{2\,y}{c^2} + \frac{2 \left( 1 - c^2\,q^2 \right)}{c^2\,y} \right].
\end{align}
We use $\int_{-q}^{q} dx\,\ln (q/|x|) = 2\,q$, the vanishing of the integral of $x\,\ln (q/|x|)$ over the symmetric range, and
\begin{align}
\int_{-q}^{q} dx\,\frac{\ln (q/|x|)}{\mathcal{R} + c\,x} = \frac{2\,q}{\mathcal{R}} \int_0^1 dt\,\frac{-\,\ln t}{1 - a^2\,t^2} = \frac{1}{c} \left[ \mathrm{Li}_2 (a) - \mathrm{Li}_2 (-a) \right], \quad a = \frac{c\,q}{\mathcal{R}}\,,
\end{align}
which follows from the series $\int_0^1 dt\,(-\ln t)\,t^{2n} = 1/(2\,n + 1)^2$. This gives $\tilde{\mathcal{J}}_{1} + \tilde{\mathcal{J}}_{-1} = 4\,\mathcal{R}\,q/c^2 + \left[ 2 \left( 1 - c^2\,q^2 \right)/c^3 \right] \Xi (c\,q)$.

For the transitions $-1 \to 1$, we substitute $t = 1 + c\,\check{D}$. Then $2\,k_2 + \check{D} = \left( \mathcal{R} + t \right)/c$ and $2\,k_1 - \check{D} = \left( \mathcal{R} - t \right)/c$, and $B = \left[ 2 - t - \left( 1 - c^2\,q^2 \right)/t \right]/c^2$. Hence
\begin{align}
\tilde{\mathcal{J}}_{0} = \frac{1}{c^3} \int_{1 - c\,q}^{1 + c\,q} dt \left[ 2 - t - \frac{1 - c^2\,q^2}{t} \right] \ln \left| \frac{\mathcal{R} + t}{\mathcal{R} - t} \right|.
\end{align}
The primitive of $\left( 2 - t \right)\ln \left| (\mathcal{R} + t)/(\mathcal{R} - t) \right|$ is $\mathcal{E} (t)$ of Eq.~\eqref{eq:kwstat-EX}. It follows from $\int dt\,\ln \left| (\mathcal{R} + t)/(\mathcal{R} - t) \right| = (\mathcal{R} + t) \ln \left| \mathcal{R} + t \right| + (\mathcal{R} - t) \ln \left| \mathcal{R} - t \right|$ and from an integration by parts for the term proportional to $t$. The primitive of $t^{-1} \ln \left| (\mathcal{R} + t)/(\mathcal{R} - t) \right|$ is $\Xi (t)$, since $\int du\,[\ln (1 + u) - \ln |1 - u|]/u = \Re\,\mathrm{Li}_2 (u) - \mathrm{Li}_2 (-u)$. Collecting the terms gives Eq.~\eqref{eq:kwstat-pi}.

For $q > k_1$, the limits are $\Sigma_a = q$ and $\Sigma_l = \max \lbrace q, 2\,k_1 - \check{D} \rbrace$. Relative to the expressions above, the integral $\tilde{\mathcal{J}}_{1}$ lacks the interval $\Sigma \in (2\,k_1 - q, q)$. In $\tilde{\mathcal{J}}_{0}$, the logarithm $\ln |2\,k_1 - \check{D}|$ is replaced by $\ln q$ for $\check{D} \in (2\,k_1 - q, q)$. The two corrections combine into $\int_{2\,k_1 - q}^{q} d\Sigma \left[ A_+ (\Sigma) + B (\Sigma) \right] \ln \left( |2\,k_1 - \Sigma|/q \right)$, and $A_+ + B = 0$ identically. Eq.~\eqref{eq:kwstat-pi} therefore holds for every $c\,q < 1$, with the absolute values in the logarithms and the real part of the dilogarithms for $t > \mathcal{R}$.

For small $q$, we expand $\mathcal{B} (u) = b_1\,u + b_3\,u^3 + \mathcal{O} (u^5)$ with
\begin{align}
b_1 = \frac{4 \left( \mathcal{R}^2 + 1 \right)}{\mathcal{R}}\,, \quad
b_3 = \frac{4}{3}\,\ln \!\left( \frac{\mathcal{R} + 1}{\mathcal{R} - 1} \right) - \frac{4}{\mathcal{R}} + \frac{4}{9\,\mathcal{R}^3}\,.
\end{align}
We used $d\Xi/dt = t^{-1} \ln \left| (\mathcal{R} + t)/(\mathcal{R} - t) \right|$ and the expansion of the integrals around $t = 1$. The coefficient $b_1$ gives $-\rho$ and $b_3$ gives the $q^2$ terms of Eq.~\eqref{eq:kwstat-small}.

For the Kohn anomaly, we write $s = \mathcal{R} - t$ with $t = 1 + c\,q$, so that $s = c \left( 2\,k_1 - q \right)$. The non-analytic terms of $\mathcal{B}$ come from $\mathcal{E} (1 + u)$ and $\Xi (1 + u)$. The first contains $\left[ (2 - \mathcal{R})\,s + s^2/2 \right] \ln |s|$. Near $t = \mathcal{R}$, we have $\Re\,\mathrm{Li}_2 (t/\mathcal{R}) = \pi^2/6 - \ln (t/\mathcal{R}) \ln |1 - t/\mathcal{R}| - \mathrm{Li}_2 (1 - t/\mathcal{R})$, which contributes $\left( s/\mathcal{R} + s^2/(2\,\mathcal{R}^2) \right) \ln |s|$ to $\Xi (t)$. The prefactor $1 - u^2$ equals $\mathcal{R} \left( 2 - \mathcal{R} \right) + 2 \left( \mathcal{R} - 1 \right) s - s^2$ at $u = \mathcal{R} - 1 - s$. The terms of order $s \ln |s|$ cancel, and the terms of order $s^2 \ln |s|$ add up to $-\left( \mathcal{R} - 1 \right) s^2 \ln |s|/\mathcal{R}$. Inserting this into Eq.~\eqref{eq:kwstat-pi} and using $c\,q \to \mathcal{R} - 1$ gives the coefficient in Eq.~\eqref{eq:kwstat-kohn}.

\subsection{Kohn singularity and Friedel oscillations}
\label{app:kwfull-friedel}

We give the steps behind Eqs.~\eqref{eq:kwfried-sing}--\eqref{eq:kwfried-coul}. Near $t = \mathcal{R}$, the identity $\Re\,\mathrm{Li}_2 (x) = \pi^2/6 - \ln x\,\ln |1 - x| - \mathrm{Li}_2 (1 - x)$ holds for $x > 0$ and the last term is analytic at $x = 1$. With $x = t/\mathcal{R}$, the function $\Xi (t)$ therefore contains the non-analytic term $-\ln (t/\mathcal{R})\,\ln \left| \mathcal{R} - t \right|$, and $\mathrm{Li}_2 (-t/\mathcal{R})$ is analytic. In Eq.~\eqref{eq:kwstat-pi}, the argument $t = 1 + u$ with $u = c\,q$ of $\mathcal{E}$ and $\Xi$ reaches $\mathcal{R}$ at $q = 2\,k_1$. All other terms are analytic there, since $1 - u = 2 - \mathcal{R} > 0$ for $\mathcal{R} < 2$ and $\mathcal{R} + t > 0$. The coefficient of $\ln \left| \mathcal{R} - 1 - u \right|$ in $\mathcal{B} (u)$ is hence
\begin{align}
\tilde{\mathcal{C}} (u) = \frac{1}{2} \left( \mathcal{R} - 1 - u \right) \left( 3 - \mathcal{R} - u \right) + \left( 1 - u^2 \right) \ln \!\left( \frac{1 + u}{\mathcal{R}} \right),
\end{align}
where the first term comes from $\mathcal{E} (1 + u)$ and the second from $-\left( 1 - u^2 \right) \Xi (1 + u)$. The function $\tilde{\mathcal{C}}$ obeys $\tilde{\mathcal{C}} (\mathcal{R} - 1) = \tilde{\mathcal{C}}' (\mathcal{R} - 1) = 0$ and $\tilde{\mathcal{C}}'' (\mathcal{R} - 1) = -\,2 \left( \mathcal{R} - 1 \right)/\mathcal{R}$. Since $\ln \left| \mathcal{R} - 1 - c\,q \right| = \ln c + \ln \left| q - 2\,k_1 \right|$, and the constant $\ln c$ is absorbed in $\Pi_{\rm reg}$, Eq.~\eqref{eq:kwstat-pi} gives Eq.~\eqref{eq:kwfried-sing} with $-\tilde{\mathcal{C}}$ in the prefactor. Using $\mathcal{R} - 1 - u = -\,c \left( q - 2\,k_1 \right)$, the leading term is $\left[ g/(16\,\pi^2\,v_0\,\mathcal{R}) \right] \left( q - 2\,k_1 \right)^2 \ln \left| q - 2\,k_1 \right|$, because $c^3\,q \to c^2 \left( \mathcal{R} - 1 \right)$ at $q = 2\,k_1$.

For $\epsilon_{2\,k_1}$, we evaluate Eq.~\eqref{eq:kwstat-EX} at $u = \mathcal{R} - 1$. The limit $t \to \mathcal{R}$ gives $\mathcal{E} (\mathcal{R}) = 4\,\mathcal{R}\,\ln (2\,\mathcal{R}) - \mathcal{R}^2$ and $\Xi (\mathcal{R}) = \pi^2/6 + \pi^2/12 = \pi^2/4$. At $t = 2 - \mathcal{R}$, we have $\mathcal{E} (2 - \mathcal{R}) = 2 \left( 1 + \mathcal{R} \right) \ln 2 + 2 \left( \mathcal{R} - 1 \right) \ln \left( 2 \left( \mathcal{R} - 1 \right) \right) - \mathcal{R} \left( 2 - \mathcal{R} \right)$. The prefactor is $1 - u^2 = \mathcal{R} \left( 2 - \mathcal{R} \right)$ and $c^3\,q^3 = \left( \mathcal{R} - 1 \right)^3$. Collecting the terms gives Eq.~\eqref{eq:kwfried-eps}, where the logarithms of $2$ cancel.

For the density, we use $\delta n (r) = \left( 2\,\pi^2\,r \right)^{-1} \int_0^\infty dq\,q\,\sin (q\,r)\,\delta n (q)$. The singular part is $\delta n_{\rm sing} (q) = A \left( q - q_0 \right)^2 \ln \left| q - q_0 \right| \mathcal{U} (q_0)/\epsilon_{2\,k_1}^2$ with $q_0 = 2\,k_1$ and $A = g/(16\,\pi^2\,v_0\,\mathcal{R})$. At large $r$, the integral is dominated by the neighbourhood of $q_0$, where $q \simeq q_0$ in the factor $q$. The Fourier transform of the generalised function $\ln |x|$ is $-\,\pi/|r|$ for $r \neq 0$. The Fourier transform of $x^2\,\ln |x|$ follows from two derivatives with respect to $r$ and equals $2\,\pi/|r|^3$. Hence $\int_0^\infty dq\,q\,\sin (q\,r)\,A \left( q - q_0 \right)^2 \ln \left| q - q_0 \right| \simeq 2\,\pi\,q_0\,A\,\sin (q_0\,r)/r^3$, and Eq.~\eqref{eq:kwfried-dn} follows. The corrections are of relative order $1/(k_1\,r)$. For the Coulomb potential, $\mathcal{U} (2\,k_1) = -\,\pi\,Z\,e^2/k_1^2$ gives Eq.~\eqref{eq:kwfried-coul}.

\subsection{Plasmon dispersion and Landau damping}
\label{app:kwfull-wdisp}

We give the steps behind Eqs.~\eqref{eq:kwplas-pi24}--\eqref{eq:kwplas-bg}. We work at fixed $\varpi$ with $\varpi \gg \mathcal{R}\,q$ and $q < k_1$, so that $\Sigma_a = 2\,k_1 - q$ and $\Sigma_l = 2\,k_1 - \check{D}$ in Eq.~\eqref{eq:kwre-N}.

For the intraband terms, we use the shift $\Sigma = 2\,k_s + x$ and $y = \mathcal{R} + c\,x$, as in Appendix~\ref{app:kwfull-static}. The two channels have the same logarithm and combine into
\begin{align}
\mathcal{N}_{1,1} + \mathcal{N}_{-1,-1} = \frac{2}{c^2} \int_{-q}^{q} dx \left[ y + \frac{1 - c^2\,q^2}{y} \right] \ell (x), \quad
\ell (x) = \ln \left| \frac{\varpi^2 - q^2\,y^2}{\varpi^2 - x^2\,y^2} \right| = -\sum_{n = 1}^{\infty} \frac{\left( q^{2\,n} - x^{2\,n} \right) y^{2\,n}}{n\,\varpi^{2\,n}}\,.
\end{align}
The terms with $n \geq 3$ contribute at order $q^7$ and beyond. The integrals needed are $\int_{-q}^{q} dx \left( q^2 - x^2 \right) = 4\,q^3/3$, $\int_{-q}^{q} dx \left( q^2 - x^2 \right) x^2 = 4\,q^5/15$, and $\int_{-q}^{q} dx \left( q^4 - x^4 \right) = 8\,q^5/5$, while odd powers of $x$ vanish. With Eq.~\eqref{eq:kwre-pi}, the intraband parts are
\begin{align}
\Pi_2^{\rm intra} = \frac{g\,\mathcal{R} \left( \mathcal{R}^2 + 1 \right)}{12\,\pi^2\,v_0\,c^2\,\varpi^2}\,, \quad
\Pi_4^{\rm intra} = \frac{g}{120\,\pi^2\,v_0} \left[ \frac{6\,\mathcal{R}^3 \left( \mathcal{R}^2 + 1 \right)}{c^2\,\varpi^4} - \frac{4\,\mathcal{R}}{\varpi^2} \right].
\end{align}

For the transitions $-1 \to 1$, we write $\ln \left| P (\check{D}) \right| - \ln \left| Q (\check{D}) \right|$ with $P = \varpi^2 - \left( 2\,k_2 + \check{D} \right)^2 \left( 1 + c\,\check{D} \right)^2$ and $Q = \varpi^2 - \left( 2\,k_1 - \check{D} \right)^2 \left( 1 + c\,\check{D} \right)^2$. Their expansions are $\ln \left| P (\check{D}) \right| = \ln \left| P_0 \right| + p_1\,\check{D} + p_2\,\check{D}^2 + \dots$, and similarly for $Q$ with $r_1$ and $r_2$, where $P_0 = \varpi^2 - 4\,k_2^2$ and $Q_0 = \varpi^2 - 4\,k_1^2$. The first-order coefficients are
\begin{align}
p_1 = -\,\frac{4\,k_2 \left( \mathcal{R} + 2 \right)}{\varpi^2 - 4\,k_2^2}\,, \quad
r_1 = -\,\frac{4\,k_1 \left( \mathcal{R} - 2 \right)}{\varpi^2 - 4\,k_1^2}\,,
\end{align}
and the second-order coefficients follow from the second derivatives. The weight is $\left( q^2 - \check{D}^2 \right)/\left( 1 + c\,\check{D} \right) = \left( q^2 - \check{D}^2 \right) \left( 1 - c\,\check{D} + c^2\,\check{D}^2 - \dots \right)$. The integrals $\int_{-q}^{q} d\check{D} \left( q^2 - \check{D}^2 \right) \left( 1 + c^2\,\check{D}^2 \right) = 4\,q^3/3 + 4\,c^2\,q^5/15$, $\int_{-q}^{q} d\check{D}\,\check{D} \left( q^2 - \check{D}^2 \right)/\left( 1 + c\,\check{D} \right) = -\,4\,c\,q^5/15 + \mathcal{O} (q^7)$, and $\int_{-q}^{q} d\check{D}\,\check{D}^2 \left( q^2 - \check{D}^2 \right) = 4\,q^5/15$ give
\begin{align}
\mathcal{N}_{-1,1} = \ln \left| \frac{P_0}{Q_0} \right| \left[ \frac{4\,q^3}{3} + \frac{4\,c^2\,q^5}{15} \right] - \frac{4\,c\,q^5}{15} \left( p_1 - r_1 \right) + \frac{4\,q^5}{15} \left( p_2 - r_2 \right) + \mathcal{O} (q^7).
\end{align}
Inserting this together with the intraband terms into Eq.~\eqref{eq:kwre-pi} gives Eq.~\eqref{eq:kwplas-pi24}. The $q^2$ term of the transitions $-1 \to 1$ is the logarithm of $\Pi_2$, and the $q^4$ term is the sum of the last four terms of $\tilde{\mathcal{P}}$.

For the dispersion, we write $\Re\epsilon^{\rm RPA} (q, \omega) = \epsilon_0 (\omega) - 4\,\pi\,e^2\,q^2\,\Pi_4 (\omega/v_0) + \mathcal{O} (q^4)$. The zero satisfies $\epsilon_0 \left( \omega_*^2 (q) \right) = 4\,\pi\,e^2\,q^2\,\Pi_4$, and expanding $\epsilon_0$ about $\omega_*^2 (0)$ with $\epsilon_0 (\omega_*^2 (0)) = 0$ gives $\omega_*^2 (q) - \omega_*^2 (0) = 4\,\pi\,e^2\,q^2\,\Pi_4/\mathcal{D}$. With $4\,\pi\,e^2\,g/(120\,\pi^2\,v_0) = \alpha/(30\,\pi)$, this is Eq.~\eqref{eq:kwplas-wdisp}. The derivative of the logarithm in $\epsilon_0$ gives the bracket of $\mathcal{D}$. For the Bohm-Gross limit, we keep the first term of $\tilde{\mathcal{P}}$ and $\mathcal{D} = \omega_p^2/\omega_*^4$. With $\varpi = \omega_*/v_0$ and $\omega_p^2 = \alpha\,v_0^2\,\mathcal{R} \left( \mathcal{R}^2 + 1 \right)/(3\,\pi\,c^2)$, we obtain $\omega_*^2 (q) - \omega_p^2 = \left( 3/5 \right) v_0^2\,\mathcal{R}^2\,q^2$.


\end{document}